\documentclass[11pt,dvipsnames]{article}
\usepackage{caption}
\usepackage{scalerel}
\usepackage{amsthm,latexsym,amsxtra,graphicx,appendix,epstopdf,feynmf,hyperref,setspace,fix-cm}
\usepackage[normalem]{ulem}
\usepackage{makeidx}
\usepackage{fullpage}
\usepackage{amsmath}
\usepackage{amssymb}
\usepackage{setspace}
\usepackage{bbm}
\usepackage{dsfont}
\usepackage{epsfig}
\usepackage[font=footnotesize,labelfont=bf,justification=centerlast,width=.94\textwidth]{caption}
\usepackage{cite}
 \usepackage{multirow}
\usepackage{array,booktabs}
\usepackage{subfigure}
\usepackage{mathtools}
\usepackage{color}
\usepackage[makeroom]{cancel}
\usepackage{tikz}
\usepackage{pgfplots}
\usepackage{bm}
\pgfplotsset{compat=1.18}

\usepackage{tensor}
\usepackage{simplewick}
\usepackage{frcursive}

\usepackage{pgfornament,multicol}

\usepackage{caption}
\usepackage{hyperref}

\usepackage[skip=0.2cm, indent=0pt]{parskip}

\hypersetup{
    colorlinks,%
    citecolor=blue,%
    filecolor=blue,%
    linkcolor=blue,%
    urlcolor=blue
}
\usepackage{float}
\usepackage{slashed}
\usepackage{tikz}
\usetikzlibrary{decorations.pathmorphing}

\usepackage{tcolorbox}
\usepackage{tabularx}
\usepackage{array}
\usepackage{colortbl}
\tcbuselibrary{skins}

\tcbset{tab2/.style={enhanced,fonttitle=\bfseries,fontupper=\normalsize\sffamily,
colback=Goldenrod!20!white,colframe=Blue!50!black,colbacktitle=Goldenrod!50!white,
coltitle=black,center title}}

\def\p{\partial}

\def\r{\rightarrow}

\def\b{\beta}

\newcommand{\bi}{\begin{itemize}}
\newcommand{\ei}{\end{itemize}}
\newcommand{\bea}{\begin{eqnarray}}
\newcommand{\eea}{\end{eqnarray}}

\def\={\, = \,}
\def\rh{r_{\text{h}}}

\def\bomega{\bm{\omega}}
\def\r{\mathfrak{r}}

\def\b{\tilde\beta}

\def\r{\mathfrak{r}}

\def\bomega{{\boldsymbol{\omega}}}

\def\XXint#1#2#3{{\setbox0=\hbox{$#1{#2#3}{\int}$}
     \vcenter{\hbox{$#2#3$}}\kern-.5\wd0}}

\def\={\, = \,}

\newcommand{\beq}{\begin{equation}}

\newcommand{\eeq}{\end{equation}}

\usepackage{tikz}
\usetikzlibrary{positioning}
\usetikzlibrary{intersections}
\usetikzlibrary{fadings} 
\usetikzlibrary{arrows.meta} 
\usetikzlibrary{arrows}

\tikzfading[name=fade out,
inner color=transparent!0,
outer color=transparent!100]

\definecolor{cherryblossompink}{rgb}{1.0, 0.72, 0.77}
\definecolor{lightblue}{rgb}{0.68, 0.85, 0.9}

\usetikzlibrary{decorations.pathmorphing}
\usetikzlibrary{decorations.pathreplacing,decorations.markings}

\usetikzlibrary{backgrounds,automata}

\newsavebox\CBox
\newcommand\hcancel[2][0.5pt]{%
  \ifmmode\sbox\CBox{$#2$}\else\sbox\CBox{#2}\fi%
  \makebox[0pt][l]{\usebox\CBox}%
  \rule[0.75\ht\CBox-#1/2]{\wd\CBox}{#1}}

\numberwithin{equation}{section}
\allowdisplaybreaks  

\begin{document}

\vspace*{2.5cm}
\begin{center}
{ \Large \textsc{Undulating Conformal Boundaries in 3D Gravity} }\\ \vspace*{0.5cm}

\vspace*{1cm}
\end{center}

\begin{center}
{\small Weam Abou Hamdan$^1$ and  Chawakorn Maneerat$^{1,2}$}
\end{center}
\begin{center}
{
\footnotesize
\vspace{0.2cm}
$^1$ Department of Mathematics, King's College London, Strand, London WC2R 2LS, United Kingdom\\
$^2$ Institute for Theoretical Physics, Universiteit van Amsterdam,
Science Park 904, 1098 XH, The Netherlands}
\end{center}
\begin{center}
{\nolinkurl{weam.abou\_hamdan@kcl.ac.uk}, \nolinkurl{c.maneerat@uva.nl}} 
\end{center}

\vspace*{0.5cm}

\vspace*{1.5cm}
\begin{abstract}
\noindent 
We consider three-dimensional Einstein gravity in Euclidean signature with a finite boundary of torus topology endowed with an induced metric of fixed conformal class and a constant trace of extrinsic curvature $K$. For vanishing, positive, and negative cosmological constant $\Lambda$, we analytically determine boundaries enclosing different patches of locally flat, de Sitter (dS$_3$), and Anti-de Sitter (AdS$_3$) spaces. We find solutions that depend non-trivially on either cycle of the torus, noting that some of them exhibit self-intersections. Adapting the Gibbons-Hawking prescription of interpreting the Euclidean gravitational path integral as a thermal partition function, we explore the rich semi-classical thermodynamic phase space of the problem. While most saddles are found to be either thermally unstable or metastable compared to those with uniform boundaries, we find inhomogeneous solutions that are thermodynamically favourable in the case of $\Lambda < 0$ and $2<K|\Lambda|^{-1/2}<3/\sqrt{2}$. Moreover, for all values of $\Lambda$, there exist patches of space with a non-contractible thermal circle and a macroscopic entropy. We comment on a recasting of the problem in terms of classical strings. We further analyse the problem in both the stretched dS$_3$ horizon limit and the AdS$_3$ boundary limit. Using the results for the latter, we find limitations to a proposed holographic dual that consists of a two-dimensional holographic conformal field theory coupled to timelike Liouville, deformed by a marginal $T\bar{T}$-type operator.

\bigskip

\end{abstract}

\newpage

\setcounter{tocdepth}{2}
\tableofcontents

\newpage

\section{Introduction}

The choice of boundary conditions in gravity has attracted considerable recent interest, particularly in relation to questions in quantum gravity. In this paper, we focus on conformal boundary conditions, whereby one fixes the conformal class of the induced metric and the trace of the extrinsic curvature $K$ at the boundary. In Euclidean signature, these boundary conditions define an elliptic problem for Einstein gravity, allowing for a sensible construction of perturbation theory, in contrast to Dirichlet boundary conditions, where the full induced metric is fixed \cite{Anderson:2006lqb}; see also \cite{Witten:2018lgb}. In Lorentzian signature, the choice of boundary conditions that leads to a mathematically well-posed initial boundary value problem for general relativity is a topic under active investigation\cite{Fournodavlos:2020wde,An:2021fcq,Fournodavlos:2021eye,Anninos:2023epi,Anninos:2024wpy,Anninos:2024xhc,An:2025rlw,An:2025gvr,Liu:2025xij}.

One main application of gravity with finite boundaries is black hole thermodynamics from a quasi-local perspective. Starting with \cite{York:1986it}, it was shown that asymptotically flat Schwarzschild black holes in four dimensions can be thermally stabilised by enclosing the horizon with a finite Dirichlet boundary. For positive cosmological constant $\Lambda>0$, de Sitter thermodynamics with Dirichlet boundaries was studied in \cite{Wang:2001gt,Draper:2022ofa,Banihashemi:2022jys,Banihashemi:2022htw}, where the cosmological horizon was shown to be thermally unstable. Building on this setup, microscopic countings of the de Sitter horizon entropy were proposed in \cite{Coleman:2021nor,Batra:2024kjl,Silverstein:2024xnr}. More recently, thermodynamics of various gravitational systems in the framework of the Euclidean gravitational path integral with conformal boundary conditions has been explored in \cite{Anninos:2023epi,Anninos:2024wpy,Banihashemi:2024yye,Banihashemi:2025qqi,Galante:2025tnt,Galante:2025emz,Anninos:2025fer,Marini:2026zjk}. Some of these works consider boundaries of topology $S^1\times S^{d-1}$, endowed with the conformal structure of a round spatial sphere times a thermal circle, together with constant $K$. The semi-classical thermal partition function is then computed from static, spherically symmetric Euclidean saddles, yielding thermodynamics with boundary extensivity at high temperature \cite{Banihashemi:2024yye,Banihashemi:2025qqi}. In three dimensions, the thermal partition functions for both $\Lambda>0$ \cite{Anninos:2024wpy} and $\Lambda<0$ \cite{Allameh:2025gsa} take the form of two-dimensional conformal field theory partition functions. In three and four dimensions, conformal boundary conditions can render cosmological horizons thermally stable \cite{Anninos:2024wpy}, perhaps allowing for a microscopic description of de Sitter space from a quasi-local point of view \cite{Anninos:2011af,Anninos:2012qw,Anninos:2022ujl,Galante:2023uyf}.

In parallel, Einstein gravity with conformal boundary conditions in Lorentzian signature has been studied in a variety of settings; see, for instance, \cite{Anninos:2011zn,Anninos:2023epi,Anninos:2024wpy,Liu:2024ymn,Anninos:2024xhc,Liu:2025xij,Galante:2025emz,Anninos:2025zgr}. A novel dynamical feature is the existence of boundary degrees of freedom, encoded in the dynamical boundary Weyl factor $\bomega$. In $d+1$ bulk dimensions, its equation of motion is given by the radial Hamiltonian constraint evaluated at the boundary \cite{Anninos:2024xhc},
\begin{equation}\label{eqn: weyl eqn}
    \mathcal{D}^m \mathcal{D}_m \bomega 
    - \frac{d-2}{2}\mathcal{D}_m \bomega \mathcal{D}^m \bomega 
    - \frac{R}{2(d-1)}
    -\frac{64\pi^2G_N^2}{2(d-1)}T^{mn}T_{mn}e^{-2(d-1)\bomega}
    +\left(K^2+\frac{2d}{d-1}\Lambda\right)\frac{e^{2\bomega}}{2d}=0\, ,
\end{equation}
where $T_{mn}$ is the analogue of the Brown--York tensor for conformal boundary conditions, satisfying tracelessness and conservation equations \cite{Odak:2021axr}. We will describe this equation in more detail below. In situations where the bulk geometry is spherically symmetric, the spherical sector of \eqref{eqn: weyl eqn} becomes a dynamical equation for the boundary spatial radius \cite{Anninos:2024wpy,Liu:2024ymn,Anninos:2024xhc,Galante:2025tnt}. The late-time behaviour of the solutions for Minkowski-filling geometries was analysed in \cite{Liu:2024ymn}. More generally, the linearised solutions include exponentially growing modes generated by physical diffeomorphisms. 

As observed in \cite{Anninos:2024wpy,Galante:2025tnt}, such exponentially growing Lorentzian solutions can have Euclidean counterparts that oscillate along the thermal circle. This motivates the question we address here: do inhomogeneous Euclidean saddles contribute to the thermal partition function, and how do they affect the thermodynamics?

We investigate this question in pure three-dimensional Einstein gravity with a boundary of torus topology, where the problem is particularly tractable. We impose that the induced metric is conformally flat and that the trace of the extrinsic curvature is constant. We then construct thermal saddles that are either \textit{non-static} or \textit{non-circular}, and compute their thermodynamic properties.\footnote{Analogous phenomena may occur in non-gravitational systems, where thermal solutions may spontaneously break the thermal $U(1)$ isometry or a spatial $U(1)$ isometry. For instance, in quantum mechanics with a wrong-sign kinetic term and a harmonic potential, equivalently with a standard kinetic term and an inverted potential, thermal solutions can break the thermal $U(1)$ symmetry to a discrete subgroup. 
See \cite{Lewkowycz:2013nqa} for an example of spontaneous breaking of thermal $U(1)$ symmetry in the context of holography. On the other hand, thermodynamic systems with spontaneously broken spatial isometries are of significant phenomenological relevance in condensed matter physics, and an attempt to construct holographic dual gravitational descriptions of such systems was initiated in \cite{Vegh:2013sk}.}

A further motivation for working in three dimensions comes from the recently proposed duality \cite{Allameh:2025gsa} between AdS$_3$ with finite boundaries obeying conformal boundary conditions and a holographic CFT$_2$ coupled to timelike Liouville theory (as analysed in \cite{Anninos:2021ene,Anninos:2024iwf,Chatterjee:2025yzo}), deformed by an exactly marginal operator of $T\bar T$ type \cite{Smirnov:2016lqw,Cavaglia:2016oda,McGough:2016lol}. The existence of inhomogeneous saddles and their on-shell action may be used as a test for this proposal.

\textbf{Organization and summary of results}

In section \ref{sec: general framework}, we present the general framework and provide a treatment of finding boundaries obeying conformal boundary conditions via the embedding method. As we focus primarily on thermodynamic properties, boundaries are chosen to have topology $S^1\times S^1$. On the boundary, we fix the induced metric to be conformally related to a flat torus metric with a conformally invariant periodicity $\b$, a constant trace of the extrinsic curvature $K$, and vanishing angular potential. We refer to $\b^{-1}$ as the conformal temperature. We allow non-trivial boundary profiles. We call those that vary along the thermal circle \textit{non-static}, while those that vary along the spatial circle \textit{non-circular}. 

Sections \ref{sec: Lambda 0}, \ref{sec: Lambda +}, and \ref{sec: Lambda -} consider the problem for zero, positive, and negative cosmological constants, respectively. In each section, we review previously studied static and circular solutions and present closed-form solutions for a novel family of non-static or non-circular boundaries. We investigate the parametric regimes for which these solutions exist and discuss their geometric and thermodynamic properties. At the end of each section, we analyse the thermodynamic phase space of the system. Note that we also find self-intersecting solutions for all values of $\Lambda$, but we do not include them in the thermodynamic analysis.\footnote{We note that, in the context of AdS-JT gravity with finite cutoff, the absence of self-intersection condition was imposed in \cite{Iliesiu:2020zld,Stanford:2020qhm} as part of the definition of the gravitational path integral.} In the negative cosmological constant case, we further analyse the limit in which the boundary approaches the asymptotic boundary of AdS$_3$, and test the holographic proposal in \cite{Allameh:2025gsa} by comparing the resulting torus partition function with the gravity result.

Appendices \ref{sec: Ellip Jacobi}, \ref{sec: du T flat 1}, and \ref{sec: self-intersection} provide further technical details of the computations in the main text.

Here is a summary of the main results found in this paper.

\textbf{Zero cosmological constant.} Here, we find a class of non-static pole patch solutions with non-zero conformal entropy. The number of these solutions at a given value of $K>0$ diverges as we take the conformal inverse temperature $\b$ to infinity, see equation \eqref{Lambda=0 number of non-static pole sols}. We also consider patches bounded by a compact Rindler horizon, where we find non-circular solutions with an entropy that matches the Bekenstein-Hawking formula. These two classes of solutions are related by a modular transformation. Moreover, all inhomogeneous solutions in the case of $\Lambda=0$ are thermally unstable, and the most thermodynamically favourable saddle is always the uniform one, transitioning from the Rindler homogeneous boundary to the pole patch homogeneous boundary at the critical temperature $\b_c=2\pi$, see figure \ref{fig: Ionshell flat}. Note that there are no solutions with $K<0$.

\textbf{Positive cosmological constant.} The three-sphere background with $\Lambda=+\frac{1}{\ell^2}$ is the only case with boundaries with either sign of $K\ell$. In the case of $K\ell>0$, the solutions are qualitatively similar to those in flat space. As for the case of $K\ell<0$, the number of solutions given some boundary data is determined by the value of $K\ell$, independently of the value of $\b$, see equation \eqref{eqn: number sol Lambda>0}. In this case, there exist metastable inhomogeneous solutions with positive specific heat, but their on-shell action is dominated by that of the solution with a homogeneous boundary, see figure \ref{fig: Ionshell dS KL<0}. Moreover, some of these solutions admit a stretched horizon limit as $K\ell \rightarrow -\infty$, whereby the boundary oscillates very close to the horizon, see section \ref{sec: stretch}. For all signs of $K\ell$, modular invariance of the partition function holds.

\textbf{Negative cosmological constant.} In the case of hyperbolic space with $\Lambda=-\frac{1}{\ell^2}$, there exist solutions only for $K\ell>0$. The most striking feature here is the existence of thermally stable saddles whose on-shell action is more thermodynamically favourable than that of the homogeneous solution with the same boundary data, see figure \ref{fig: Ionshell AdS K<3/sqrt{2}}. This occurs in the parametric regime $2<K\ell<3/\sqrt{2}$. Once again, we check modular invariance $\mathcal{Z}(\b,K)=\mathcal{Z}(\frac{4\pi^2}{\b},K)$. We also comment on the limit $K\ell \rightarrow 2^+$, where the boundary oscillates near the infinite conformal boundary of AdS, see section \ref{sec: AdS limit}. We then use these results to test the proposal for a holographic dual in \cite{Allameh:2025gsa}, which is defined by a two-dimensional conformal field theory coupled to timelike Liouville and deformed by a marginal $T\bar T$-type operator. By comparing the on-shell action of the inhomogeneous bulk saddles with that of their counterparts in the boundary theory, we find a mismatch in the case where the boundary radius has large oscillations, which on the boundary theory means that the effective deformation parameter in the boundary theory is not parametrically small, see section \ref{sec: holographic comparison}.

\section{General framework}\label{sec: general framework}

In this section, we provide a general framework for three-dimensional Einstein gravity in Euclidean signature with finite boundaries obeying conformal boundary conditions.

\subsection{Gravity with conformal boundary conditions}\label{sec: GR CBC}

We consider general relativity in Euclidean signature with an arbitrary cosmological constant $\Lambda$ in three dimensions. Let $\mathcal{M}$ be a spacetime manifold with a boundary $\Gamma\equiv \p\mathcal{M}$. The Euclidean action is given by
\begin{equation}\label{eqn: action}
    I = -\frac{1}{16 \pi G_N}\int_\mathcal{M} d^3 x \sqrt{\det g_{\mu\nu}} \left(R-2\Lambda\right)-\frac{1}{16 \pi G_N}\int_\Gamma d^2\sigma \sqrt{\det g_{ij}}\,K\,,
\end{equation}
where $G_N$ is Newton's constant. The manifold $\mathcal{M}$ is endowed with a metric $g_{\mu\nu}$, and $g_{ij}$ is the induced metric on $\Gamma$. We adopt the notation in which the Greek indices $\mu,\nu,\dots=1,2,3$ refer to bulk indices, while Latin indices $i,j,\dots=1,2$ are used for boundary indices. We use $x^\mu$ to denote the bulk coordinate, while $\sigma^i$ refers to boundary coordinates. The trace $K$ of the extrinsic curvature $K_{ij}$ of the boundary is defined as
\begin{equation}\label{eqn: def K}
    K \equiv g^{ij}K_{ij} \, , \qquad K_{ij} \equiv \frac{1}{2}\mathcal{L}_{n^\mu} g_{ij} \, ,
\end{equation}
where $n^\mu$ is the unit normal vector of the boundary, satisfying $n^\mu n_\mu=1$, taken to be outward-pointing. For a non-vanishing cosmological constant, we take $\Lambda=\pm \tfrac{1}{\ell^2}$, where $\ell$ denotes the dS (AdS) characteristic length for a positive (negative) value of $\Lambda$. 

At the boundary, we impose conformal boundary conditions, defined as holding fixed both the conformal structure of the induced metric and the trace of the extrinsic curvature of the boundary,
\begin{equation}\label{eqn: def conf BCs}
    \text{Conformal boundary conditions}\quad : \quad \{[\left.g_{ij}\right|_{\Gamma}]_{\text{conf}}\, , \left.K\right|_{\Gamma}\} \, , \qquad \text{fixed} \, .
\end{equation}
The boundary term in \eqref{eqn: action} leads to a well-defined variational principle upon imposing conformal boundary conditions \cite{An:2021fcq , Odak:2021axr}. In three dimensions, this term is a half of the standard Gibbons-Hawking-York term for Dirichlet boundary conditions.

In practice, it is useful to work with a conformal representative of the chosen conformal structure $[g_{ij}]_{\text{conf}}$, which we denote as $\tilde g_{ij}$. As such, the induced metric on the boundary is written as
\begin{equation}\label{eqn: ind metr condition}
    \left.ds^2\right|_{\Gamma} = g_{ij}d\sigma^i d\sigma^j = e^{2\bomega} \tilde g_{ij} d\sigma^i d\sigma^j \, ,
\end{equation}
where $\bomega$ is the Weyl factor of the boundary associated with $\tilde g_{ij}$. Following the Brown-York prescription of defining a quasi-local stress tensor \cite{Brown:1992br}, varying the action \eqref{eqn: action} on-shell with respect to $\tilde g_{ij}$ leads to the conformal stress tensor $T_{ij}$, defined as
\begin{equation}\label{eqn: def conf Tij}
    T_{ij}\equiv \left.- \frac{1}{8\pi G_N}\left(K_{ij}-\frac{1}{2}g_{ij} K\right)\right|_{\Gamma}\, .
\end{equation}
It follows from this definition that $T_{ij}$ is traceless with respect to $\tilde g_{ij}$. 

The equations of motion arising from the action in \eqref{eqn: action} are given by the Einstein field equations,
\begin{equation}\label{eqn: Einstein eqn}
    R_{\mu\nu}-\frac{1}{2}Rg_{\mu\nu}+\Lambda g_{\mu\nu}=0 \, .
\end{equation}
The projection of the Einstein equations on the boundary leads to the radial Hamiltonian and momentum constraint equations. In terms of the data $(\tilde g_{ij},K, \bomega, T_{ij})$, the Hamiltonian constraint is given by the Lichnerowicz-York equation \cite{York:1972sj, Lichnerowicz:1944, Choquet-Bruhat:2009xil}, 
\begin{equation}\label{eqn: Hamil constr}
    \tilde{\mathcal{D}}^i \tilde{\mathcal{D}}_i \bomega - \frac{\tilde{\mathcal{R}}}{2}-32 \pi^2 G_N^2T^{ij}T_{ij} \,e^{-2\bomega}+\frac{1}{4}\left(K^2+4\Lambda\right) e^{2\bomega} =0\,,
\end{equation}
where $\tilde{\mathcal{D}}_i$ and $\tilde{\mathcal{R}}$ are, respectively, the covariant derivative and the Ricci scalar built from $\tilde g_{ij}$, and the indices are raised and lowered using $\tilde g_{ij}$. In the case of a constant $K$, the momentum constraint is given by
\begin{equation}\label{eqn: Momentum constr}
    \tilde{\mathcal{D}}^iT_{ij} = 0\, ,
\end{equation}
which implies that $T_{ij}$ is covariantly conserved with respect to $\tilde g_{ij}$. 

We also note that the Ricci scalar $\mathcal{R}$ of the induced metric $g_{ij}$ is given by $\mathcal{R}= e^{-2\bomega}(\tilde{\mathcal{R}}-2\tilde{\mathcal{D}}^i\tilde{\mathcal{D}}_i \bomega )$. Using \eqref{eqn: Hamil constr}, we find
\begin{equation}
    \mathcal{R}= -64 \pi^2 G_N^2 T^{ij}T_{ij}e^{-4\bomega}+\frac{1}{2}(K^2+4\Lambda) \, .
\end{equation}
We note that for $T_{ij}=0$, the boundary has a constant intrinsic curvature.

\subsection{Conformal thermodynamics}\label{ref: thermo}

The main quantity we would like to compute is the leading semi-classical limit of the Euclidean gravitational path integral $\mathcal{Z}$, 
\begin{equation}\label{eqn: def Z}
    \mathcal{Z}\approx \sum_{g^{*}_{\mu\nu}} e^{-I [g^*_{\mu\nu}]} \, , \qquad \text{as}\quad G_N \to 0 \, , 
\end{equation}
where the action $I$ is given by \eqref{eqn: action}. The sum is over all possible solutions to \eqref{eqn: Einstein eqn} with a boundary $\Gamma$ equipped with the prescribed conformal boundary data $(\tilde g_{ij},K)$. In the following, we consider a boundary with torus topology $S^1\times S^1$, and refer to the corresponding quantity \eqref{eqn: def Z} as the torus partition function.

At the boundary $\Gamma$, we impose
\begin{equation}\label{eqn: bdry metric S1xS1}
    \left.ds^2\right|_{\Gamma} = e^{2\bomega}(du^2+\r^2 d\varphi^2)\, , \qquad K=\text{constant} \, ,
\end{equation}
where $\r$ is an arbitrary positive number, and the Weyl factor $e^{\bomega}$ is left unfixed. The boundary coordinates are identified, as per the topology, under 
\begin{equation}\label{eqn: bdry coord iden}
    (u,\varphi)\sim(u+\beta,\varphi)\sim (u,\varphi+2\pi)\, .
\end{equation}
In what follows, we allow the Weyl factor to depend non-trivially on the boundary coordinates in such a way that the boundary is smooth. For \eqref{eqn: bdry metric S1xS1}, this requires $\bomega(u,\varphi)$ to be a smooth function on the torus, i.e., it must satisfy a periodicity condition 
\begin{equation}\label{eqn: periodic weyl}
    \bomega(u,\varphi) = \bomega(u+\beta,\varphi)=\bomega(u,\varphi+2\pi) \, .
\end{equation}
We refer to boundaries with a constant $\bomega$ as homogeneous or uniform.

The circle parameterised by $u$ is interpreted as the thermal circle, while $\varphi$ can be seen as a coordinate of a spatial circle. The dimensionless ratio $\b\equiv \beta/\mathfrak{r}$ is a conformally invariant quantity characterizing the global structure of \eqref{eqn: bdry metric S1xS1}, and is interpreted as an inverse conformal temperature. The resulting torus partition function is then a function of $\b$ and $K$. Following the Gibbons-Hawking prescription \cite{Gibbons:1976ue}, it is interpreted as a thermal partition function in the conformal canonical ensemble \cite{Anninos:2024wpy,Galante:2025tnt}.

Since by definition \eqref{eqn: def Z} depends only on the conformal class of the flat metric on a torus, it is expected to be invariant under the transformation
\begin{equation}\label{eqn: modular transf}
\b \to \frac{4\pi^2}{\b} \, .
\end{equation}
In the context of a Euclidean two-dimensional conformal field theory on a torus, this is known as a modular S transformation. In the following, we will show that \eqref{eqn: def Z} is invariant under \eqref{eqn: modular transf} within the class of Euclidean saddles we consider.

It is useful to examine thermodynamic properties of the individual saddle solutions that appear in \eqref{eqn: def Z}. By evaluating the on-shell action $I_\text{on-shell}(\b,K)$ of a particular solution, one can compute the conformal energy, conformal entropy, and specific heat at fixed $K$ using the thermodynamic relations,
\begin{equation}\label{eqn: thermo relation}
    E_\text{conf} \equiv  \left.\p_{\b}\right|_{K} I_\text{on-shell} \, , \quad \mathcal{S}_\text{conf} \equiv \left.\left(\b \p_{\b}-1\right)\right|_{K} I_\text{on-shell} \, , \quad C_K \equiv -\left.\b^2 \p_{\b}^2\right|_{K} I_\text{on-shell} \, .
\end{equation}
The sign of $C_K$ can be used to determine the thermal stability (or lack thereof) of the corresponding saddle solution. Namely, the solution is said to be thermally stable (unstable) if $C_K$ is positive (negative).

\subsection{Embedding perspective}\label{sec: embedding}

One of the tasks in computing \eqref{eqn: def Z} is to start from a prescribed boundary $\Gamma$ and determine an in-filling geometry that solves the Einstein equations \eqref{eqn: Einstein eqn} while reproducing the specified conformal boundary data on $\Gamma$. Instead, given the nature of three-dimensional gravity, we begin with a bulk solution of \eqref{eqn: Einstein eqn} and seek a two-dimensional surface $\Gamma$ embedded in the bulk that satisfies the conformal boundary conditions. This perspective allows us to reduce the problem of solving \eqref{eqn: Einstein eqn} subject to the conformal boundary conditions to a conformal embedding problem with a prescribed trace of the extrinsic curvature. In the following, we demonstrate this embedding problem and derive conditions imposed by fixing the conformal structure and the trace of the extrinsic curvature. The rest of this section holds for any value of the cosmological constant.

In three dimensions, the local geometry of a solution to \eqref{eqn: Einstein eqn} is completely fixed by the cosmological constant. We denote the bulk solution by
\begin{equation}
    ds^2 = g_{\mu\nu}dx^\mu dx^\nu \, ,
\end{equation}
where $\{x^\mu\}$ is some choice of coordinates. 

Now, we seek a two-dimensional surface $\Gamma$ which has the prescribed conformal representative $\tilde g_{ij}$ and an extrinsic curvature with fixed constant trace $K$. We describe the location of $\Gamma$ with embedding functions,
\begin{equation}
    \left.x^\mu\right|_{\Gamma} = X^\mu (\sigma^i) \, ,
\end{equation}
where $\sigma^i$ denote boundary coordinates. The induced metric is given by
\begin{equation}\label{eqn: embed ind metric}
    \left.ds^2\right|_\Gamma=g_{ij}d\sigma^i d\sigma ^j= g_{\mu\nu}(X)\p_i X^{\mu} \p_j X^\nu d\sigma^i d\sigma^j \, .
\end{equation}
The unit normal vector $n^\mu$ can be computed by
\begin{equation}\label{eqn: embed unit vec}
    \left.n^\mu\right|_\Gamma = \frac{1}{2}\frac{\epsilon^{\mu}{}_{\nu\rho}\epsilon^{ij}\p_i X^\nu \p_j X^\rho}{\sqrt{\det g_{kl}}} \, , 
\end{equation}
where $\epsilon_{\mu\nu\rho}$ and $\epsilon_{ij}$ denote the volume form of bulk and boundary, respectively. It is straightforward to show that $g_{\mu\nu}n^\mu n^\nu =1$.  Using the definition \eqref{eqn: def K} of the extrinsic curvature, we find
\begin{equation}\label{eqn: embed Kij}
    K_{ij} = \p_j X^\mu \left(\p_i n_\mu-\Gamma_{\mu\nu}^\rho \p_iX^\nu n_\rho\right)\,.
\end{equation}
Using \eqref{eqn: embed ind metric} and \eqref{eqn: embed unit vec}, the right hand side of \eqref{eqn: embed Kij} becomes a non-linear combination of $X^\mu$ and its derivatives. 

We now derive the equations of motion of $X^{\mu}(\sigma^i)$ that arise from imposing conformal boundary conditions. Using \eqref{eqn: embed ind metric}, requiring that the induced metric is \eqref{eqn: ind metr condition} leads to 
\begin{equation}\label{eqn: embed conf class cond}
    g_{\mu\nu}\p_iX^\mu \p_j X^\nu - \frac{1}{2} \tilde g_{ij} \tilde g^{kl}g_{\mu\nu}\p_k X^\mu \p_{l} X^\nu =0 \, ,
\end{equation}
where the corresponding Weyl factor is given by
\begin{equation}
    e^{2\bomega} = \frac{1}{2} g_{\mu\nu}\tilde g^{ij}\p_i X^\mu \p_j X^\nu \, .
\end{equation}
Using \eqref{eqn: embed Kij}, fixing the trace of the extrinsic curvature to be the prescribed $K$ imposes a non-linear second-order partial differential equation,
\begin{equation}\label{eqn: embed K}
    K = g^{ij}\p_j X^\mu \left(\p_i n_\mu-\Gamma_{\mu\nu}^\rho \p_iX^\nu n_\rho\right)\,.
\end{equation}
Equations \eqref{eqn: embed conf class cond} and \eqref{eqn: embed K} constitute a local conformal embedding of $\Gamma$ with the prescribed trace of the extrinsic curvature $K$. For $\Gamma$ to be a boundary, we further require that the embedding can be extended globally. This leads to the condition that no self-intersection should occur, i.e., 
\begin{equation}\label{eqn: global emb}
    X^\mu(\sigma_1^i)  \neq X^\mu(\sigma^i_2) \, , \qquad \text{for any distinct pair of }\sigma_1^i,\sigma_2^i \in \Gamma\,.
\end{equation}

\section{Zero cosmological constant}\label{sec: Lambda 0}

We first study conformal thermodynamics for three-dimensional gravity with zero cosmological constant.
We begin by a brief review of the Euclidean bulk solutions to Einstein's equations in three dimensions with zero cosmological constant \cite{Deser:1983tn}.

Solutions to \eqref{eqn: Einstein eqn} with zero cosmological constant are locally given by three-dimensional flat space. Working in cylindrical coordinates, we have a metric of the form
\begin{equation}\label{eqn: flat metric}
    ds^2 = d\tau^2+dr^2+r^2 d\phi^2 \, , \qquad \Lambda =0\, ,
\end{equation}
where $r\in (0,\infty)$ and $\phi\sim\phi+2\pi$. We choose the coordinate $\tau$ to be identified under $\tau\sim \tau+\beta_\tau$, where $\beta_\tau$ is an arbitrary positive number. The topology of the manifold is therefore $S^1\times \mathbb{R}^2$.

Taking $\tau \to i t$, we obtain the three-dimensional Minkowski metric written in cylindrical coordinates, which covers the whole spacetime. Taking $\phi \to i t$ instead, we arrive at the Rindler metric with a compact Rindler horizon of size $\beta_\tau$ situated at $r=0$. This exchange of the bulk thermal and spatial circles leads to the equivalence,
\begin{equation}\label{eqn: gloMink Rind}
    \text{global Minkowski}\longleftrightarrow\text{compact horizon Rindler} \, .
\end{equation}
In the following, we fix the choice of bulk thermal circle based on the boundary thermal circle.

We now consider solutions with a boundary obeying the boundary conditions \eqref{eqn: bdry metric S1xS1}, employing the embedding method. We review both pole patch and Rindler patch solutions, defined as spacetimes which do and do not contain the compact Rindler horizon, in sections \ref{Lambda = 0 pole patch} and \ref{Lambda = 0 Rindler patch} respectively. In each section, we look for non-static and non-circular boundaries. For each of these solutions, we compute their thermodynamic quantities in the conformal canonical ensemble defined in \ref{ref: thermo}. Finally, in section \ref{sec: flat thermo}, we combine these results and explore the thermodynamic phase space of the system.

\subsection{Pole patch} \label{Lambda = 0 pole patch}

We first study the conformal thermodynamics of the pole patch solutions. These are solutions endowed with the bulk metric
\begin{equation}\label{eqn: bulk metric pole}
    ds^2 = d\tau^2+dr^2 + r^2 d\phi^2\, , \qquad \Lambda=0 \, ,
\end{equation}
where $\tau\sim\tau+\beta_\tau$ for some arbitrary positive number $\beta_\tau$ and  $\phi \sim \phi+2\pi$. The spacetime region of interest includes the origin $r=0$ and has a boundary situated at the radial coordinate $\left.r\right|_{\Gamma}$, which will be specified later. At the boundary, we impose the conformal boundary conditions \eqref{eqn: bdry metric S1xS1}.

\subsubsection{Static and circular pole patch}\label{Lambda = 0 static and circular pole patch}

The first class of boundaries is a family of homogeneous boundaries, described by constant-$r$ surfaces. These boundaries and their thermodynamic properties may be analysed by taking the flat space limit of the $S^3$ solution in \cite{Anninos:2024wpy}.

We judiciously describe the boundary as
\begin{equation}\label{eqn: flat homo bdry}
    \left.\tau\right|_{\Gamma}=\frac{u}{K\r} \, , \qquad \left.r\right|_{\Gamma}=\frac{1}{K} \, , \qquad \left.\phi\right|_{\Gamma} = \varphi \, ,
\end{equation}
where $K>0$ is a positive constant. The region of flat space is defined by $r\in[0,\tfrac{1}{K})$. 
The unit normal vector is given by $n^\mu\p_\mu = \p_r$. 

It is straightforward to check that the boundary obeys the conformal boundary conditions \eqref{eqn: bdry metric S1xS1}. In particular, the Weyl factor and conformal stress tensor are given by
\begin{equation}\label{eqn: flat bomega Tij homo}
    \bomega= \log\frac{1}{K\r} \, , \qquad T_{ij} d\sigma^i d\sigma^j = -\frac{1}{16 \pi G_N K\r^2}\left(-du^2+\r^2 d\varphi^2\right) \, .
\end{equation}
The constant Weyl factor implies that the boundary is intrinsically flat. 

For consistency of the global structure of bulk and boundary coordinates, we have $ \beta_\tau = \b/K$. Since $\beta_\tau$ is a free parameter, it implies that the solution exists for all $\b>0$ and $K>0$.

\textbf{Conformal thermodynamics.} To compute the thermodynamic properties of the solution, we evaluate the on-shell action \eqref{eqn: action}. Plugging in \eqref{eqn: flat metric} and  \eqref{eqn: flat homo bdry}, we obtain
\begin{equation}\label{eqn: flat I pole}
    I_\text{on-shell}= - \frac{\b}{8 G_N K} \, .
\end{equation}
Using the thermodynamic relations \eqref{eqn: thermo relation}, the corresponding conformal energy, entropy, and specific heat are given by
\begin{equation}
    \label{Lambda=0 static and circular thermo}
    E_\text{conf}= - \frac{1}{8G_NK} \, , \qquad \mathcal{S}_\text{conf}=C_K= 0 \, .
\end{equation}
The energy is independent of $\b$, and the vanishing of entropy and specific heat is consistent with the absence of a horizon.

\subsubsection{Non-static pole patch} \label{Lambda = 0 non-static pole patch}

The second class of solutions has a boundary which varies along the boundary coordinate $u$. We parameterise the boundary by
\begin{equation}\label{eqn: flat bdry 1}
    \left.\tau\right|_{\Gamma}=\tau(u)\, , \qquad \left.r\right|_{\Gamma}=r(u) \, , \qquad \left.\phi\right|_{\Gamma}=\varphi\, .
\end{equation}
The patch of flat space is defined to be a region of \eqref{eqn: flat metric} with $r\in [0,r(u))$. The unit normal vector of the boundary is given by
\begin{equation}\label{eqn: flat n^mu}
    n^\mu \p_\mu = \frac{-\p_u r \p_\tau + \p_u \tau\p_r}{\sqrt{(\p_ur)^2+(\p_u \tau)^2}} \, .
\end{equation}
The outward-pointing condition on $n^\mu$ requires that $\p_u\tau>0$. The $u$-dependence of the boundary radius reflects the non-static nature of the solution.

\textbf{Problem.} The conformal boundary conditions impose restrictions on $\tau(u)$ and $r(u)$ via the conditions \eqref{eqn: embed conf class cond} and \eqref{eqn: embed K}. In particular, the conformal class condition reads
\begin{equation}\label{eqn: Lambda=0 eq1}
    (\r\p_u\tau)^2+(\r\p_ur)^2=r^2 \,,
\end{equation}
while the trace of the extrinsic curvature condition reads
\begin{equation}\label{eqn: Lambda=0 eq2}
    \frac{(\p_ur)^2 \p_u \tau + (\p_u\tau)^3+r\p_u r \p_u^2 \tau-r \p_u \tau \p_u^2 r }{r\left((\p_u\tau)^2+(\p_u r)^2\right)^{3/2}} = K \, ,
\end{equation}
where $K$ is the prescribed constant trace of the extrinsic curvature. 

In the following, we find solutions to \eqref{eqn: Lambda=0 eq1} and \eqref{eqn: Lambda=0 eq2} and analyse their properties. 
Then, we show that imposing the global embedding condition \eqref{eqn: global emb} further restricts the space of solutions. 

\begin{figure}[H]
        \centering
         \subfigure[$K^2V_\text{eff}$ versus $Kr$]{
                \includegraphics[scale=0.37]{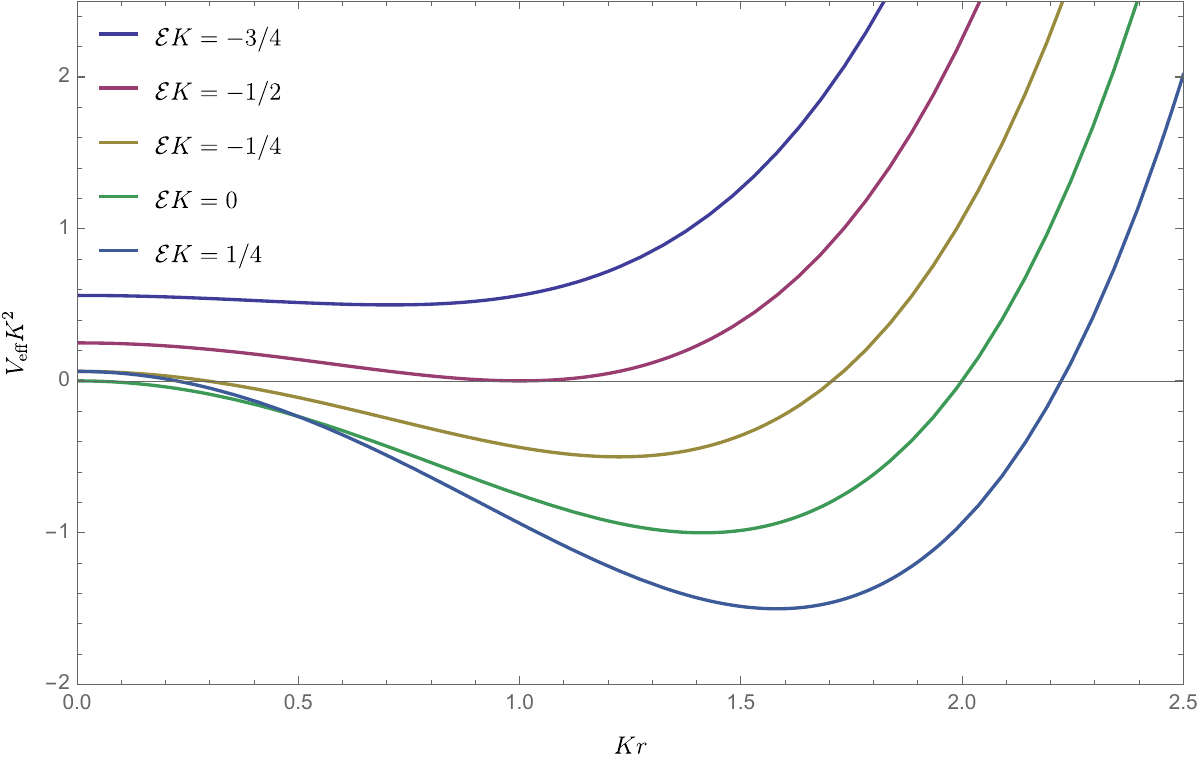}\label{fig: Veff flat}}  \quad
                 \subfigure[$K r_\pm$ versus $\mathcal{E}K$]{
                \includegraphics[scale= 0.37]{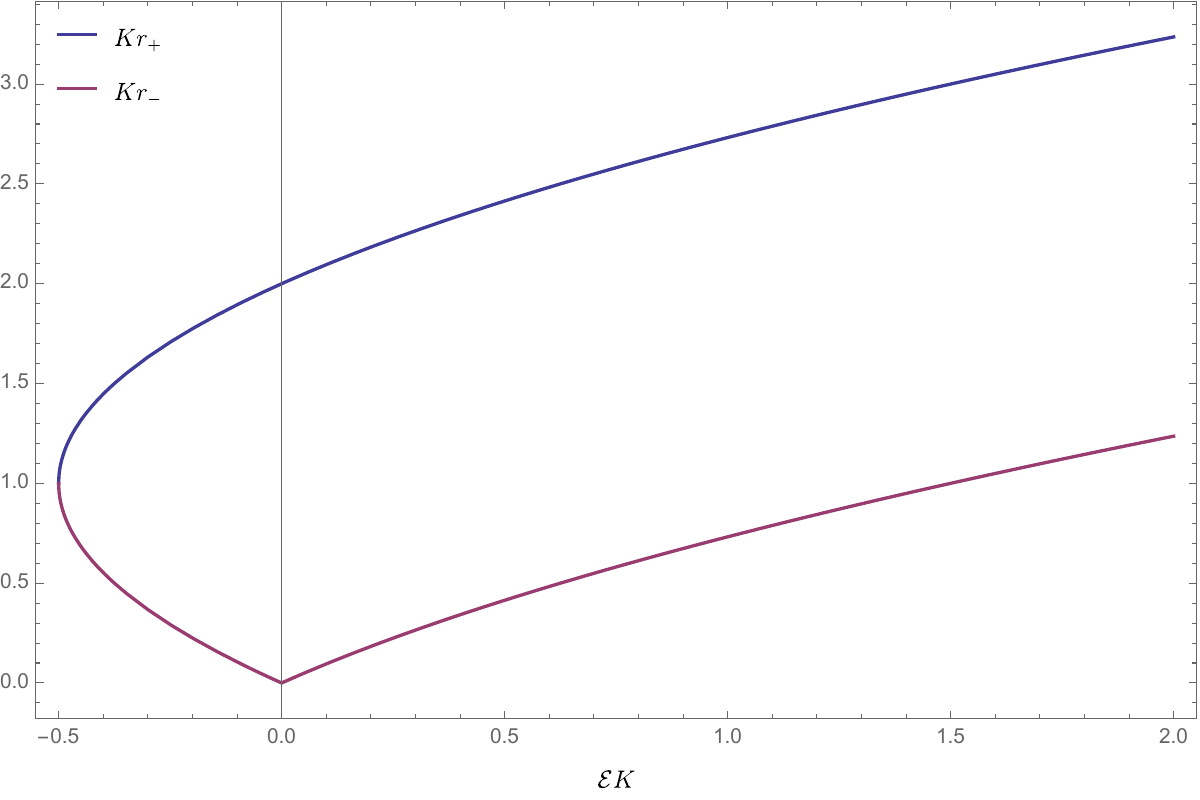} \label{fig: rpm flat}}                           
                \caption{A plot of $V_\text{eff}$ as a function of $r$  for $\mathcal{E}K=-\tfrac{3}{4},-\tfrac{1}{2},-\tfrac{1}{4},0,\tfrac{1}{4}$ (left) and a plot of $r_\pm$ as a function of $\mathcal{E}$ (right). When $\mathcal{E}K=-\tfrac{3}{4}$, the potential is always above zero, and hence no solution exists. For $\mathcal{E}K=-\tfrac{1}{2}$, the potential intersects zero once, at which the solution becomes $u$-independent. For $\mathcal{E}K=-\tfrac{1}{4},0,\tfrac{1}{4}$, there exists an interval where the potential is below zero. This leads to a solution oscillating between the intersection of the potential and zero.} 
\end{figure}

\textbf{Solution.} To solve \eqref{eqn: Lambda=0 eq1} and \eqref{eqn: Lambda=0 eq2}, we first decouple $\tau(u)$ and $r(u)$ by implementing \eqref{eqn: Lambda=0 eq1} on \eqref{eqn: Lambda=0 eq2}. As a result, we have
\begin{equation}\label{eqn: du T as e}
    \frac{d\mathcal{E}}{du}=0 \, , \qquad \mathcal{E}\equiv\frac{1}{2}Kr^2-\r \p_u \tau  \, ,
\end{equation}
which implies that the introduced dimension-length quantity $\mathcal{E}$ is an integration constant. The derivation of this equation is illustrated in appendix \ref{sec: du T flat 1}. Using this to replace $\p_u \tau$ in \eqref{eqn: Lambda=0 eq1}, we obtain an ordinary first-order differential equation,
\begin{equation}\label{eqn: flat bdry eqn 2}
    (\r\p_ur)^2 + V_\text{eff}(r)=0 \, , \qquad V_\text{eff}(r)= \frac{1}{4}K^2 r^4 - \left(1+\mathcal{E} K\right)r^2 + \mathcal{E}^2 \, .
\end{equation}
A plot of $V_\text{eff}(r)$ for various $\mathcal{E}$ is shown in figure \ref{fig: Veff flat}. 

We now solve \eqref{eqn: flat bdry eqn 2}. It is instructive to first analyse its qualitative features. This equation of motion mimics that of a classical particle subject to an effective potential $V_\text{eff}(r)$ in one dimension. The classical trajectory of \eqref{eqn: flat bdry eqn 2} is dictated by the zeros of $V_\text{eff}$. In particular, $r(u)$ is bounded and oscillates between $r_\pm$, the two positive roots of $V_\text{eff}(r_\pm)=0$,
\begin{equation}\label{eqn: flat R_pm}
    r_-<r(u)<r_+ \, , \qquad r_\pm = \sqrt{\frac{2+2\mathcal{E} K \pm 2\sqrt{1+2\mathcal{E} K}}{K^2}} \, .
\end{equation}
The plot of $r_\pm$ as a function of the dimensionless parameter $\mathcal{E}K$ is shown in figure \ref{fig: rpm flat}. The reality condition of $r_\pm$ imposes a bound
\begin{equation}
    \label{EK>-1/2 flat}
    \mathcal{E}K \geq -\frac{1}{2} \, .
\end{equation}
At $\mathcal{E} K=-\tfrac{1}{2}$, the two roots coincide, $r_+=r_-=\frac{1}{K}$, while $\mathcal{E} K=0$ leads to the vanishing of the smaller root, $r_-=0$.

The closed-form solution to \eqref{eqn: flat bdry eqn 2} is given by 
\begin{equation}\label{eqn: flat R sol}
    r(u) = r_+ \text{dn}\left(\frac{Kr_+ }{2}\frac{u-u_0}{\r} \, \bigg| \,m\right) \, ,
\end{equation}
where $m\equiv1-\frac{r_-^2}{r_+^2}$. dn$(x|m)$ denotes the Jacobi elliptic dn function, $u_0$ is an integration constant.

Inserting \eqref{eqn: flat R sol} in \eqref{eqn: du T as e} and integrating over $u$, we find 
\begin{equation}\label{eqn: flat T sol}
    \tau(u) = \tau_0-\mathcal{E} \frac{u-u_0}{\r}+r_+E\!\left(\mathrm{am}\left(\frac{K r_+}{2}\frac{u-u_0}{\mathfrak{r}};m\right) \bigg| \, m\right)\, ,
\end{equation}
where $E$ denotes the incomplete elliptic integral of the second kind, and $\tau_0$ is an integration constant. Relevant properties of dn and $E$ are reviewed in appendix \ref{sec: Ellip Jacobi}. Solutions \eqref{eqn: flat R sol} and \eqref{eqn: flat T sol} are uniquely determined upon fixing $\mathcal{E}$, $K$, $u_0$, and $\tau_0$, where the last two respectively correspond to shifting the coordinates $u$ and $\tau$. Using translational symmetry, we will set these to zero from now on. Therefore, by measuring $r(u)$ and $\tau(u)$ in the unit of $K$, we find that the only parameter describing the solution is $\mathcal{E} K$. 

\begin{figure}[H]
    \centering

    \begin{minipage}[c]{0.48\textwidth}
        \centering
        \includegraphics[scale=0.32]{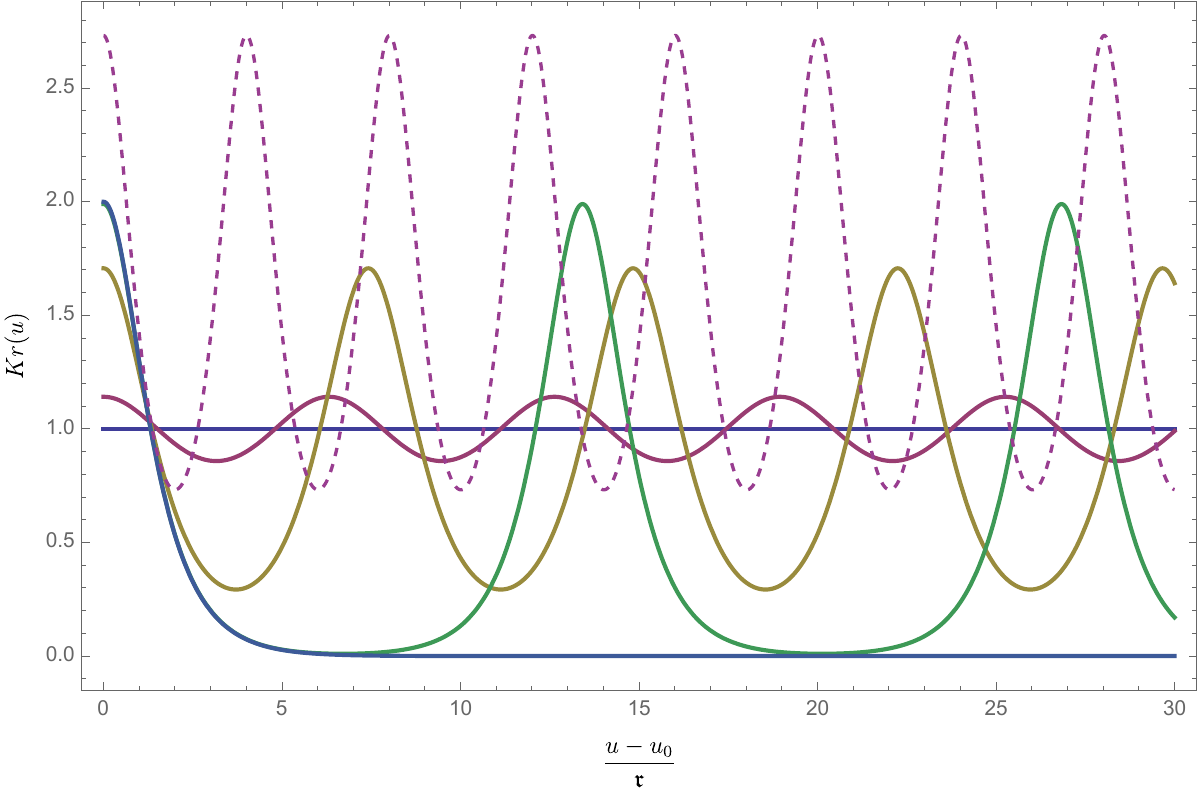}\label{fig: Rsol flat}

        \vspace{0.5em}

        \includegraphics[scale=0.32]{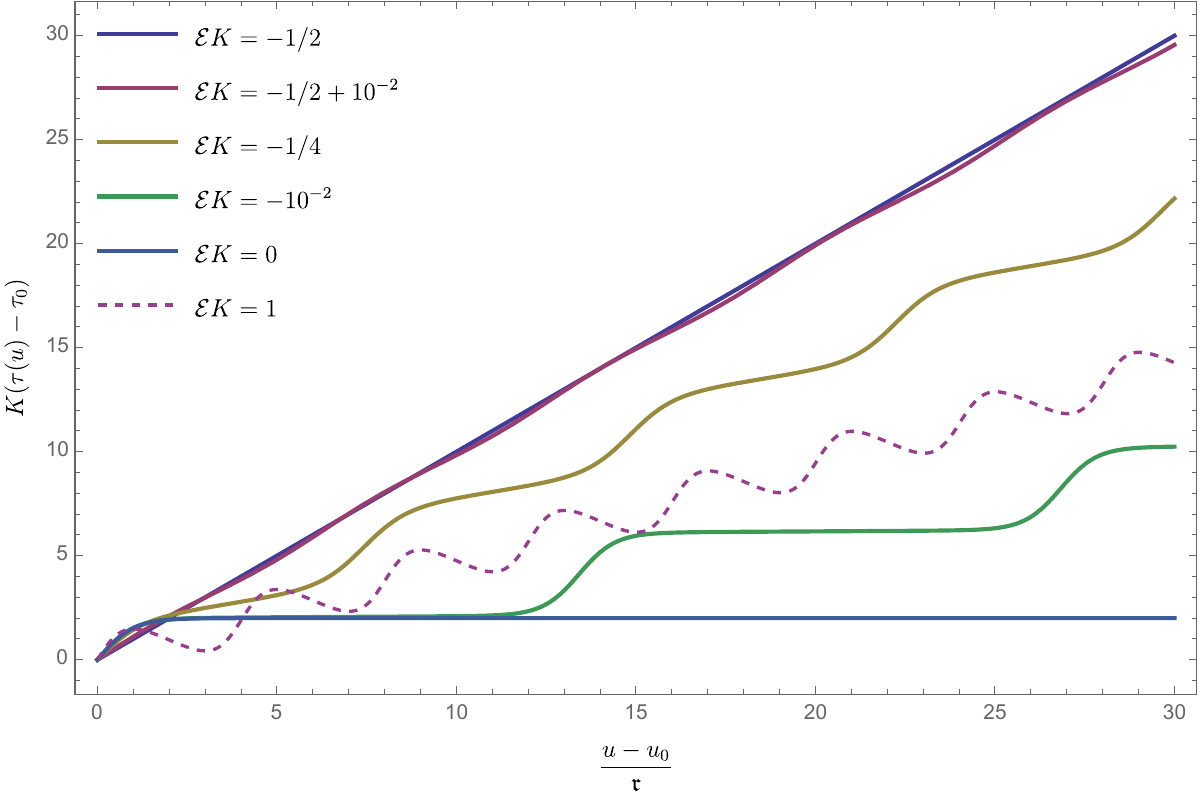}\label{fig: Tsol flat}
    \end{minipage}
    \hfill
    \begin{minipage}[c]{0.48\textwidth}
        \centering
        \includegraphics[scale=0.34]{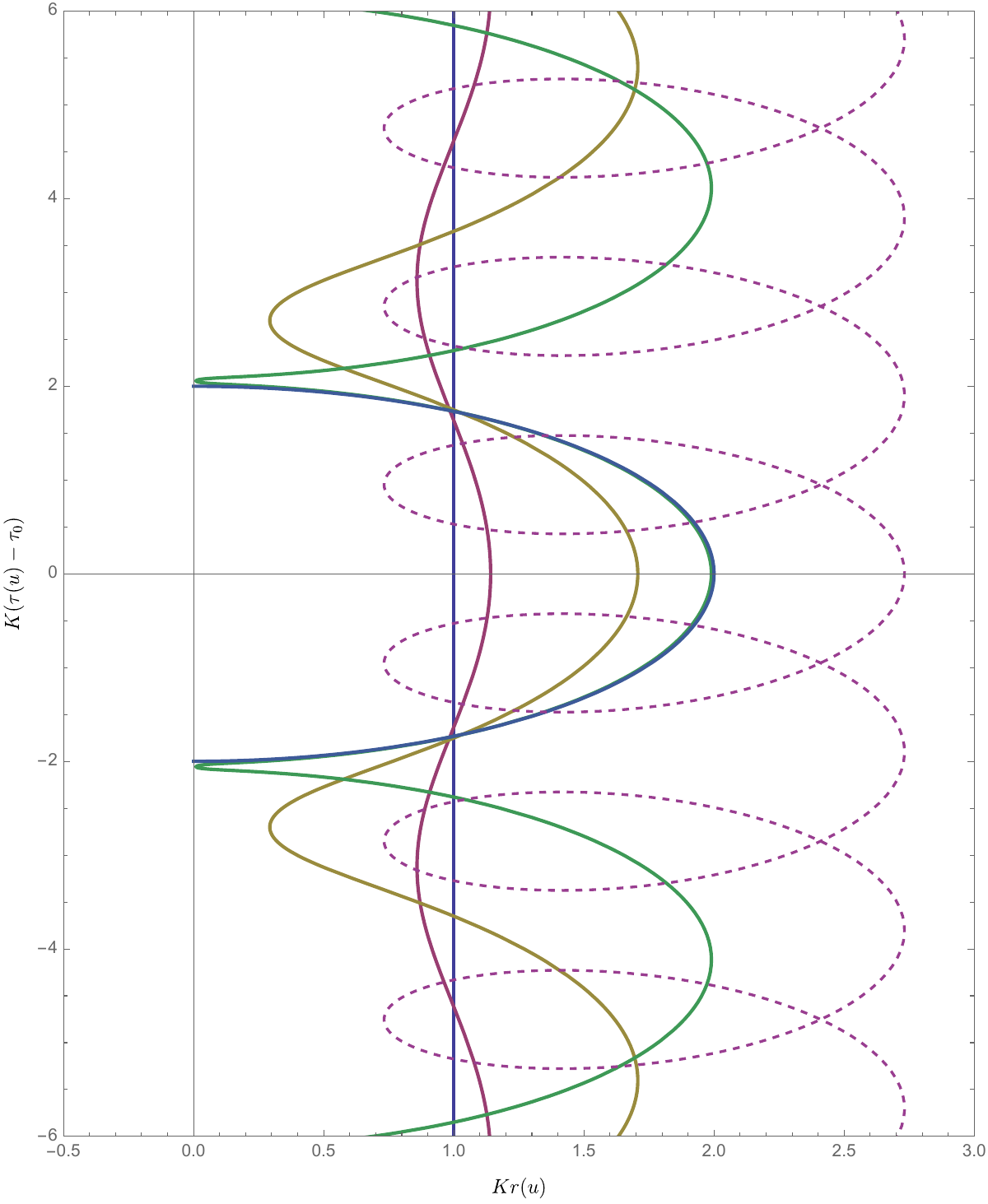}\label{fig: RTsol flat}
    \end{minipage}

    \caption{Plots of $Kr$ (left top) and $K(\tau(u)-\tau_0)$ (left bottom) versus $\tfrac{u-u_0}{\r}$ and a plot of the boundary trajectory in the $r-\tau$ plane for $\mathcal{E}K=-\tfrac{1}{2},-\tfrac{1}{2}+10^{-2},-10^{-2},0$ depicted in solid lines and $\mathcal{E}K=1$ depicted in a dashed line.}\label{fig: Rsol Tsol flat}
\end{figure}

The solution obeys the periodic condition,
\begin{equation}\label{eqn: periodic flat R T}
    r\left(u+ \frac{4n \r \mathcal{K}(m)}{Kr_+} \right)=r(u) \, , \quad \tau\left(u+ \frac{4n \r \mathcal{K}(m)}{Kr_+}  \right) = \tau(u)+2n\left(-\frac{2\mathcal{E}\mathcal{K}(m)}{Kr_+}  +r_+E(m)\right)\,,
\end{equation}
where $n$ is an arbitrary integer, and $\mathcal{K}(m)$ and $E(m)$ are respectively complete elliptic integral of the first and second kind. In particular, $\tau(u)$ shifts by an integer multiple of a function of $\mathcal{E}$ and $K$ each time $u$ completes the corresponding integer number of periods. We emphasise that, since $m$ is a function of $\mathcal{E}K$, the period of the solution is determined by $\mathcal{E}K$. A plot of $(\tau(u),r(u))$ for various value of $\mathcal{E}K$ is illustrated in figure \ref{fig: Rsol Tsol flat}. 

Identifying \eqref{eqn: flat R sol} and \eqref{eqn: flat T sol} as the location of the boundary, we obtain a family of boundaries obeying the conformal boundary conditions, labelled by the dimensionless parameter $\mathcal{E}K$. We note that, in mathematical literature, these surfaces are known as unduloid surfaces, see \cite{Delaunay1841,Osserman1986}. The Weyl factor and conformal stress tensor are given by
\begin{equation}\label{eqn: flat Weyl}
    \bomega(u) = \log \frac{r}{\r}\, , \qquad T_{ij}d\sigma^id\sigma^j=\frac{\mathcal{E}}{8 \pi G_N\r^2}\left(-du^2+\r^2 d\varphi^2\right)\, .
\end{equation}
We note that $T_{ij}$ is a constant tensor fixed entirely by $\mathcal{E}$. 

\textbf{Simple solutions.} At special values of $\mathcal{E}K$, solutions \eqref{eqn: flat R sol} and \eqref{eqn: flat T sol} can be written in terms of elementary functions. Using the asymptotic expansions in \eqref{eqn: asymp dn}, \eqref{eqn: asymp am}, and \eqref{eqn: asymp ellipticE}, solutions with $\mathcal{E}K\to-\tfrac{1}{2}$ are given by
\begin{equation}\label{eqn: flat e=-1/2 sol}
    r(u) = \frac{1}{K}+\frac{\sqrt{2\mathcal{E}K+1}}{K}\cos\frac{u}{\r} + \mathcal{O}(2\mathcal{E}K+1)\, , \quad \tau(u)=\frac{u}{K \r}+\frac{\sqrt{2\mathcal{E}K+1}}{K}\sin\frac{u}{\r}+\mathcal{O}(2\mathcal{E}K+1) \, .
\end{equation}
The leading term reproduces the homogeneous boundary in \eqref{eqn: flat homo bdry}. The first corrections are given by sine and cosine functions, reflecting the oscillatory behaviour of the solutions. The amplitude is controlled by the small parameter $2\mathcal{E}K+1$, where its non-analytic power reflects the fact that no solution exists in a small neighbourhood of $\mathcal{E}K<-\tfrac{1}{2}$.

Another simple solution can be obtained in the limit of $\mathcal{E}K\to0$. Using again the asymptotic expansions in \eqref{eqn: asymp dn}, \eqref{eqn: asymp am}, and \eqref{eqn: asymp ellipticE}, we find
\begin{equation}\label{eqn: eK->0 flat sol}
    r(u) = \frac{2}{K\cosh{\frac{u}{\r}}}+\frac{\mathcal{E}K}{K \cosh{\frac{u}{\r}}}\left(1-\frac{u}{\r}\tanh\frac{u}{\r}\right)+\mathcal{O}((\mathcal{E}K)^2) \, , \quad \tau(u)=r(u)\sinh\frac{u}{\r}+\mathcal{O}((\mathcal{E}K)^2)\,.
\end{equation}
At the leading order, the solution describes a round two-sphere boundary of radius $\tfrac{2}{K}$ as can be seen from the equation
\begin{equation}
    r(u)^2+\tau(u)^2=\frac{4}{K^2}+\mathcal{O}(\mathcal{E}K)\, .
\end{equation}
The first correction describes a deviation away from the sphere solution. We note that, due to the term $\tfrac{u}{\r}\tanh \frac{u}{\r}$, the correction is not a smooth function on the sphere.

\textbf{Self-intersection.} Let us comment on a salient feature of the perturbative solution \eqref{eqn: eK->0 flat sol}. When $\mathcal{E}K>0$, we find that
\begin{equation}
     r(u_\text{int})=r(-u_\text{int}) \, , \qquad \tau(u_\text{int})=\tau(-u_\text{int}) \, , \qquad u_\text{int}\equiv \frac{2\r}{\mathcal{E}K} \, ,
\end{equation}
which violates the global embedding condition \eqref{eqn: global emb}. This implies that the boundary contains a self-interesting curve at $u=\pm u_\text{int}$ and $\varphi \in (0,2\pi)$. In contrast, no such self-intersection occurs for $\mathcal{E}K<0$. One might be concerned that the above argument relies on evaluating the solution at large $u$ which potentially violates the small $\mathcal{E}K$ expansion. In appendix \ref{sec: self-intersection}, we provide a proof that the self-intersection is related to the non-monotonicity of $\tau(u)$ and generically occurs for any finite $\mathcal{E}K>0$ solution. As a result, the space of permissible solutions must obey
\begin{equation}\label{eqn: flat no self inter}
    -\frac{1}{2} < \mathcal{E} K <0 \, .
\end{equation}

Below, we study conformal thermodynamics of the non-static pole patch solutions assuming \eqref{eqn: flat no self inter}.

\textbf{Conformal thermodynamics.}
We start by analysing the conformal boundary data of the solution.
The global structure of the bulk imposes periodicity of the bulk coordinates. For this to be compatible with the periodicity of the boundary \eqref{eqn: bdry coord iden}, we require
\begin{equation}\label{eqn: flat periodic cond}
    \tau(u+\beta)=\tau(u)+\beta_\tau \, , \qquad r(u+\beta) = r(u) \, .
\end{equation}
Using the periodic property of the solutions, \eqref{eqn: periodic flat R T}, we obtain
\begin{equation}\label{eqn: flat wobbly conf data}
    \b = \frac{4n\mathcal{K}(m)}{Kr_+} \, , \qquad \beta_\tau = 2n \left(-\frac{2\mathcal{E}\mathcal{K}(m)}{Kr_+}+r_+ E(m)\right) \, .
\end{equation}
The integer $n$ measures the number of times the thermal circle winds around the $\tau$-circle, and hence is dubbed the winding number. For a fixed $n$, these equations can, in principle, be inverted to find $\mathcal{E}K$ and $\beta_\tau$ as a function of $\b$ and $K$. In the regime where the boundary does not exhibit self-intersection, the inverse is one-to-one.

For any fixed $n$, knowing that the allowed solutions have $-\frac{1}{2}<\mathcal{E}K<0$, the inverse conformal temperature obtained from \eqref{eqn: flat wobbly conf data} is bounded from below,
\begin{equation}\label{eqn: flat bound b}
    \b>2n\pi \, .
\end{equation}
Specifically, $\b$ attains its minimum as the solution approaches the static case, $\mathcal{E}K\to -\tfrac{1}{2}$,  and increases with $\mathcal{E} K$ until it diverges logarithmically as $\mathcal{E} K \to 0^{-}$, 
\begin{equation}
    \b = \begin{cases}
        2n \pi+\frac{n\pi}{2}\left(2\mathcal{E}K+1\right) + \mathcal{O}((2\mathcal{E}K+1)^2) \, , \qquad &\text{as} \quad \mathcal{E}K\to-\tfrac{1}{2} \, , \\
        -2n \log\frac{-\mathcal{E}K}{8} +\mathcal{O}(\mathcal{E}K\log{(-\mathcal{EK}})) \, , \qquad &\text{as} \quad \mathcal{E}K\to 0^{-}\, . 
    \end{cases}
\end{equation}
The bound \eqref{eqn: flat bound b} implies that the number of coexisting non-static solutions $\mathfrak{n}$ for a given $\b$ and $K$ is given by 
\begin{equation}
    \label{Lambda=0 number of non-static pole sols}
    \mathfrak{n}_{\text{non-static pole}} = \left\lceil \frac{\tilde{\beta}}{2\pi}\right\rceil-1 \, ,
\end{equation}
where $\left\lceil \cdot\right\rceil$ is the ceiling function. Consequently, the number of coexisting solutions grows as $\b \rightarrow \infty$.

To analyse the thermodynamic properties of these solutions, we evaluate the on-shell action,
\begin{equation}
    I_\text{on-shell}= - \frac{\mathcal{E}\b+\beta_\tau}{4G_N}=- \frac{nr_+ E(m)}{2G_N} \, .
\end{equation} 
Using \eqref{eqn: flat wobbly conf data} to obtain $\mathcal{E} K$ as a function of $\b$, the on-shell action is a function of $\b$ and $K$.

We now compute the thermodynamic properties. Using \eqref{eqn: thermo relation}, we find the corresponding conformal energy and entropy,
\begin{equation}
    E_\text{conf} = \frac{\mathcal{E}}{4G_N} \, , \qquad \mathcal{S}_\text{conf} = \frac{2\mathcal{E} \b+\beta_\tau}{4G_N}\, .
\end{equation}
Indeed, these quantities satisfy $\delta E_\text{conf}=\b^{-1}\delta \mathcal{S}_\text{conf}$ while holding $K$ fixed. 

The conformal energy $E_\text{conf}$ is equal to the parameter $\mathcal{E}$ measured in the unit of $4G_N$, in agreement with the computation of the conformal stress tensor \eqref{eqn: flat Weyl}. In the parameter regime where self-intersection does not occur, $E_\text{conf}$ is negative definite and is bounded from below by that of the static pole patch solution at $\mathcal{E}K=-\frac{1}{2}$. In the low temperature limit, the energy goes to $0^{-}$.

Away from the static limit, the conformal entropy $\mathcal{S}_\text{conf}$ is non-vanishing even without any horizon present in the bulk. In the low-temperature limit, the entropy tends to a constant value,
\begin{equation}
    \mathcal{S}_\text{conf}=\frac{n}{G_N K} + \mathcal{O}( \tilde{\beta} e^{-\frac{\b}{2n}}) \, ,
\end{equation}
where the first correction is exponentially suppressed in the inverse temperature. Combined with the energy, we find that the on-shell action is purely entropic and is given by an integer multiple of the two-sphere on-shell action,
\begin{equation}
    \label{two-sphere on-shell action flat}
    I^{(\text{two-sphere})}_{\text{on-shell}}=- \frac{1}{G_NK} \, \, .
\end{equation}

The specific heat at fixed $K$ is given by
\begin{equation}
    C_K=\frac{4\mathcal{E}(1+2\mathcal{E}K)\mathcal{K}(m)^2 n}{G_N K^2 r_+ \left(Kr_+^2E(m)+2\mathcal{E}\mathcal{K}(m) \right)}<0 \, .
\end{equation}
The negative definiteness of $C_K$ implies that the solution is always thermally unstable. In the static limit $\mathcal{E}K \rightarrow -\frac{1}{2}$, $C_K$ converges to a finite value $C_K\to-\tfrac{n \pi}{G_N K}$ that differs from the result \eqref{Lambda=0 static and circular thermo} using the static boundary. And it converges to zero in the low-temperature two-sphere limit $\mathcal{E} \rightarrow 0^{-}$.

\subsubsection{Non-circular pole patch} \label{Lambda = 0 noncircular pole patch}

Now we study a class of solutions with a boundary that varies along the boundary $\varphi$-direction. We describe the boundary as
\begin{equation}\label{eqn: flat bdry 2}
    \left.\tau\right|_{\Gamma}=\frac{u}{K\r}\, , \qquad \left.r\right|_{\Gamma}=r(\varphi) \, , \qquad \left.\phi\right|_{\Gamma}=\phi(\varphi)\, .
\end{equation}
The patch of flat space is defined as the region of $r\in [0,r(\varphi))$. The unit normal vector is given by
\begin{equation}
    n^\mu \p_\mu = \frac{r\p_\varphi \phi\p_r -r\p_\varphi r\p_\phi}{\sqrt{(r\p_\varphi \phi)^2+(\p_\varphi r)^2}} \, .
\end{equation}
Requiring that $n^\mu$ is outward-pointing leads to a condition $\p_\varphi \phi >0$.

\textbf{Problem.} The conformal boundary conditions impose restrictions on \eqref{eqn: flat bdry 2} via the conditions \eqref{eqn: embed conf class cond} and \eqref{eqn: embed K}. In particular, the conformal class condition leads to
\begin{equation}\label{eqn: Lambda=0 eq3}
    (r\p_\varphi \phi)^2 + (\p_\varphi r)^2 =\frac{1}{K^2}\, ,
\end{equation}
while the trace of the extrinsic curvature condition leads to
\begin{equation}\label{eqn: Lambda=0 eq4}
    \frac{2 (\p_\varphi r)^2 \p_\varphi \phi + r^2 (\p_\varphi \phi)^3+r \p_\varphi r\p_\varphi^2 \phi - r \p_\varphi \phi \p_\varphi^2 r}{\left((r\p_\varphi \phi)^2+(\p_\varphi r)^2\right)^{3/2}} = K \, ,
\end{equation}
where $K$ is the prescribed constant trace of the extrinsic curvature. 

Now, we solve \eqref{eqn: Lambda=0 eq3} and \eqref{eqn: Lambda=0 eq4} and explicitly show that they describe boundaries that are related to the homogeneous boundary by the isometries of flat space.

\textbf{Solutions.} First, we observe that by implementing \eqref{eqn: Lambda=0 eq3}, \eqref{eqn: Lambda=0 eq4} can be simplified to a total derivative,
\begin{equation}
    \frac{d\mathcal{E}}{d\varphi}=0 \, , \qquad \mathcal{E}\equiv  \frac{1}{2}Kr^2 - K r^2 \p_\varphi \phi \, ,
\end{equation}
which allows us to identify $\mathcal{E}$ as a constant of motion. A derivation of this equation is similar to the derivation of \eqref{eqn: du T as e}, which is provided in appendix \ref{sec: du T flat 1}. Using this to replace $\p_\varphi \phi$ in \eqref{eqn: Lambda=0 eq3}, we obtain a first-order differential equation of $r(\varphi)$,
\begin{equation}\label{eqn: du T flat 22}
    (\p_\varphi y)^2 + y^2  =1+2\mathcal{E}K\, , \qquad y(\varphi)\equiv \frac{1}{2}K^2r(\varphi)^2-(1+\mathcal{E}K) \, .
\end{equation}
This equation takes the form of a one-dimensional harmonic oscillator where $y(\varphi)$ is the displacement and $1+2\mathcal{E}K$ is the energy. It follows that a solution exists only when $\mathcal{E}K>-\tfrac{1}{2}$. The general solution is given by
\begin{equation}\label{eqn: R flat phi sol}
    \frac{1}{2}K^2r(\varphi)^2= 1+\mathcal{E}K+\sqrt{1+2\mathcal{E}K}\sin\left(\varphi-\varphi_0\right) \, , 
\end{equation}
where $\varphi_0$ is an integration constant. Plugging this to \eqref{eqn: du T flat 22} and solving for $\phi(\varphi)$, we find
\begin{equation}\label{eqn: F flat phi sol}
    -\mathcal{E}K\tan\left(\phi(\varphi)-\phi_0-\frac{\varphi-\varphi_0}{2}\right) =  \sqrt{1+2\mathcal{E}K}+ \left(1+\mathcal{E}K\right)\tan\left(\frac{\varphi-\varphi_0}{2}\right)\, ,
\end{equation}
where $\phi_0$ is an integration constant. Note that $\phi(\varphi)$ must be judiciously endowed with piecewise integration constants for the solution to be smooth.

It is straightforward to show that solutions \eqref{eqn: R flat phi sol} and \eqref{eqn: F flat phi sol} satisfy
\begin{equation}
    \left(r(\varphi)\cos (\phi(\varphi)-\phi_0)-x_0\right)^2+\left(r(\varphi)\sin(\phi(\varphi)-\phi_0)-y_0\right)^2=\frac{1}{K^2} \, , 
\end{equation}
where
\begin{equation}
    x_0=\text{sign}(\mathcal{E}K)\frac{1+2 \mathcal{E}K}{\sqrt{2+2\mathcal{E}K}} \, , \quad y_0=-\text{sign}(\mathcal{E}K)\frac{\sqrt{1+2\mathcal{E}K}}{\sqrt{2+2\mathcal{E}K}}\, .
\end{equation}
This equation describes a circle of radius $\tfrac{1}{K}$ with the origin at $(x_0,y_0)$ of a constant $\tau$-slice. For $\mathcal{E}K=-\tfrac{1}{2}$, we recover the homogeneous boundary. Consequently, any solutions \eqref{eqn: R flat phi sol} and \eqref{eqn: F flat phi sol} describe the homogeneous boundary shifted by a global translation of the two-dimensional $r-\phi$ plane, and thereby are geometrically equivalent patches of flat space.

Another way to see this is to consider the intrinsic and extrinsic geometry of the boundary. Using \eqref{eqn: Lambda=0 eq3} and \eqref{eqn: Lambda=0 eq4}, we find that the Weyl factor and conformal stress tensor are given by
\begin{equation}\label{eqn: flat 2 bomega Tij}
    \bomega=\log\frac{1}{K \r}\,, \qquad T_{ij}d\sigma^id\sigma^j=-\frac{1}{16 \pi G_NK \mathfrak{r}^2}\left(-du^2+\r^2d\varphi^2\right)\, .
\end{equation}
We observe that these quantities do not depend on $r(\varphi)$ and $\phi(\varphi)$ and are identical to the ones found in the static case, see \eqref{eqn: flat bomega Tij homo}. Furthermore, when combined with the prescribed conformal structure and the trace of the extrinsic curvature, we find that both intrinsic and extrinsic geometry of this solution are identical to the homogeneous boundary described in section \ref{Lambda = 0 static and circular pole patch}. As such, the boundary found from solving \eqref{eqn: Lambda=0 eq3} and \eqref{eqn: Lambda=0 eq4} can be at most different from the homogeneous boundary by isometry of the flat background, and hence there are no non-circular pole patch solutions,
\begin{equation}
    \mathfrak{n}_{\text{non-circular pole}}=0 \, .
\end{equation}

\subsection{Rindler patch} \label{Lambda = 0 Rindler patch}

In this section, we study the conformal thermodynamics of the Rindler patch solutions. These are defined as solutions endowed with the bulk metric
\begin{equation}\label{eqn: bulk metric rindler}
    ds^2 = \bar r^2d\bar \tau^2 + d\bar r^2 + d\bar\phi^2 \, , \qquad \Lambda=0 \, ,
\end{equation}
where $\bar \tau \sim \bar\tau+2\pi$ and $\bar\phi\sim\bar \phi+\beta_{\bar\phi}$ for some arbitrary positive number $\beta_{\bar\phi}$. The spacetime region of interest encompasses the compact horizon $r=0$, which has a proper size of $\beta_{\bar\phi}$, and is bounded by the boundary situated at $\left. \bar r\right|_\Gamma$, which will be specified later. At the boundary, we impose the conformal boundary conditions \eqref{eqn: bdry metric S1xS1}.

We note that the metric \eqref{eqn: bulk metric rindler} is geometrically equivalent to \eqref{eqn: bulk metric pole}. This can be seen by a trivial coordinate transformation,
\begin{equation}\label{eqn: Rindler pole transf}
    \bar \tau = \phi \, , \qquad \bar \phi=\tau \, , \qquad \bar r=r \, ,
\end{equation}
together with the parameter identification $\beta_\tau = \beta_{\bar \phi}$. As such, we exploit this equivalence to find Rindler patch solutions with a variety of boundaries using results from the pole patch solutions. We note however that the resulting solutions are physically distinct as they lead to different physical observables, such as the on-shell action.

\subsubsection{Static and circular Rindler patch} \label{Lambda=0 static and circular Rindler}

The first class of boundaries is a family of homogeneous boundaries, described by constant-$\bar r$ surfaces. These boundaries and their thermodynamic properties are first analysed by taking the flat space limit of the dS$_3$ solution in \cite{Anninos:2024wpy}.

We judiciously describe the boundary as
\begin{equation}\label{eqn: flat homo bdry 2}
    \left.\bar\tau\right|_{\Gamma}= \frac{2\pi u}{\beta}\, , \qquad \left. \bar r\right|_{\Gamma}=\frac{1}{K} \, , \qquad \left.\bar \phi\right|_{\Gamma} = \frac{2\pi \r \varphi}{\beta K} \, .
\end{equation}
where $K>0$ is a positive constant. The region of flat space is defined by $\bar r\in[0,\tfrac{1}{K})$. The unit normal vector is given by $n^\mu\p_\mu = \p_{\bar r}$. 

It is straightforward to check that the boundary solves the conformal boundary conditions \eqref{eqn: bdry metric S1xS1}. In particular, the Weyl factor and conformal stress tensor are given by
\begin{equation}
    \bomega= \log \frac{2\pi}{\beta K}\, , \qquad T_{ij} d\sigma^i d\sigma^j = \frac{\pi}{4 \beta^2 K G_N}\left(-du^2+\r^2d\varphi^2\right) \, .
\end{equation}
The constant Weyl factor again implies that the boundary is intrinsically flat.

Imposing periodicity of the bulk and boundary coordinates, we obtain
\begin{equation}
    \label{eqn: bdry data homo flat 2}
    \b=\frac{4\pi^2}{K \beta_{\bar \phi}}\, ,
\end{equation}
Since $\beta_{\bar \phi}$ is a free parameter, the solution exists for all $\b$ and $K>0$.

\textbf{Conformal thermodynamics.} Evaluating the on-shell action \eqref{eqn: action}, we find
\begin{equation}\label{eqn: flat I rindler}
    I_\text{on-shell}= - \frac{\pi^2}{2G_N K \b} \, .
\end{equation}
Using the thermodynamic relations \eqref{eqn: thermo relation}, the corresponding conformal energy, entropy, and specific heat are given by
\begin{equation}
    E_\text{conf}= \frac{\pi^2\mathfrak{c}_\text{flat}}{3\b^2} \, , \qquad \mathcal{S}_\text{conf}=C_K=\frac{2\pi^2\mathfrak{c}_\text{flat}}{3 \b}\, , \qquad \mathfrak{c}_\text{flat}\equiv \frac{3}{2G_NK} \, .
\end{equation}
Using \eqref{eqn: bdry data homo flat 2}, we find that the entropy agrees with the standard area law of the Rindler horizon, $\tfrac{A}{4G_N}=\tfrac{\beta_{\bar \phi}}{4G_N}$. The positive definiteness of $C_K$ implies that the solution is thermally stable.  These quantities reproduce the flat space limit of dS$_3$ conformal thermodynamic quantities obtained in \cite{Anninos:2024wpy}, where $\mathfrak{c}_\text{flat}$ is interpreted as the number of effective degrees of freedom.

\subsubsection{Non-static Rindler patch} \label{Lambda=0 non-static Rindler}

Now we consider non-static Rindler patches. These are solutions with a boundary varying non-trivially on $u$. We parameterise the boundary as
\begin{equation}
    \left.\bar \tau\right|_{\Gamma} = \bar\tau\!\left(u\right)\, , \qquad \left.\bar r\right|_{\Gamma} = \bar r\!\left(u\right) \, , \qquad \left.\bar \phi\right|_{\Gamma} = \frac{\beta_{\bar \phi}\varphi}{2\pi} \, .
\end{equation} 
The $n^{\bar r}$ component of the unit normal vector has the same sign as $\partial_{u}\bar \tau$, and so we only consider solutions with $\partial_{u}\bar \tau>0$ for the outward-pointing condition to hold.

\textbf{Problem.} The conformal boundary conditions impose the following conditions on the boundary location. In particular, the conformal class condition leads to
\begin{equation}
    \label{induced metric condition non-static Rindler Lambda=0}
    (\partial_{u}\bar r)^2+\bar r^2(\partial_{u}\bar \tau)^2=\frac{\beta_{\bar \phi}^2}{4\pi^2 \mathfrak{r}^2} \, , 
\end{equation} 
while the trace of the extrinsic curvature condition leads to
\begin{equation}
    \label{extrinsic curvature condition non-static Rindler Lambda=0}
    \frac{\bar r \partial_u \bar \tau \left(\bar r (\partial_u \bar \tau)^2-\partial_u^2 \bar r\right)+\partial_u \bar r(\bar r  \partial_u^2 \bar \tau+2 \partial_u \bar r \partial_u \bar \tau)}{\left((\partial_u \bar r)^2+\bar r^2 (\partial_u \bar \tau)^2\right)^{3/2}} = K \, .
\end{equation}
In the following, we solve these equations.

\textbf{Solutions.} It is straightforward to show that the general solution to equations \eqref{induced metric condition non-static Rindler Lambda=0} and \eqref{extrinsic curvature condition non-static Rindler Lambda=0} is given by
\begin{equation}
    \bar\tau(u)=\phi\left(\frac{\beta_\phi K u}{2\pi\r}\right)\, , \qquad \bar r(u)=r\left(\frac{\beta_\phi K u}{2\pi\r}\right)\, ,
\end{equation}
where the functions $\phi(\varphi)$ and $r(\varphi)$ are given by \eqref{eqn: F flat phi sol} and \eqref{eqn: R flat phi sol}. 

Following the argument in section \ref{Lambda = 0 noncircular pole patch}, we conclude that there is no non-trivial non-static Rindler patch solution.

\subsubsection{Non-circular Rindler patch} \label{Lambda=0 non-circular Rindler}

Now we consider the third class of Rindler patches, which are those with a boundary varying along $\varphi$. These are dubbed non-circular Rindler patch solutions. We parameterise the boundary as
\begin{equation}
    \left.\bar \tau\right|_{\Gamma} = \frac{2\pi u}{\beta}\, , \qquad \left.\bar r\right|_{\Gamma} = \bar r\!\left(\varphi\right) \, , \qquad \left.\bar \phi\right|_{\Gamma} = \bar \phi(\varphi) \, .
\end{equation}
The $n^{\bar r}$ component of the unit normal vector has the same sign as $\partial_{\varphi}\bar \phi$, and so we only consider solutions with $\partial_{\varphi} \bar \phi>0$, in order to satisfy the outward-pointing condition.

\textbf{Problem.} Now we impose the conformal boundary conditions. These lead to the following equations of $\bar r(\varphi)$ and $\bar \phi(\varphi)$,
\begin{equation}\label{eqn: non cir rindler eqn1}
    \frac{4\pi^2 \r^2}{\beta^2}\bar r^2 = (\p_\varphi \bar \phi)^2+(\p_\varphi \bar r)^2\, ,
\end{equation}
and
\begin{equation}\label{eqn: non cir rindler eqn2}
    \frac{(\p_\varphi \bar \phi)^3+\bar r \p_\varphi \bar r \p_\varphi^2 \bar \phi + \p_\varphi \bar \phi(\p_\varphi \bar r)^2-\p_\varphi \bar \phi \bar r \p_\varphi^2 \bar r}{\bar r \left((\p_\varphi \bar \phi)^2+(\p_\varphi \bar r)^2\right)^{3/2}}=K\,.
\end{equation}

\textbf{Solutions.} The general solution to \eqref{eqn: non cir rindler eqn1} and \eqref{eqn: non cir rindler eqn2} is given by
\begin{equation}
    \bar r(\varphi) = r\!\left(\frac{2\pi \r^2 \varphi}{\beta}\right) \, , \qquad \bar \phi (\varphi) = \tau\!\left(\frac{2\pi \r^2 \varphi}{\beta}\right)\, ,
\end{equation}
where the functions $r(u)$ and $\tau(u)$ are given by \eqref{eqn: flat R sol} and \eqref{eqn: flat T sol} respectively. The solution is labelled by the dimensionful parameter $\mathcal{E}$.

The Weyl factor and conformal stress tensor are given by
\begin{equation}
    \bomega(\varphi)=\log \frac{2\pi \bar r(\varphi)}{\beta} \, , \qquad T_{ij}d\sigma^i d\sigma^j = - \frac{\pi \mathcal{E}}{2\beta^2 G_N}\left(-du^2+\r^2 d\varphi^2\right) \, .
\end{equation}
The periodicity structure is given by
\begin{equation}\label{eqn: flat wobbly conf data 2}
    \b = \frac{\pi^2 Kr_+ }{n\mathcal{K}(m)} \, , \qquad \beta_{\bar \phi} = 2n \left(-\frac{2\mathcal{E}\mathcal{K}(m)}{Kr_+}+r_+ E(m)\right) \, ,
\end{equation}
where all the parameters are defined in the same way as before.

\textbf{Conformal thermodynamics.} The on-shell action is given by
\begin{equation}
    I_\text{on-shell}= - \frac{\pi^2 \mathcal{E}}{G_N \b}- \frac{\beta_{\bar \phi}}{4G_N} \, .
\end{equation}
Using the thermodynamic relations, we obtain the conformal energy and entropy,
\begin{equation}
    E_\text{conf} = - \frac{\pi^2 \mathcal{E}}{G_N \b^2} \, , \qquad \mathcal{S}_\text{conf}= \frac{\beta_{\bar \phi}}{4G_N} \, .
\end{equation}

The specific heat at fixed $K$ is given by
\begin{equation}\label{specific heat Lambda=0 non-cir}
    C_K = - \frac{2\pi^2 \mathcal{E}}{G_N \b} +\frac{4 n \mathcal{E}(1+2\mathcal{E} K)\mathcal{K}(m)^2}{r_+G_NK\left(2\mathcal{E}K\mathcal{K}(m)+r_+^2K^2E(m)\right)}<0 \, .
\end{equation}
The negative definiteness implies that the solution is always thermally unstable. The static limit yields a finite value of $C_K$, i.e. $C_K \to - \tfrac{n\pi}{2 G_NK}$ as $\b\to \tfrac{2\pi}{n}$, while it tends to zero in the high temperature limit.

\subsection{Thermodynamic phase space}\label{sec: flat thermo}

Here, we combine results from the previous sections and study thermodynamic properties of the total system. The main quantity of interest is the torus partition function $\mathcal{Z}(\b,K)$ which, in the semi-classical limit, includes contributions from all permissible classical solutions,
\begin{equation}\label{eqn: torus Z flat total}
    \mathcal{Z}(\b,K) = e^{-I_\text{on-shell}^{(\text{hom. pole})}}+e^{-I_\text{on-shell}^{(\text{hom. Rindler})}}+\sum_{n} e^{-I_{n,\text{on-shell}}^{\text{(non-static pole)}}}+\sum_{n} e^{-I_{n,\text{on-shell}}^{\text{(non-circular Rindler)}}} \, .
\end{equation}
Treating $\mathcal{Z}(\b,K)$ in a saddle point approximation, only the contribution with the lowest on-shell action dominates the partition function. We call a saddle which dominates the partition function and has a positive specific heat a stable configuration, while a sub-dominant one that has a positive specific heat is called a meta-stable configuration. A saddle with a negative specific heat is called unstable. 
In the following, we explain all the terms that appear in \eqref{eqn: torus Z flat total}.
\begin{itemize}
    \item The first and second contributions come from the static pole patch and Rindler patch solutions, whose on-shell actions are given in \eqref{eqn: flat I pole} and \eqref{eqn: flat I rindler},
    \begin{equation}
        I_\text{on-shell}^{(\text{pole})}=I_\text{on-shell}^{(\text{Rindler})}\bigg|_{\b \rightarrow \frac{4\pi^2}{\b}}= -\frac{\b}{8 G_N K} \, , 
    \end{equation}
    These solutions exist for all $\b$ and $K>0$. 
    
    \item The third and fourth contributions consist of a finite sum of non-static pole patch and non-circular Rindler patch solutions. The number of these solutions, found in \eqref{Lambda=0 number of non-static pole sols}, depends only on $\b$ and is given by
    \begin{equation}
        \label{number of sols non-circular Rindler}
        \mathfrak{n}_{\text{non-static pole}}=\mathfrak{n}_{\text{non-circular Rindler}}\bigg|_{\b \rightarrow \frac{4\pi^2}{\b}}=\left\lceil \frac{\b}{2\pi}\right\rceil-1 \, .
    \end{equation}
    Each solution is characterised by a pair $(\mathcal{E},n)$ that solves, in the case of the pole patch, the $\b$ equation in \eqref{eqn: flat wobbly conf data}, and in the case of the Rindler patch, the $\b$ equation in \eqref{eqn: flat wobbly conf data 2}. Their on-shell actions are given by
    \begin{equation}
        I_\text{on-shell}^{(\text{non-static pole})}=I_\text{on-shell}^{(\text{non-circular Rindler})}\bigg|_{\b \rightarrow \frac{4\pi^2}{\b}} = - \frac{nr_+ E(m)}{2G_N} \, ,
    \end{equation}
    where $m=1-\frac{r_-^2}{r_+^2}$ and $r_{\pm}$ are functions of $\mathcal{E}$ and $K$ given by \eqref{eqn: flat R_pm}. The non-static pole patch solutions exist only for $\b>2\pi$ and $K>0$, while the non-circular Rindler patch solutions exist only for $\b<2\pi$ and $K>0$. 
\end{itemize}

\begin{figure}[H]
        \centering
         
                \includegraphics[scale=0.5]{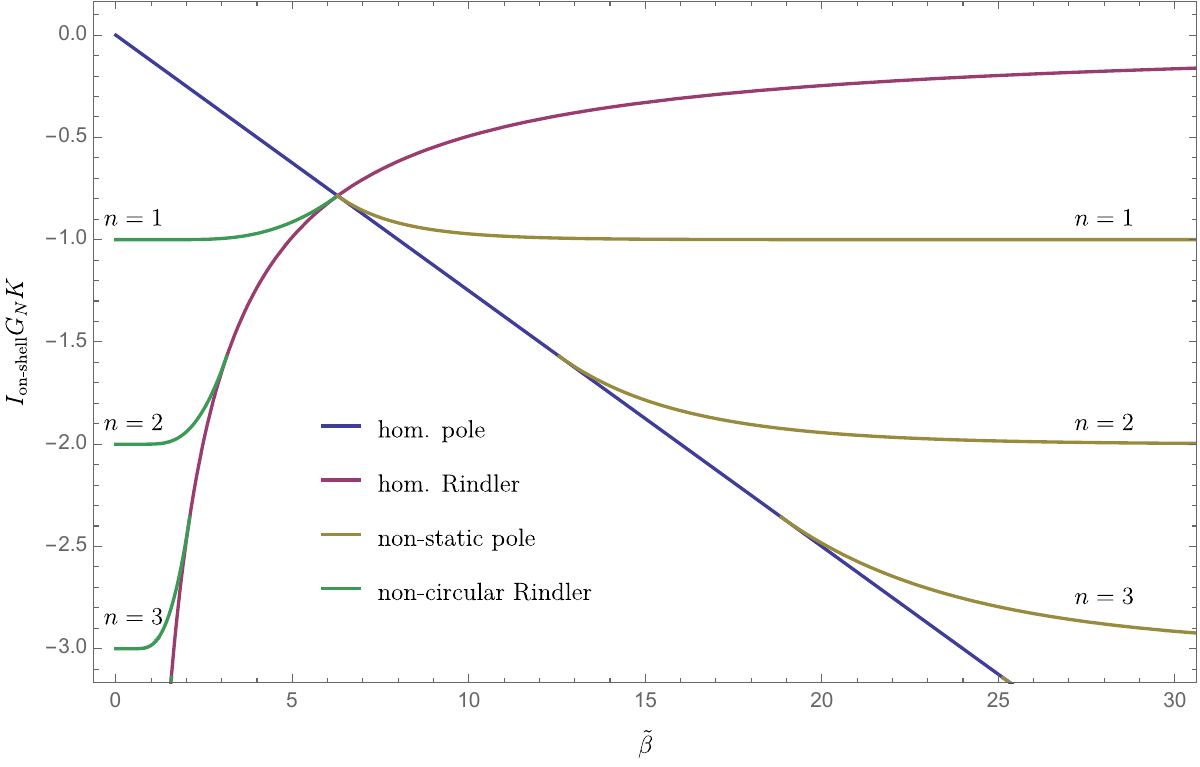}
                \caption{A plot of $I_\text{on-shell}$ of various contributions versus $\b$.} \label{fig: Ionshell flat}
\end{figure}

Now we describe thermodynamic phases of the total system. A plot of the on-shell action of various contributions is depicted in figure \ref{fig: Ionshell flat}. It turns out that $K$ only appears as an overall factor of the on-shell action, which together with $G_N$ forms a dimensionless quantity. As such, the following discussion holds for any value of $K>0$. 

In the low temperature regime, the stable configuration is given by the pole patch solution. By heating up the system, there is a critical inverse temperature at $\b_c\equiv 2\pi$. Across this temperature, there is a change of the dominant saddle from the pole patch to the Rindler patch. As a result, the system undergoes a first order phase transition, where the energy exhibits a discontinuous jump,
\begin{equation}
    E_\text{conf}=\begin{cases}
        -\frac{1}{8G_NK} \, , \qquad &  \b>\b_c \, , \\
        \frac{\pi^2}{2G_NK \b^2} \, , \qquad &  0<\b<\b_c \, .
    \end{cases}
\end{equation}
For $\b < \b_c$, the Rindler patch remains a stable configuration in the high-temperature regime.

\textbf{Zoo of unstable configurations.} The total number $\mathfrak{n}$ of coexisting solutions depends on the conformal inverse temperature. In particular, we find
\begin{equation}
    \mathfrak{n} = \begin{cases}
        2+\left(\left\lceil\frac{\b}{\b_c}\right\rceil-1\right) \, , \qquad & \b>\b_c \, , \\
        2+\left(\left\lceil\frac{\b_c}{\b}\right\rceil -1\right) \, , \qquad & \b<\b_c \, ,
    \end{cases}
\end{equation}
where the $2$ above counts the static pole patch and Rindler patch solutions. It turns out that all the non-static pole patches and non-circular Rindler patches are sub-dominant with a negative specific heat, and are therefore unstable configurations. As $\tilde{\beta} \rightarrow n \tilde{\beta}_c$ (or $\tilde{\beta} \rightarrow  \frac{\tilde{\beta}_c}{n}$), the on-shell action, energy, and entropy of these solutions match those of the static pole patch (static Rindler patch) at these temperatures. However, the specific heat does not respect this limit and is discontinuous,
\begin{equation}
    \begin{cases}
        C_K^{(\text{hom. pole})} &\overset{\b\to n\b_c}{\longrightarrow} 0 \, ,\\
        C_K^{(n-\text{pole})}&\overset{\b\to n\b_c}{\longrightarrow} -\frac{n\pi}{G_NK} \,  ,
    \end{cases}
     \qquad 
    \begin{cases}
        C_K^{(\text{hom. Rindler})} &\overset{\b\to \frac{1}{n}\b_c}{\longrightarrow} \frac{n\pi}{2G_NK} \, ,\\
        C_K^{(n-\text{Rindler})}&\overset{\b\to \frac{1}{n}\b_c}{\longrightarrow} -\frac{n\pi}{2G_NK} \, ,
    \end{cases}
\end{equation}
for any positive integer $n$.

\section{Positive cosmological constant}\label{sec: Lambda +}

In this section, we study the gravitational path integral in the case of positive cosmological constant. We first briefly review the Euclidean bulk solutions to Einstein's equations in three dimensions with a positive cosmological constant \cite{Deser1984-sj}.

Solutions to \eqref{eqn: Einstein eqn} with $\Lambda>0$ are locally given by the round three-sphere metric of radius $\Lambda^{-1/2}$. We consider a choice of coordinates where the metric is given by
\begin{equation}\label{eqn: dS metric}
    ds^2 = \frac{\rh^2-r^2}{\ell^2} \, d\tau^2 + \frac{\ell^2}{\rh^2-r^2} \, dr^2+r^2 d\phi^2 \, , \qquad \Lambda = \frac{1}{\ell^2}\, ,
\end{equation}
where $\rh$ is an arbitrary positive constant, $r\in \left(0,\rh\right)$, and $\phi \sim \phi+2\pi$. 

For $\rh=\ell$, the metric \eqref{eqn: dS metric} leads to an everywhere-smooth geometry provided that the coordinate $\tau$ is identified under $\tau \sim \tau+2\pi \ell$. Without this identification, e.g. imposing that $\tau \sim \tau + \beta_\tau$ for some positive parameter $\beta_\tau \neq 2\pi \ell$, the solution has a conical singularity at $r=\ell$.

For $\rh\neq \ell$, the metric \eqref{eqn: dS metric} develops two conical singularities located at $r=0$ and $r=\rh$. The singularity at $r=\rh$ can be eliminated by identifying the coordinate $\tau\sim\tau+\tfrac{2\pi \ell^2}{\rh}$, while the singularity at $r=0$ is always present for any $\rh\neq \ell$. 

These two cases are related by a coordinate transformation. Applying the diffeomorphism
\begin{equation}\label{eqn: dS coord transf}
    \frac{\bar r^2}{\ell^2}=1-\frac{r^2}{\rh^2} \, , \qquad \bar \tau = \rh \phi \, , \qquad \bar \phi = \frac{\rh \tau}{\ell^2}\, ,
\end{equation}
the metric \eqref{eqn: dS metric} is mapped to
\begin{equation}\label{eqn: dS metric2}
    ds^2 = \frac{\ell^2-\bar r^2}{\ell^2} \, d\bar\tau^2 + \frac{\ell^2}{\ell^2-\bar r^2} \, d\bar r^2+\bar r^2 d\bar \phi^2 \, , \qquad \Lambda = \frac{1}{\ell^2}\, ,
\end{equation}
which is the original metric \eqref{eqn: dS metric} with $\rh = \ell$. Choosing the periodicity of $\tau$ such that only the singularity at $r=0$ is present, the barred coordinates obey $\bar\phi \sim \bar \phi +2\pi$ and $\bar \tau \sim \bar \tau + 2 \pi \rh$, and the singularity is now mapped to $\bar r = \ell$. This exercise shows the geometric equivalence between \eqref{eqn: dS metric} with $\rh=\ell$ and identifications $\tau \sim \tau + \beta_\tau$ and $\phi \sim \phi+2\pi$ for any $\beta_\tau$ and \eqref{eqn: dS metric} with identifications $\tau \sim \tau+\tfrac{2\pi \ell^2}{\rh}$ and $\phi\sim \phi+2\pi$. The switching between the Euclidean temporal and spatial coordinates in \eqref{eqn: dS coord transf} will play an important role in the modular invariance of the torus partition function considered below.

We note that by taking $\tau \to i t$, the metric \eqref{eqn: dS metric} with $\rh= \ell$ describes the static patch of dS$_3$, a region which is causally accessible to an observer sitting at $r=0$. The circle $r=\ell$ is the one-dimensional cosmological horizon experienced by the observer. 

We now consider solutions with a boundary obeying the boundary conditions \eqref{eqn: bdry metric S1xS1}, employing the embedding method. We review both pole patch and cosmic patch solutions, defined as spacetimes which do and do not contain the cosmological horizon, in sections \ref{Lambda>0 pole patch} and \ref{Lambda>0 cosmic patch} respectively. In each section, we look for non-static and non-circular boundaries. For each of these solutions, we compute their thermodynamic quantities in the conformal canonical ensemble defined in \ref{ref: thermo}. Finally, in section \ref{Lambda>0 thermodynamic phase space}, we combine these results and explore the thermodynamic phase space of the system.

\subsection{Pole patch} \label{Lambda>0 pole patch}

We first study the conformal thermodynamics of the pole patch solutions. These are solutions endowed with the bulk metric
\begin{equation}\label{eqn: bulk metric pole de Sitter}
    ds^2 = \frac{\ell^2-r^2}{\ell^2} \,d\tau^2+\frac{\ell^2}{\ell^2-r^2} \, dr^2 + r^2 d\phi^2\, , \qquad \Lambda=\frac{1}{\ell^2} \, ,
\end{equation}
where $\tau\sim\tau+\beta_\tau$ for some arbitrary positive number $\beta_\tau$ and  $\phi \sim \phi+2\pi$. The spacetime region of interest includes the origin $r=0$ and is bounded by the boundary situated at the radial coordinate $\left.r\right|_{\Gamma}$, which will be specified later. As this region does not contain the horizon, we do not fix $\beta_{\tau}$. At the boundary, we impose the conformal boundary conditions \eqref{eqn: bdry metric S1xS1}.

\subsubsection{Static and circular pole patch}\label{Lambda>0 static and circular pole}

For any constant value for the trace of the extrinsic curvature $K$, there exists a pole patch solution parameterised by
\begin{equation}
    \label{Lambda>0 homogeneous boundary pole patch}
    \left.\tau\right|_{\Gamma}=\frac{\ell}{2\mathfrak{r}}\left(\sqrt{K^2\ell^2+4}-K\ell\right)u \, , \qquad \left.r\right|_{\Gamma}=\ell \sqrt{\frac{1}{2}-\frac{K\ell}{2\sqrt{K^2\ell^2+4}}} \, , \qquad \left.\phi\right|_{\Gamma} = \varphi \, ,
\end{equation}
where the bulk is defined over $r \in [0,\left.r\right|_{\Gamma})$. The unit normal vector $n^{\mu}$ is chosen to be pointing outward, i.e., $n^{r}>0$, using which we evaluate the induced metric on the boundary,
\begin{equation}
    \label{induced metric homogeneous pole patch Lambda>0}
    \left.ds^2\right|_{\Gamma}= \frac{\ell^2}{2\r^2}\left(1-\frac{K\ell}{\sqrt{K^2\ell^2+4}}\right)\left(du^2+\r^2d\varphi^2\right) \, ,
\end{equation}
which is intrinsically flat, and the conformal stress tensor,
\begin{equation}
    \label{conformal stress tensor homogeneous pole patch Lambda > 0}
    T_{ij}d\sigma^{i}d\sigma^{j}=-\frac{\ell\left(\sqrt{K^2\ell^2+4}-K\ell\right)}{32\pi G_N\mathfrak{r}^2} \left(-du^2+\mathfrak{r}^2d\varphi ^2\right) \, .
\end{equation}
Unlike the flat case, these solutions exist for all real values of $K\ell$.

\textbf{Conformal thermodynamics.} The periodicities of the bulk and boundary coordinates are related by
\begin{equation}
    \label{eqn: bdry data homo flat}
    \tilde{\beta}=\left(\sqrt{K^2\ell^2+4}+K\ell\right)\frac{\beta_{\tau}}{2\ell} \, .
\end{equation}
The on-shell action of the homogeneous pole patch solution is given by
\begin{equation}\label{eqn: Lambda>0 I pole}
    I_\text{on-shell}=  -\frac{\tilde{\beta} \ell}{16 G_N}\left(\sqrt{K^2\ell^2+4}-K\ell\right)\, .
\end{equation}
Using the thermodynamic relations \eqref{eqn: thermo relation}, the corresponding conformal energy, entropy, and specific heat are given by
\begin{equation}
    E_\text{conf}=-\frac{\ell}{16 G_N}\left(\sqrt{K^2\ell^2+4}-K\ell\right) \, , \qquad \mathcal{S}_\text{conf}=C_K= 0 \, .
\end{equation}

\subsubsection{Non-static pole patch}\label{Lambda>0 non-static pole}

We now consider the class of pole patch solutions with a boundary radius that varies with the thermal boundary coordinate $u$. In particular, we parameterise the boundary by
\begin{equation}
    \label{Lambda>0 bdry 1}
    \left.\tau\right|_{\Gamma}=\tau(u)\, , \qquad \left.r\right|_{\Gamma}=r(u) \, , \qquad \left.\phi\right|_{\Gamma}=\varphi\, .
\end{equation}
The bulk is defined over the coordinate range $r \in [0,\left.r\right|_{\Gamma})$. The unit normal vector is given by
\begin{equation}\label{eqn: Lambda>0 n^mu}
    n^\mu \p_\mu = \frac{-\partial_{u}r \partial_{\tau}+\partial_{u}\tau \partial_{r}}{\sqrt{\frac{\ell^2-r^2}{\ell^2}(\partial_{u}r)^2+\frac{\ell^2}{\ell^2-r^2}(\partial_{u}\tau)^2}} \, ,
\end{equation}
where $\p_u \tau>0$ so that $n^\mu$ is pointing-outward.

\textbf{Problem.} The conditions \eqref{eqn: embed conf class cond} and \eqref{eqn: embed K}, derived from the conformal boundary conditions, impose constraints on $\tau(u)$ and $r(u)$: one from the form of the induced metric,
\begin{equation}
    \label{eqn: Lambda>0 eq1}
    \frac{1}{f}(\mathfrak{r}\partial_{u}r)^2+f(\mathfrak{r}\partial_{u}\tau)^2=r^2 \, ,
\end{equation}
and from the constant trace of the extrinsic curvature  $K$,
\begin{equation}
    \label{eqn: Lambda>0 eq2}
    \frac{3 r (\partial_{u}r)^2 (\partial_{u}\tau) f'+r f^2 (\partial_{u}\tau)^3 f'+2 f \left(r \partial_{u}r \partial_{u}^2\tau+\partial_{u}\tau \left((\partial_{u}r)^2-r \partial_{u}^2r\right)\right)+2 f^3 (\partial_{u}\tau)^3}{2 r \sqrt{\frac{1}{f}(\partial_{u}r)^2+f (\partial_{u}\tau)^2} \left(f^2 (\partial_{u}\tau)^2+(\partial_{u}r)^2\right)}=K \, .
\end{equation}
where $f(r)=\frac{\ell^2-r^2}{\ell^2}$ and its derivative are evaluated at $r(u)$. 

Notice that any solution $(\tau(u),r(u))$ implies the existence of another solution $(-\tau(u),r(u))$ which, because of time-reversal symmetry in \eqref{eqn: dS metric}, is physically identical. While it seems from \eqref{eqn: Lambda>0 eq2} that the latter has the opposite $K$, one must also flip the orientation of the normal vector $n^\mu$, and so the two solutions have the same extrinsic curvature.

In what follows, we first find general solutions to equations \eqref{eqn: Lambda>0 eq1} and \eqref{eqn: Lambda>0 eq2}, then impose the embedding condition \eqref{eqn: global emb}, which restricts the space of solutions.

\textbf{Solution.} One can show, following a proof similar to the one in appendix \ref{sec: du T flat 1} for the flat case, that equations \eqref{eqn: Lambda>0 eq1} and \eqref{eqn: Lambda>0 eq2} can be rewritten as a first-order ODE of the form
\begin{equation} \label{constant of motion Lambda>0}
    \frac{d\mathcal{E}}{du}=0 \, , \qquad \mathcal{E}\equiv\frac{1}{2}Kr^2-\frac{\ell^2-r^2}{\ell^2} \,\mathfrak{r}\partial_{u}\tau  \, ,
\end{equation}
where $\mathcal{E}$ is an integration constant with length dimension. We now use this and equation \eqref{eqn: Lambda>0 eq1} to write an ordinary differential equation in $r(u)$, given by
\begin{equation}
    \label{EOM Lambda>0}
    (\r\p_ur)^2 + V_\text{eff}(r)=0 \, , \qquad V_\text{eff}(r)\equiv \left(\frac{K^2}{4}+\frac{1}{\ell^2}\right)r^4 - \left(1+\mathcal{E}K\right)r^2 + \mathcal{E}^2\, .
\end{equation}
This is the equation of motion of a classical particle moving in a potential well $V_{\text{eff}}(r)$, oscillating between positions $r_{-}$ and $r_{+}$ given by
\begin{equation}
    \label{R plus and minus Lambda>0}
    r_-<r(u)<r_+ \, , \qquad \frac{r_\pm}{\ell} = \sqrt{\frac{2+2\mathcal{E}K\pm2\sqrt{1-\frac{4\mathcal{E}^2}{\ell^2 }+2\mathcal{E}K}}{K^2\ell^2+4}} \, .
\end{equation}
For these values to be real, the constant of motion $\mathcal{E}$ is restricted to a compact interval:
\begin{equation}
    \label{E plus minus Lambda>0}
    \mathcal{E}_- \leq\mathcal{E}\leq \mathcal{E}_+\, , \qquad \mathcal{E}_\pm \equiv \frac{\ell}{4}\left(K\ell \pm \sqrt{K^2\ell^2+4}\right)\, .
\end{equation}
The two roots $r_{\pm}$ coincide at both $\mathcal{E}_{\pm}$. Moreover, $\mathcal{E}=0$ leads to the vanishing of the smaller root $r_-$, while $\mathcal{E}=\mathcal{E}_0 \equiv \frac{1}{2}K\ell^2$ takes $r_+$ to its maximum value $r_+=\ell$.

By taking $K\ell\to+\infty$ and $\tfrac{\mathcal{E}}{\ell}\to0$ while keeping $\mathcal{E}K$ fixed, equations \eqref{R plus and minus Lambda>0} and \eqref{E plus minus Lambda>0} reduce to those of flat space in \eqref{eqn: flat R_pm} and \eqref{EK>-1/2 flat}.

The closed-form solution to \eqref{EOM Lambda>0} for arbitrary $\mathcal{E}$ in the range \eqref{E plus minus Lambda>0} is given by
\begin{equation}
    \label{general sol for r Lambda>0}
    r(u)=r_+  \text{dn}\left(\frac{r_+\sqrt{K^2\ell^2+4}}{2\ell}\frac{u-u_0}{\r} \, \bigg| \, m\right)\, ,
\end{equation}
where $\mathrm{dn}(x \, | \, m)$ is the Jacobi elliptic dn function, $m=1-\frac{r_-^2}{r_+^2}$, and $u_0$ is an integration constant. It follows from \eqref{constant of motion Lambda>0} that
\begin{equation}
    \label{general sol for tau(u) Lambda>0}
    \tau(u) = \tau_0- \frac{K\ell^2(u-u_0)}{2\r}+\frac{(K\ell^2-2\mathcal{E})\Pi\left(\frac{r_+^2-r_-^2}{r_+^2-\ell^2} \, ; \, \text{am}(x|m) \, \big| \, m\right)}{\left(\frac{\ell^2-r_+^2}{\ell^2}\right)\frac{r_+}{\ell}\sqrt{K^2\ell^2+4}}\, ,
\end{equation}
where $x=\tfrac{r_+\sqrt{K^2\ell^2+4}}{2\ell}\tfrac{u-u_0}{\mathfrak{r}}$, $\Pi(n \, ; \, z \, | \, m)$ is the incomplete elliptic integral of the third kind, $\mathrm{am}(x;m)$ is the Jacobi amplitude function, and $\tau_0$ is an integration constant. In what follows, we shift the $\tau$ and $u$ coordinates in such a way to set $\tau_0=0$ and $u_0=0$.

This solution has a periodic structure, given by
\begin{equation}
    \label{Periodicity properties Lambda>0}
    \begin{cases}
        r\!\left(u+\frac{4\r\ell n \mathcal{K}(m)}{r_+ \sqrt{K^2\ell^2+4}}\right) = r \, , \\
        \tau \!\left(u+\frac{4\r\ell n \mathcal{K}(m)}{r_+ \sqrt{K^2\ell^2+4}}\right) \! =\tau(u) + \frac{2n\ell^2}{r_{+}\sqrt{K^2\ell^2+4}} \!\left(\!-K\ell \mathcal{K}(m)+\frac{\ell\left(K\ell^2-2\mathcal{E}\right)}{\ell^2-r_+^2}\Pi\left(\frac{r_+^2-r_-^2}{r_+^2-\ell^2} \, \bigg| \, m\right)\right) \, ,
    \end{cases}
\end{equation}
where $n$ is an arbitrary integer, $\mathcal{K}(m)$ is the complete elliptic integral of the first kind, and $\Pi(n \, | \, m)$ is the complete elliptic integral of the third kind. These solutions are evidently analogous to the unduloids found in flat space, and so the parametric plots $(\tau(u),r(u))$ in this case are qualitatively similar to the ones in figure \ref{fig: Rsol Tsol flat}.

Identifying \eqref{general sol for r Lambda>0} and \eqref{general sol for r Lambda>0} as the location of the boundary, we obtain a family of boundaries obeying the conformal boundary conditions, labelled by the dimensionless parameter $\mathcal{E}K$. The Weyl factor and conformal stress tensor are given by
\begin{equation}\label{eqn: Lambda>0 pole Weyl}
    \bomega(u) = \log \frac{r}{\r}\, , \qquad T_{ij}d\sigma^id\sigma^j=\frac{\mathcal{E}}{8 \pi G_N\r^2}\left(-du^2+\r^2 d\varphi^2\right)\, .
\end{equation}

\textbf{Self-intersections and regime of validity.} As in the flat case, this solution violates the global embedding condition \eqref{eqn: global emb} for some values of $\mathcal{E}$.

For $K\ell>0$, there are no self-intersections for $\mathcal{E}_{-} \leq \mathcal{E} \leq 0$. Self-intersections then emerge over $0<\mathcal{E}<\mathcal{E}_0$, and disappear over $\mathcal{E}_0 \leq \mathcal{E} \leq \mathcal{E}_{+}$.\footnote{Note that there are also solutions with $K\ell=0$ that are qualitatively similar to the $K\ell>0$ solutions.}

For $K\ell<0$, there are no self-intersections for $\mathcal{E}_{-} \leq \mathcal{E} \leq \mathcal{E}_0$. Then, self-intersections appear over $\mathcal{E}_0 < \mathcal{E} < 0$, and disappear over $0 \leq \mathcal{E} \leq \mathcal{E}_{+}$.

Crucially, the non-self-intersecting solutions that appear over $\max(0,\mathcal{E}_0) < \mathcal{E} \leq \mathcal{E}_{+}$ have $\partial_{u}\tau < 0$, which means that an outward-pointing vector should have an orientation opposite to the one chosen in \eqref{eqn: Lambda>0 n^mu}. This means that the trace of the extrinsic curvature of these solutions is really $-K$ and not $K$. Because of time-reversal symmetry, these solutions are identical to those with the opposite $\partial_{u} \tau>0$ and hence the opposite $\mathcal{E}$. 

Therefore, the independent physical solutions have
\begin{equation}
    \begin{cases}
        \mathcal{E}_- \leq \mathcal{E}<0 \, , \qquad K\ell>0 \, ,\\
        \mathcal{E}_- \leq \mathcal{E} \leq \mathcal{E}_0 \, , \qquad K\ell<0 \, .
    \end{cases}
\end{equation}

\textbf{Simple solutions.} There are particular values of $\mathcal{E}$ where the solutions simplify. 

For instance, the $\mathcal{E}=\mathcal{E}_{-}$ solution is exactly the homogeneous one in \eqref{Lambda>0 homogeneous boundary pole patch}. 

At exactly $\mathcal{E}=\mathcal{E}_0$, the expression for $\tau(u)$ reduces to $\tau(u)=\tau_0-\frac{K\ell^2}{2}\frac{u-u_0}{\mathfrak{r}}$.\footnote{Note that, although this is a solution to the equations of motion, it is not a smooth limit $\mathcal{E} \rightarrow \mathcal{E}_0$ of \eqref{general sol for tau(u) Lambda>0}.}

And at $\mathcal{E}=0$, the smaller root $r_{-}$ vanishes, and the boundary has a sphere topology. This solution therefore does not contribute to the thermal partition function we are studying.

\textbf{Conformal thermodynamics.} The global structure of the bulk must be consistent with the periodicity structure of the boundaries found above, which requires that
\begin{equation}\label{eqn: Lambda>0 periodic cond}
    \tau(u+\beta)=\tau(u)+\beta_\tau \, , \qquad r(u+\beta) = r(u) \, .
\end{equation}
This boundary data was found in \eqref{Periodicity properties Lambda>0} to be
\begin{equation}
    \label{conformal periodicity}
    \b = \frac{4\ell n \mathcal{K}(m)}{r_{+}\sqrt{K^2\ell^2+4}} \,, \qquad \beta_\tau = \frac{2n\ell^2}{r_{+}\sqrt{K^2\ell^2+4}} \left(-K\ell \mathcal{K}(m)+\frac{\ell\left(K\ell^2-2\mathcal{E}\right)}{\ell^2-r_+^2}\Pi\left(\frac{r_+^2-r_-^2}{r_+^2-\ell^2} \, \bigg| \, m\right)\right)  \, .
\end{equation}
Moreover, there are different ranges of validity of $\mathcal{E}$ for the thermodynamic problem depending on the sign of $K\ell$. This imposes the following constraints on the conformal periodicity.

\textbf{Positive $K\ell$.} Given $K\ell>0$ and fixed winding number $n$, the conformal periodicity $\tilde{\beta}$ increases monotonically as a function of $\mathcal{E}$ over the allowed interval $(\mathcal{E}_{-},0)$. Then, for every $n \in \mathbb{N}^{*}$, $\b$ is bounded from below,
\begin{equation}
    \label{lower bound on beta Lambda>0}
    \tilde{\beta} > \tilde{\beta}^{(-)}_{n} \, \, \, \, , \, \, \, \, \tilde{\beta}^{(-)}_{n} \equiv n \,\sqrt{\frac{8\pi^2}{K^2\ell^2+4-K\ell \sqrt{K^2\ell^2+4}}}  \, ,
\end{equation}
and unbounded from above since $\tilde{\beta} \rightarrow \infty$ as $\mathcal{E} \rightarrow 0^{-}$. Accordingly, given some $\tilde{\beta}>0$ and $K\ell>0$, the number of non-static pole patch solutions, which are characterised by discrete values of $(\mathcal{E},n)$, is
\begin{equation}
    \label{number of non-static pole-patch sols Lambda>0 KL>0}
    \mathfrak{n}_{{\text{non-static pole, }K\ell>0}}= \left\lceil\frac{\tilde{\beta}}{\tilde{\beta}^{(-)}_{n=1}} \right \rceil-1 \, \, .
\end{equation}
Notably, the number of solutions grows with $\tilde{\beta}$, just like in flat space.

\textbf{Negative $K\ell$.} Given $K\ell<0$ and fixed winding number $n$, the conformal periodicity $\tilde{\beta}$ increases monotonically as a function of $\mathcal{E}$ over the allowed interval $(\mathcal{E}_{-},\mathcal{E}_0)$. Then, for each $n \in \mathbb{N}^{*}$, it is again bounded from below by \eqref{lower bound on beta Lambda>0} but also bounded from above,
\begin{equation}
    \label{bounds on beta KL<0 Lambda>0}
    \tilde{\beta}^{(-)}_{n}<\tilde{\beta} \leq \tilde{\beta}^{(0)}_{n} \, \, \, \, , \, \, \, \, \tilde{\beta}_{n}^{(0)} \equiv \frac{4n}{\sqrt{K^2\ell^2+4}}\,\mathcal{K}\!\left(\frac{4}{K^2\ell^2+4}\right) \, .
\end{equation}
Then, given some $\tilde{\beta}>0$ and $K\ell<0$, the number of non-static pole patch solutions is
\begin{equation}
    \label{number of non-static pole-patch sols Lambda>0 KL<0}
    \mathfrak{n}_{\text{non-static pole, }K\ell<0} = \left\lceil \frac{\tilde{\beta}}{\tilde{\beta}^{(-)}_{n=1}} \right\rceil - \left \lceil \frac{\tilde{\beta}}{\tilde{\beta}_{n=1}^{(0)}} \right \rceil \, .
\end{equation}
The number of solutions tends to grow with $\tilde{\beta}$, at a rate that grows as $K \ell \rightarrow 0^{-}$ where $\tilde{\beta}^{(0)}_{n=1} \rightarrow \infty$.

\medskip

We now evaluate the on-shell action of each of these solutions. Starting from \eqref{eqn: action} and plugging in the bulk metric \eqref{eqn: dS metric} and the induced metric \eqref{eqn: bdry metric S1xS1} with $\omega(u)=\log\frac{r}{\mathfrak{r}}$, one can then use the relation \eqref{constant of motion Lambda>0} to write the action in terms of $\tilde{\beta}$ and $\beta_{\tau}$ as\footnote{Interestingly, the solution with $\mathcal{E}=\mathcal{E}_0$, whose thermodynamics we study only when $K\ell<0$, has $\beta_{\tau}=-\frac{1}{2}K\ell^2 \tilde{\beta}$ and therefore a vanishing on-shell action.}
\begin{equation}
    \label{on-shell action Lambda > 0, pole patch}
    I_{\text{on-shell}}=-\frac{1}{4G_N} \left(\tilde{\beta}\mathcal{E}+\beta_{\tau} \right) = \frac{\ell n \left(K \ell^2-2 \mathcal{E}\right) \left(\mathcal{K}(m) \left(r_{+}^2-\ell^2\right)+\ell^2 \Pi \left(\left.\frac{r_{+}^2-r_{-}^2}{r_{+}^2-\ell^2}\right| m \right)\right)}{2 G_N r_{+} \sqrt{K^2 \ell^2+4} \left(r_{+}^2-\ell^2\right)} \, .
\end{equation}
We then use the thermodynamic relations \eqref{eqn: thermo relation} to find the conformal energy and entropy,
\begin{equation}
    \label{lambda > 0 pole patch thermo}
    E_{\text{conf}}=\frac{\mathcal{E}}{4G_N} \, \, , \, \, \mathcal{S}_{\text{conf}}=\frac{2\tilde{\beta}\mathcal{E}+\beta_{\tau}}{4G_N} \, \, .
\end{equation}
As in the flat case, the constant of motion $\mathcal{E}$ plays the role of a conformal energy in units of $4G$. The specific heat is given by
 \begin{equation}
     \label{specific heat Lambda>0 non-static}
     C_K=\frac{4 \ell n \mathcal{E}(1-\frac{4\mathcal{E}^2}{\ell^2}+2\mathcal{E} K)\mathcal{K}(m)^2}{r_+G_N\sqrt{K^2\ell^2+4}\left(2\mathcal{E}\left(K-\frac{4\mathcal{E}}{\ell^2}\right)\mathcal{K}(m)+r_+^2\left(K^2+\frac{4}{\ell^2}\right)E(m)\right)}<0 \, .
 \end{equation}
The negative definiteness of $C_K$ indicates that all these non-static solutions, similar to their counterparts in flat space, are thermally unstable.

\textbf{Spherical boundary limit.} In the $\mathcal{E} \rightarrow 0^{-}$ limit, where $\b=-\log\left(\mathcal{E}^2 \left(K^2+\frac{4}{\ell^2} \right) \right)+\mathcal{O}(\mathcal{E})$ we recover the on-shell action of the spherical boundary reported in \cite{Anninos:2024wpy}, plus corrections,
\begin{equation}
    \label{lambda > 0 pole patch e->0}
    I_{\text{on-shell}}=-\frac{\ell}{2G_N}\arctan \left(\frac{2}{K\ell}\right)-\frac{\mathcal{E}}{2G_N}+\mathcal{O}(\mathcal{E}^2) \, .
\end{equation}

\subsubsection{Non-circular pole patch}\label{Lambda>0 non-circular pole}

Now we study a class of solutions with a boundary that varies with respect to the boundary coordinate $\varphi$. We parameterise this boundary by
\begin{equation}\label{eqn: Lambda>0 bdry non-circular pole}
    \left.\tau\right|_{\Gamma}=\frac{\beta_{\tau}u}{\beta} \, , \qquad \left.r\right|_{\Gamma}=r(\varphi) \, , \qquad \left.\phi\right|_{\Gamma}=\phi(\varphi)\, .
\end{equation}
We consider the region $r \in [0,r(\varphi))$. The $n^{r}$ component of the unit normal vector has the same sign as $\partial_{\varphi}\phi$, and so we only take into consideration solutions with $\partial_{\varphi}\phi>0$, so that the outward-pointing condition is satisfied.

\textbf{Problem.} The conformal boundary conditions impose restrictions on \eqref{eqn: Lambda>0 bdry non-circular pole} via the equations \eqref{eqn: embed conf class cond} and \eqref{eqn: embed K}. The condition on the conformal class is given by
\begin{equation}\label{eqn: Lambda>0 non-circular pole eq1}
    \frac{\beta_{\tau}^2 \mathfrak{r}^2}{\beta^2}f=\frac{ (\partial_{\varphi}r)^2}{f}+r^2(\partial_{\varphi}\phi)^2 \, ,
\end{equation}
while the condition on the trace of the extrinsic curvature gives
\begin{equation}\label{eqn: Lambda>0 non-circular pole eq2}
    \frac{2 r \partial_{\varphi}\phi (\partial_{\varphi}r)^2  f'+f \left(r^3 (\partial_{\varphi}\phi)^3 f'-2 r \partial_{\varphi}\phi \partial_{\varphi}^2r +2 r \partial_{\varphi}^2 \phi \partial_{\varphi}r+4 \partial_{\varphi}\phi (\partial_{\varphi}r)^2\right)+2 r^2 f^2 (\partial_{\varphi}\phi)^3}{2 \left(r^2 f (\partial_{\varphi}\phi)^2+(\partial_{\varphi}\phi)^2\right)^{3/2}} = K \, ,
\end{equation}
where $f(r)=1-\frac{r^2}{\ell^2}$ and its derivative are evaluated at $r(\varphi)$. 

In what follows, we first find general solutions to equations \eqref{eqn: Lambda>0 non-circular pole eq1} and \eqref{eqn: Lambda>0 non-circular pole eq2}, then impose the embedding condition, which restricts the space of solutions.

\textbf{Solution.} One can show, following a proof similar to the one in appendix \ref{sec: du T flat 1} for the flat case, that equations \eqref{eqn: Lambda>0 non-circular pole eq1} and \eqref{eqn: Lambda>0 non-circular pole eq2} can be rewritten as a first-order ODE of the form
\begin{equation}\label{eqn: conserved quantity non-circular pole Lambda>0}
    \frac{d\mathcal{E}}{d\varphi}=0 \, , \qquad \mathcal{E}\equiv -\frac{K \beta_{\tau}^2\mathfrak{r}^2}{2\beta^2}\left(1-\frac{r^2}{\ell^2}\right)-\frac{\beta_{\tau}\mathfrak{r}}{\ell^2 \beta} \, r^2 \partial_{\varphi}\phi  \, ,
\end{equation}
where $\mathcal{E}$ is an integration constant of dimension length. We now use this and equation \eqref{eqn: Lambda>0 non-circular pole eq1} to write an ODE in the variable $y\equiv\tfrac{\beta_\tau}{2\pi}\sqrt{1-\tfrac{r^2}{\ell^2}}$, given by
\begin{equation}
    \frac{\beta^2}{4\pi^2\r^2}(\p_\varphi y)^2 +V_\text{eff}(y)  =0 \, , \qquad V_\text{eff}(y)\equiv\left(\frac{K^2}{4}+\frac{1}{\ell^2}\right)y^4 - \left(\frac{\beta_\tau^2}{4\pi^2\ell^2}-\frac{\mathcal{E} K \beta^2}{4\pi^2\r^2}\right)y^2 + \left(\frac{\mathcal{E}\beta^2}{4\pi^2\r^2}\right)^2 \, .
\end{equation}
The solution can be written in terms of the elliptic Jacobi dn function,
\begin{equation}\label{eqn: non cir pole r sol}
    y(\varphi)= \frac{\beta_\tau}{2\pi}\sqrt{1-\frac{r(\varphi)^2}{\ell^2}}=y_+ \text{dn}\left(\frac{y_+\sqrt{K^2\ell^2+4}}{2\ell}\frac{2\pi \r \varphi}{\beta} \, \bigg| \, m\right) \, ,
\end{equation}
where the roots of the effective potential $y_\pm$ and the parameter $m$ are given by
\begin{equation}\label{eqn: ypm non cir pole}
    y_\pm \equiv \ell \sqrt{\frac{\frac{\beta_\tau^2}{2\pi^2\ell^2}-\frac{\mathcal{E}K\beta^2}{2\pi^2\r^2}\pm\frac{\beta_\tau}{\pi\ell}\sqrt{\frac{\beta_\tau^2}{4\pi^2\ell^2}-\frac{\mathcal{E}^2\beta^4}{\pi^2\beta_\tau^2\r^4}-\frac{\mathcal{E}K\beta^2}{2\pi^2\r^2}}}{K^2\ell^2+4}} \, , \qquad m\equiv1-\frac{y_-^2}{y_+^2}\, .
\end{equation}
Plugging \eqref{eqn: non cir pole r sol} back in \eqref{eqn: conserved quantity non-circular pole Lambda>0} and integrating over $\varphi$, we obtain
\begin{equation}\label{eqn: non cir pole phi sol}
    \phi(\varphi)=\phi_0 +\frac{K \beta_\tau \r(\varphi-\varphi_0)}{2\beta} - \frac{\left(\frac{\mathcal{E}\beta^2}{2\pi^2\r^2}+\frac{\beta_\tau^2K}{4\pi^2}\right)\Pi\left(\frac{y_+^2-y_-^2}{y_+^2-\frac{\beta_\tau^2}{4\pi^2}};\text{am}(x|m)|m\right)}{y_+ \frac{\beta_\tau}{2\pi\ell}\left(1-\frac{4\pi^2y_+^2}{\beta_\tau^2}\right)\sqrt{K^2\ell^2+4}} \, ,
\end{equation}
where $x = \frac{\pi y_+ \mathfrak{r}  \sqrt{K^2 \ell^2+4}}{\beta \ell}\varphi$.

For reality, the parameter $\mathcal{E}$ must lie in the interval
\begin{equation}
    \label{E plus minus Lambda>0 non-circular dS pole}
    \mathcal{E}_- \leq\mathcal{E}\leq \mathcal{E}_+\, , \qquad \mathcal{E}_\pm \equiv \frac{\beta_{\tau}^2\mathfrak{r}^2}{4\ell \beta^2}\left(-K\ell \pm \sqrt{K^2\ell^2+4} \right) \, .
\end{equation}
Note that $y_{-}$ saturates its lower bound $0$ at $\mathcal{E}=0$, which means that the boundary reaches the horizon radius. And $y_{+}$ saturates its upper bound $\frac{\beta_\tau}{2\pi}$ at the particular value $\mathcal{E}_0 \equiv -\frac{\beta_{\tau}^2 K \mathfrak{r}^2}{2 \beta ^2}$, indicating that the boundary reaches the origin.

The solution admits a periodic structure,
\begin{equation}
    \label{Periodicity properties Lambda>0 non-circular dS pole}
    \begin{cases}
        r\!\left(\varphi+\frac{2n\beta \ell \mathcal{K}(m)}{\pi y_+ \mathfrak{r}\sqrt{K^2\ell^2+4}}\right) = r(\varphi) \, , \\
        \phi \!\left(\varphi+\frac{2n\beta \ell \mathcal{K}(m)}{\pi y_+ \mathfrak{r}\sqrt{K^2\ell^2+4}}\right) \! =\phi(\varphi)+n\left[\frac{\beta_{\tau}  K \ell \mathcal{K}\left(m\right)}{\pi  y_{+} \sqrt{K^2 \ell^2+4}}-\frac{4 \pi  \ell \Pi \left(\frac{y_{+}^2-y_{-}^2}{y_{+}^2-\frac{\beta_{\tau} ^2}{4 \pi ^2}} \, \big| \, m\right) \left(\frac{\beta ^2 \mathcal{E}}{2 \pi ^2 \mathfrak{r}^2}+\frac{\beta_{\tau} ^2 K}{4 \pi ^2}\right)}{\beta_{\tau}  y_{+} \sqrt{K^2 \ell^2+4} \left(1-\frac{4 \pi ^2 y_{+}^2}{\beta_{\tau}^2}\right)}\right] \, ,
    \end{cases}
\end{equation}
where $n$ is an arbitrary positive integer.

The Weyl factor and conformal stress tensor are given by
\begin{equation}
    \bomega(\varphi) = \log \left(\sqrt{1-\frac{r^2}{\ell^2}}\frac{\beta_\tau}{\beta}\right) \, , \qquad T_{ij}d\sigma^i d\sigma^j =\frac{\mathcal{E}}{8\pi G_N \r^2}\left(-du^2+\r^2d\varphi^2\right)\, .
\end{equation}

\textbf{Self-intersection and regime of validity.} These solutions are only valid for certain ranges of the parameter $\mathcal{E}$. This arises from the fact that $\phi(\varphi)$ needs to be monotonically increasing with $\varphi$ as per the outward-pointing condition for the normal vector. Accordingly, the physical solutions have
\begin{equation}
    \begin{cases}
        \mathcal{E}_{-} < \mathcal{E} < \mathcal{E}_0 \, , \qquad K\ell>0 \, , \\
        \mathcal{E}_{-}<\mathcal{E}<0 \, , \, \, \qquad K\ell<0 \, .
    \end{cases}
\end{equation}

\textbf{Simple solutions.} There are particular values of $\mathcal{E}$ where the solution simplifies.

At $\mathcal{E}=\mathcal{E}-$, the solution is precisely the homogeneous one in \eqref{Lambda>0 homogeneous boundary pole patch}.

At $\mathcal{E}=\mathcal{E}_0$, the expression in \eqref{eqn: non cir pole phi sol} reduces to $\phi(\varphi)=\phi_0+\frac{K\beta_\tau \mathfrak{r}(\varphi-\varphi_0)}{2\beta}$.

There is no smooth solution at $\mathcal{E}=0$.

\textbf{Global smoothness condition.} In fact, to ensure that the global structure is well-defined, there are two more conditions we need to impose: the periodicities in $\phi$ and $\varphi$, which have been evaluated in \eqref{Periodicity properties Lambda>0 non-circular dS pole} must both be equal to $2\pi$. Accordingly:
\begin{equation}
    \label{periodicities non-circular pole dS}
    2\pi \equiv \frac{2n\beta \ell \mathcal{K}(m)}{\pi y_+ \mathfrak{r}\sqrt{K^2\ell^2+4}} \,, \qquad 2\pi=n\left[\frac{\beta_{\tau}  K \ell \mathcal{K}\!\left(m\right)}{\pi  y_{+} \sqrt{K^2 \ell^2+4}}-\frac{4 \pi  \ell \, \Pi\! \left(\frac{y_{+}^2-y_{-}^2}{y_{+}^2-\frac{\beta_\tau^2}{4 \pi ^2}} \, \big| \, m\right) \left(\frac{\beta ^2 \mathcal{E}}{2 \pi ^2 \mathfrak{r}^2}+\frac{\beta_{\tau} ^2 K}{4 \pi ^2}\right)}{\beta_{\tau}  y_{+} \sqrt{K^2 \ell^2+4} \left(1-\frac{4 \pi ^2 y_{+}^2}{\beta_{\tau}^2}\right)}\right] \, .
\end{equation}
To solve these equations, we find it convenient to implement a change of variable $\mathcal{E}=\frac{\mathfrak{e}\beta_{\tau}^2 \mathfrak{r}^2}{\beta^2}$. The first equation in \eqref{periodicities non-circular pole dS} then determines $\beta_{\tau}$ in terms of the boundary data $\tilde{\beta}$,
\begin{equation}
    \label{beta tau in terms of beta non-circular dS pole}
    \beta_{\tau} = \frac{\sqrt{2} \ell n \mathcal{K}\!\left(m\right)}{\pi \sqrt{1+\left(\sqrt{1-2\ell^2\mathfrak{e} (2 \mathfrak{e}+K)}-\mathfrak{e} K \ell^2\right)}} \, \tilde{\beta} \, .
\end{equation}
Note that $m$ becomes independent of $\beta$, $\beta_{\tau}$, and $\mathfrak{r}$ when written in terms of $\mathfrak{e}$, and so $\beta_{\tau}$ is directly proportional to $\tilde{\beta}$. As for the  second equation in \eqref{periodicities non-circular pole dS}, it becomes entirely independent of $\beta_{\tau}$, $\beta$,  and $\mathfrak{r}$, which means that constrains the space of solutions into particular values of $(\mathfrak{e},n)$.

To find the number of solutions, we first observe that the function $\beta_{\phi}-2\pi$ is monotonically decreasing as a function of $\mathfrak{e}$ for all values of $K\ell$ and $n$. We then evaluate the value this function takes at the boundaries of the allowed interval of $\mathfrak{e}$, which depends on the sign of $K\ell$.

\textbf{Positive $K\ell$.} For $K\ell>0$, where $\mathfrak{e}$ is restricted to lie in the interval $\frac{1}{4\ell}\left(-K\ell-\sqrt{K^2\ell^2+4} \right)<\mathfrak{e}<-\frac{K}{2}$, we find that there are no solutions at all. In particular, the value of $\beta_{\phi}$ at $\mathfrak{e}_{-} \equiv \frac{1}{4\ell}\left(-K\ell-\sqrt{K^2\ell^2+4} \right)$ is given by
\begin{equation}
    \label{beta minus ds non-circular pole}
    \beta_{\phi} \, \bigg|_{\mathfrak{e}=\mathfrak{e}_-}= n \pi \sqrt{\frac{2 K \ell}{\sqrt{K^2 \ell^2+4}}+2} \, ,
\end{equation}
while the value at $\mathfrak{e}_0 \equiv -\frac{K}{2}$ is given by
\begin{equation}
    \beta_{\phi} \, \bigg|_{\mathfrak{e}=\mathfrak{e}_0}=\left(\frac{2 K \ell  \mathcal{K}\left(\frac{4}{K^2 \ell^2+4}\right)}{\sqrt{K^2 \ell^2+4}}+\pi \right) n \, .
\end{equation}
As the two values $\beta_{\phi} \, |_{\mathfrak{e}=\mathfrak{e}_-}-2\pi$ and $\beta_{\phi} \, |_{\mathfrak{e}=\mathfrak{e}_0}-2\pi$ have the same sign for any $K \ell>0$ and positive integer $n$,\footnote{The results are the same for $K\ell=0$.} there are no solutions $(\mathfrak{e},n)$ to the equation $\beta_{\phi}-2\pi=0$,
\begin{equation}
    \mathfrak{n}_{\text{non-circular pole, }K\ell>0}=0 \, .
\end{equation}

\begin{figure}[H]
        \centering
         
                \includegraphics[scale=0.5]{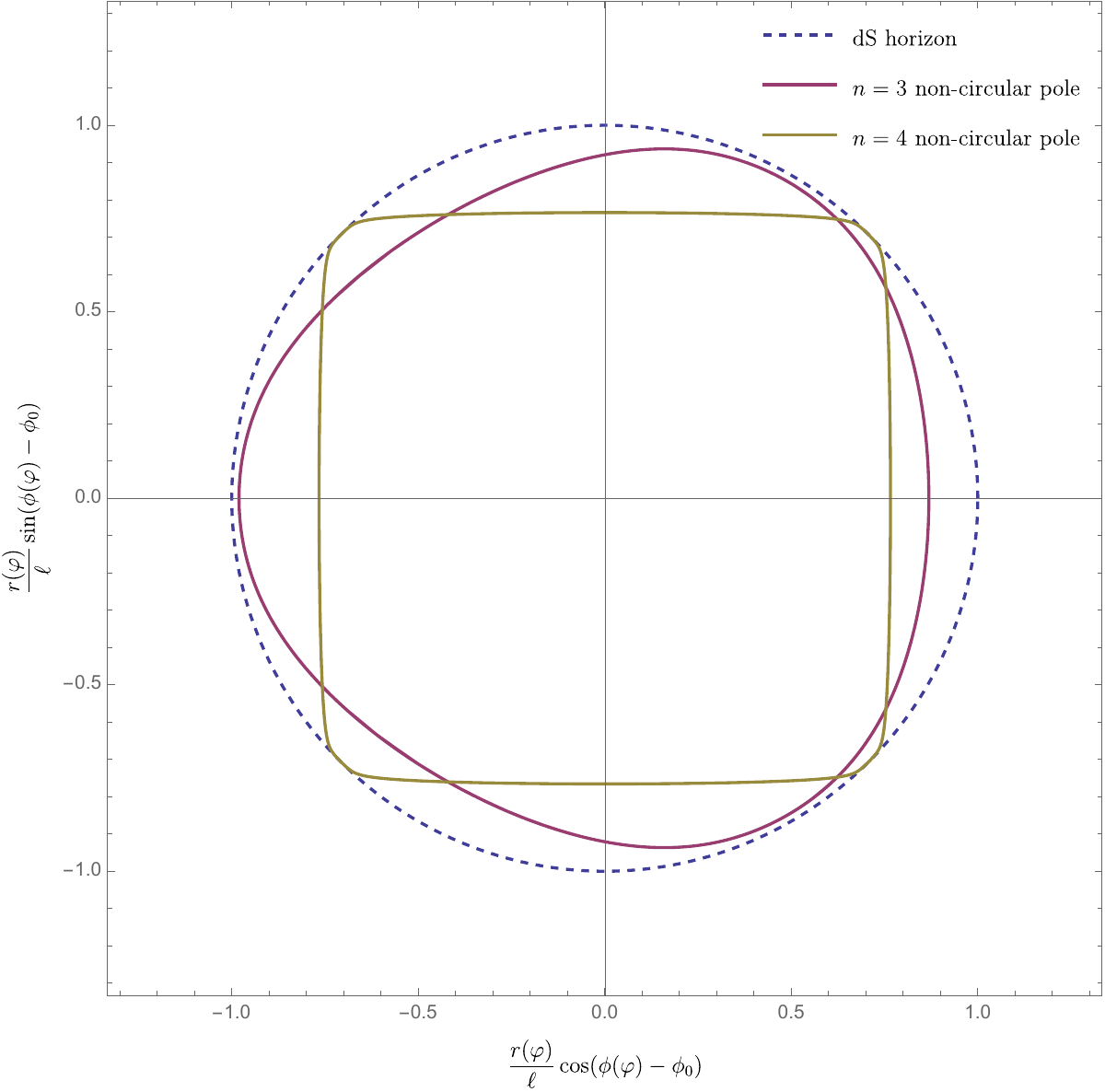}
                \caption{A parametric plot of $(\tfrac{r(\varphi)}{\ell}\cos(\phi(\varphi)-\phi_0),\tfrac{r(\varphi)}{\ell}\sin(\phi(\varphi)-\phi_0))$ of non-circular pole patches for $K\ell=-2.301$. There are two allowed solutions depicted by solid lines, whose $(\mathfrak{e},n)$ are given by $(-0.148,3)$ and $(-0.017,4)$. The dashed line is the de Sitter horizon $r=\ell$.} \label{fig: rphi plot dS KL<0}
\end{figure}

\textbf{Negative $K\ell$.} For $K\ell<0$, $\mathfrak{e}$ is restricted to lie in the interval $\frac{1}{4\ell}\left(-K\ell-\sqrt{K^2\ell^2+4} \right)<\mathfrak{e}<0$. Once again, we evaluate $\beta_{\phi}-2\pi$ at the ends of the interval. The value of $\beta_{\phi}$ at $\mathfrak{e}_-$ is the same one found in the $K\ell>0$ case, given in \eqref{beta minus ds non-circular pole}, while the minimum value is given by
\begin{equation}
    \beta_{\phi} \, \bigg|_{\mathfrak{e}=0} = -2n \arctan\!\left(\frac{2}{K\ell}\right) \, .
\end{equation}
In this case, then, given some value of $K\ell<0$, the number of non-circular pole patch solutions is given by
\begin{equation}
    \label{number of non-circular pole patch solutions dS}
    \mathfrak{n}_{\text{non-circular pole, }K\ell<0}=\left\lfloor-\frac{\pi }{\mathrm{arccot}\!\left(\frac{K \ell}{2}\right)} \right\rfloor-\left\lfloor \frac{2}{\sqrt{2+\frac{2 K \ell}{\sqrt{K^2 \ell^2+4}}}} \right\rfloor \, .
\end{equation}
The number of solutions diverges as $K \ell \rightarrow -\infty$. They exist for all values of $\tilde{\beta}>0$. A parametric plot of $(r(\varphi)\cos{\phi(\varphi)},r(\varphi)\sin{\phi(\varphi)})$ of all permissible solutions for a particular value of $K\ell$ is depicted in figure \ref{fig: rphi plot dS KL<0}.

\textbf{Conformal thermodynamics.} Given some $K\ell<0$ and $\tilde{\beta}$ with some solution characterised by $(\mathfrak{e},n)$ (or $\mathcal{E}
=\frac{\mathfrak{e}\beta_{\tau}^2\mathfrak{r}^2}{\beta^2}$), the on-shell action is given by
\begin{equation}
    \label{on-shell action de Sitter non-circular pole pach}
    I_{\text{on-shell}}=\frac{\tilde{\beta}\mathcal{E}}{4G_N}=\frac{2  \mathfrak{e} \ell^2 n^2 \mathcal{K}\left(\frac{2 \sqrt{-4 \mathfrak{e}^2-2 K \mathfrak{e}+\frac{1}{\ell^2}} \ell}{-\mathfrak{e} K \ell^2+\sqrt{-4 \mathfrak{e}^2-2 K \mathfrak{e}+\frac{1}{\ell^2}} \ell+1}\right)^2}{\left(1-\mathfrak{e}K\ell^2+ \sqrt{-4 \mathfrak{e}^2\ell^2-2 \mathfrak{e} K\ell^2+1}\right)G_N}\frac{\b}{4\pi^2} \, ,
\end{equation}
from which one deduces the conformal energy, entropy, and specific heat,
\begin{equation}
    \label{thermo non-circular dS pole}
    E_{\text{conf}}=\frac{\mathcal{E}}{4G_N} \, , \qquad \mathcal{S}_{\text{conf}}=C_K=0 \, .
\end{equation}
The vanishing entropy and specific heat is consistent with the linear-in-$\b$ behaviour of the action.

\subsection{Cosmic patch} \label{Lambda>0 cosmic patch}

In this section, we study the conformal thermodynamics of the cosmic patch solutions. These are defined as solutions endowed with the bulk metric
\begin{equation}\label{eqn: bulk metric cosmic de Sitter}
    ds^2 = \frac{\rh^2-\bar r^2}{\ell^2} \, d\bar \tau^2 + \frac{\ell^2}{\rh^2-\bar r^2} \, d\bar r^2 + \bar r^2 d\bar\phi^2 \, , \qquad \Lambda = \frac{1}{\ell^2} \, ,
\end{equation}
where $\rh$ is the horizon radius, and the global structure is given by $\bar \tau \sim \bar\tau+\frac{2\pi \ell^2}{\rh}$ and $\bar\phi\sim\bar \phi+2\pi$. To distinguish with the metric of the pole patch, \eqref{eqn: bulk metric pole de Sitter}, we put bar on the coordinates. The spacetime region of interest lies between the horizon $\bar r=\rh$ and the boundary situated at $\left.\bar r\right|_\Gamma$, which will be specified in what follows. At the boundary, we impose the conformal boundary conditions \eqref{eqn: bdry metric S1xS1}.

We note that the metric \eqref{eqn: bulk metric cosmic de Sitter} is geometrically equivalent to \eqref{eqn: bulk metric pole de Sitter}. This can be seen by the coordinate transformation,
\begin{equation}\label{eqn: cosmic pole transf}
    \bar \tau = \frac{\ell^2 \phi}{\rh} \, , \qquad \bar \phi=\frac{\tau}{\rh} \, , \qquad \frac{\bar r^2}{\rh^2}=1-\frac{r^2}{\ell^2} \, ,
\end{equation}
together with the parameter identification $\beta_\tau = 2\pi \rh$. As such, we exploit this equivalence to find cosmic patch solutions with a variety of boundaries using results from the pole patch solutions. We note, however, that the resulting solutions are physically distinct as they lead to different physical observables, such as the on-shell action.

\subsubsection{Static and circular cosmic patch}\label{Lambda>0 static and circular cosmic}

Here, we do not fix the value of $\rh$ as the origin lies outside the cosmic patch. Moreover, for any constant value $K$ for the trace of the extrinsic curvature, there exists a cosmic patch solution parameterised by
\begin{equation}
    \label{Lambda>0 homogeneous boundary cosmic patch}
    \left.\bar \tau\right|_{\Gamma}=\frac{\ell}{2\mathfrak{r}}\left(K\ell+\sqrt{K^2\ell^2+4}\right)u \, , \qquad \left.\bar r\right|_{\Gamma}=\rh \sqrt{\frac{1}{2}+\frac{K\ell}{2\sqrt{K^2\ell^2+4}}} \, , \qquad \left.\bar\phi\right|_{\Gamma} = \varphi \, ,
\end{equation}
where the bulk is defined over $\bar r \in (\left.\bar r\right|_{\Gamma},\rh]$. The unit normal vector $n^{\mu}$ points outward, which in this case means $n^{\bar r}<0$, and this gives an intrinsically flat induced metric,
\begin{equation}
    \label{induced metric homogeneous cosmic patch Lambda>0}
    \left.ds^2\right|_{\Gamma}= \frac{\rh^2}{2\r^2}\left(1+\frac{K\ell}{\sqrt{K^2\ell^2+4}}\right)\left(du^2+\r^2d\varphi^2\right) \, ,
\end{equation}
and the conformal stress tensor,
\begin{equation}
    \label{conformal stress tensor homogeneous cosmic patch Lambda > 0}
    T_{ij}d\sigma^{i}d\sigma^{j}=\frac{\rh^2 \left(K\ell+\sqrt{K^2\ell^2+4}\right)}{32 \pi  G_N \ell \mathfrak{r} ^2} \left(-du^2+\mathfrak{r}^2d\varphi ^2\right) \, .
\end{equation}

\textbf{Conformal thermodynamics.} In the cosmic patch, regularity at the cosmological horizon $r_h$ dictates that the coordinate $\tau$ has the identification $\tau \sim \tau + \frac{2 \pi \ell^2}{\rh}$. Following the parameterisation in \eqref{Lambda>0 homogeneous boundary cosmic patch}, this global structure determines a relation between $r_h$ and the conformal inverse temperature,
\begin{equation}
    \label{cosmic patch boundary data}
    \tilde{\beta}=\frac{\pi \ell}{\rh}\left(\sqrt{K^2\ell^2+4}-K\ell\right) \, \, .
\end{equation}
The on-shell action of this solution is given by
\begin{equation}
    \label{on-shell action homogeneous cosmic patch}
    I_\text{on-shell} =-\frac{\pi \rh}{4G_N}  \, ,
\end{equation}
and the thermodynamic quantities, whose evaluation requires writing $\rh$ as a function of $\tilde{\beta}$ via \eqref{cosmic patch boundary data} are
\begin{equation}
    \label{thermo homogeneous cosmic patch}
    E_\text{conf}=\frac{\pi^2\mathfrak{c}_\text{dS}}{3\b^2} \, , \qquad \mathcal{S}_\text{conf}=C_K=\frac{2\pi^2\mathfrak{c}_\text{dS}}{3\b} \, , \qquad \mathfrak{c}_\text{dS}\equiv \frac{3\ell \left(\sqrt{K^2\ell^2+4}-K\ell\right)}{4G_N}\, .
\end{equation}
The conformal entropy takes the Bekenstein-Hawking form as a quarter horizon area in units of $G$. Moreover, one can check that the on-shell action \eqref{on-shell action homogeneous cosmic patch} of the cosmic patch solution can be obtained from that of the pole patch solution in \eqref{eqn: Lambda>0 I pole} by a modular transformation $\tilde{\beta} \rightarrow \frac{4\pi^2}{\tilde{\beta}}$. As discussed in \cite{Anninos:2024wpy}, the dimensionless parameter $\mathfrak{c}_\text{dS}$ is interpreted as counting the number of effective degrees of freedom.

\subsubsection{Non-static cosmic patch}\label{Lambda>0 non-static cosmic}

Now we consider non-static cosmic patch solutions, defined as the cosmic patch with a boundary varying with $u$,
\begin{equation}
    \label{eqn: non sta cosmic}
    \left.\bar\tau\right|_{\Gamma}=\bar\tau(u)\, , \qquad \left.\bar r\right|_{\Gamma}=\bar r(u) \, , \qquad \left.\bar \phi\right|_{\Gamma}=\varphi\, .
\end{equation}
The radial component of the normal vector has a sign opposite to that of $\partial_{u}\bar \tau$, and since we are in the cosmic patch, the outward-pointing condition requires that $\p_u \bar\tau>0$.

\textbf{Problem.} The conditions \eqref{eqn: embed conf class cond} and \eqref{eqn: embed K}, derived from the conformal boundary conditions, impose equations on $\bar \tau(u)$ and $\bar r(u)$. In particular, the conformal class and the trace of the extrinsic curvature conditions are given by those in the pole patch analysis, \eqref{eqn: Lambda>0 eq1} and \eqref{eqn: Lambda>0 eq2}, with $f(r)=\tfrac{\rh^2-r^2}{\ell^2}$ and replacing $K\to-K$ and $(\tau(u),r(u))\to(\bar \tau(u),\bar r(u))$.

In what follows, we first find general solutions obeying the conformal boundary conditions, then impose the embedding condition, which restricts the space of solutions.

\textbf{Solutions.} Based on the observation made above, the general solutions are given by
\begin{equation}\label{eqn: non-sta cosmic sol}
    \bar r(u)= \rh\sqrt{1-\frac{r\!\left(\frac{u\beta}{2\pi \r^2}\right)}{\ell^2}^2} \, , \qquad \bar \tau(u)=\frac{\ell^2}{\rh} \phi\!\left(\frac{u\beta}{2\pi \r^2}\right) \, ,
\end{equation}
where the functions $r(\varphi)$ and $\phi(\varphi)$ are given by \eqref{eqn: non cir pole r sol} and \eqref{eqn: non cir pole phi sol} with $\beta_\tau = 2\pi \rh$. The solutions are labelled by a dimensionful parameter $\mathcal{E}$, which is constrained by reality and the absence of self-intersections to belong to
\begin{equation}
    \begin{cases}
        \mathcal{E}_{-} \leq \mathcal{E} \leq \mathcal{E}_0 \, , \qquad K\ell>0,\\
        \mathcal{E}_{-} \leq \mathcal{E} < 0, \, , \qquad K\ell<0.
    \end{cases}
\end{equation}
where $\mathcal{E}_- = - \tfrac{\pi^2\rh^2 \r^2}{\ell \beta^2}\left(K\ell+\sqrt{K^2\ell^2+4}\right)$ and $\mathcal{E}_0= - \tfrac{2\pi^2 r_h^2 \r^2 K}{\beta^2}$. Using \eqref{eqn: non-sta cosmic sol}, the Weyl factor and conformal stress tensor are given by
\begin{equation}
    \bomega(u) = \log \frac{\bar r(u)}{\r}\, , \qquad T_{ij}d\sigma^i d\sigma^j = -\frac{\mathcal{E}\beta^2}{32\pi^3G_N\r^4}\left(-du^2+\r^2d\varphi^2\right) \, .
\end{equation}

\textbf{Global smoothness condition.} Now we consider the periodicity compatibility between the bulk and boundary coordinates. The periodic structure of \eqref{eqn: non-sta cosmic sol} is determined from \eqref{Periodicity properties Lambda>0 non-circular dS pole}. Now we require that $u\sim u+\beta$ and $\bar \tau\sim\bar \tau+\tfrac{2\pi \ell^2}{\rh}$. These result in 
\begin{equation}\label{periodicities non-static cosmic dS}
    \b = \frac{4n \ell \mathcal{K}(m)}{y_+ \sqrt{K^2\ell^2+4}}\, , \qquad 2\pi=n\left[\frac{2\rh  K \ell \mathcal{K}\!\left(m\right)}{y_{+} \sqrt{K^2 \ell^2+4}}-\frac{2  \ell \, \Pi \left(\frac{y_{+}^2-y_{-}^2}{y_{+}^2-\rh^2} \, \big| \, m\right) \left(\frac{\beta ^2 \mathcal{E}}{2 \pi ^2 \mathfrak{r}^2}+K\rh^2\right)}{ \rh  y_{+} \sqrt{K^2 \ell^2+4} \left(1-\frac{ y_{+}^2}{\rh^2}\right)}\right]\,  ,
\end{equation}
where $y_+$ and $m$ are given by \eqref{eqn: ypm non cir pole} with $\beta_\tau =2\pi \rh$ and $n$ is an arbitrary positive integer. The second equation takes the same form as in \eqref{periodicities non-circular pole dS}. Therefore, we solve this equation using the same procedure as in the analysis of the non-circular pole patch solutions. We first consider a change of variable from $\mathcal{E}$ to $\mathfrak{e}$ via $\mathcal{E}=\tfrac{4\pi^2\mathfrak{e}\rh^2 \r^2}{\beta^2}$. The resulting equation is independent of $\rh$ and $\beta$, and becomes a transcendental equation determining $(\mathfrak{e},n)$ for a given value of $K\ell$. For $K\ell>0$, no solution of $(\mathfrak{e},n)$ exists,
\begin{equation}
    \mathfrak{n}_{\text{non-static cosmic, }K\ell>0}=0 \, .
\end{equation}
For $K\ell<0$, the number of allowed solutions $N$ is identical to \eqref{number of non-circular pole patch solutions dS}, 
\begin{equation}
    \mathfrak{n}_{\text{non-static cosmic, }K\ell<0}=\left\lfloor-\frac{\pi }{\mathrm{arccot}\!\left(\frac{K \ell}{2}\right)} \right\rfloor-\left\lfloor \frac{2}{\sqrt{2+\frac{2 K \ell}{\sqrt{K^2 \ell^2+4}}}} \right\rfloor \, .
\end{equation}
The allowed value of $(\mathfrak{e},n)$ can only be determined numerically. Plug in those value in the first equation in \eqref{periodicities non-static cosmic dS}, we obtain $\rh$ as a function of $\b$ and $K\ell$. In particular, $\rh$ is inversely proportional to $\b$.

\textbf{Conformal thermodynamics.} Evaluating the on-shell action using the solution \eqref{eqn: non-sta cosmic sol}, we obtain
\begin{equation}
    \label{lambda>0 non-static cosmic patch on-shell action}
    I_\text{on-shell}=  \frac{\b^3 \mathcal{E}}{16\pi^2G_N}=\frac{2  \mathfrak{e} \ell^2 n^2 \mathcal{K}\left(\frac{2 \sqrt{-4 \mathfrak{e}^2-2 K \mathfrak{e}+\frac{1}{\ell^2}} \ell}{-\mathfrak{e} K \ell^2+\sqrt{-4 \mathfrak{e}^2-2 K \mathfrak{e}+\frac{1}{\ell^2}} \ell+1}\right)^2}{\left(1-\mathfrak{e}K\ell^2+ \sqrt{-4 \mathfrak{e}^2\ell^2-2 \mathfrak{e} K\ell^2+1}\right)G_N}\frac{1}{\b} \,.
\end{equation}
Since $(\mathfrak{e},n)$ is determined purely from $K\ell$, the on-shell action is linear in the conformal temperature. Applying the thermodynamic relations, we find the conformal energy, entropy, and specific at fixed $K$,
\begin{equation}
    \label{non-static cosmic patch thermo}
    E_\text{conf} = -\frac{\b^2\mathcal{E}}{16\pi^2 G_N} \, , \qquad \mathcal{S}_\text{conf}=C_K= -\frac{\b^3 \mathcal{E}}{8\pi^2G_N} \, .
\end{equation}
Since only solutions with $\mathcal{E}<0$ are allowed, the specific heat is positive definite, implying thermal stability.

\subsubsection{Non-circular cosmic patch}\label{Lambda>0 non-circular cosmic}

In this section, we consider non-circular cosmic patch solutions. These are defined as the cosmic patch with a boundary varying with $\varphi$,
\begin{equation}
    \label{eqn: non cir cosmic}
    \left.\bar\tau\right|_{\Gamma}=\frac{2\pi \ell^2u}{\rh\beta}\, , \qquad \left.\bar r\right|_{\Gamma}=\bar r(\varphi) \, , \qquad \left.\bar \phi\right|_{\Gamma}=\bar \phi(\varphi)\, .
\end{equation}
Requiring that $n^{\mu}$ points outward, which means $n^{\bar r}>0$, leads to the condition $\partial_{\varphi}\bar \phi>0$.

\textbf{Problem.} Imposing the conformal boundary conditions lead to conditions on $\bar r(\varphi)$ and $\bar \phi(\varphi)$. These are given by
\eqref{eqn: Lambda>0 non-circular pole eq1} and \eqref{eqn: Lambda>0 non-circular pole eq2} with $f(r)=\tfrac{\rh^2-\bar r^2}{\ell^2}$ and $\beta_\tau=2\pi \rh$ and replacing $(r(\varphi),\phi(\varphi))\to (\bar r(\varphi),\bar \phi(\varphi))$.

\textbf{Solutions.} The general solution is given by
\begin{equation}
    \bar r(\varphi) = \rh \sqrt{1-\frac{r\left(\frac{2\pi \r^2 \varphi}{\beta}\right)^2}{\ell^2}} \, , \qquad \bar \phi(\varphi) = \frac{\tau\left(\frac{2\pi \r^2 \varphi}{\beta}\right)}{\rh} \, ,
\end{equation}
where the functions $r(u)$ and $\tau(u)$ are those from the non-static pole patch solution, \eqref{general sol for r Lambda>0} and \eqref{general sol for tau(u) Lambda>0}. As a consequence, the solutions are labelled by a dimensionful parameter $\mathcal{E}$, which upon imposing the reality and the absence of self-intersection conditions is constrained to obey
\begin{equation}
    \begin{cases}
        \mathcal{E}_-\leq \mathcal{E}<0 \, , \qquad &K\ell>0 \, , \\
        \mathcal{E}_- \leq \mathcal{E} \leq \mathcal{E}_0 \, , \qquad &K\ell<0\, ,
    \end{cases}
\end{equation}
where $\mathcal{E}_- = \tfrac{\ell}{4}(K\ell-\sqrt{K^2\ell^2+4})$ and $\mathcal{E}_0=\tfrac{1}{2}K\ell^2$.

The Weyl factor and conformal stress tensor are given by
\begin{equation}
    \bomega(\varphi) = \log \frac{2\pi r\!\left(\frac{2\pi \r^2 \varphi}{\beta}\right)}{\beta} \, , \qquad T_{ij}d\sigma^i d\sigma^j = - \frac{\pi \mathcal{E}}{2\beta^2G_N}\left(-du^2+\r^2d\varphi^2\right) \, .
\end{equation}

The conformal boundary data can be obtained via requiring the periodicity compatibility between the bulk and boundary coordinates. In particular, the identifications $\varphi \sim \varphi+2\pi$ and $\phi\sim\phi+2\pi$ impose that
\begin{equation}
    \label{conformal periodicity non-cir cosmic}
    \frac{4\pi^2}{\b} = \frac{4\ell n \mathcal{K}(m)}{r_{+}\sqrt{K^2\ell^2+4}} \,, \qquad 2\pi \rh = \frac{2n\ell^2}{r_{+}\sqrt{K^2\ell^2+4}} \left(-K\ell \mathcal{K}(m)+\frac{\ell\left(K\ell^2-2\mathcal{E}\right)}{\ell^2-r_+^2}\Pi\left(\frac{r_+^2-r_-^2}{r_+^2-\ell^2} \, \bigg| \, m\right)\right)  \, ,
\end{equation}
where $r_\pm$ are functions of $\mathcal{E}$ and $K\ell$, given by \eqref{R plus and minus Lambda>0}, and $m=1-\frac{r_-^2}{r_+^2}$. Consequently, these equations fix $\rh$ and $\mathcal{E}$ as a function of $\b$ and $K\ell$.

\textbf{Conformal thermodynamics.} Evaluating the on-shell action, we obtain
\begin{equation}
    I_\text{on-shell}= - \frac{1}{4G_N} \left(\frac{4\pi^2 \mathcal{E}}{\b}+2\pi \rh\right) \, .
\end{equation}
Applying the thermodynamic relations, the conformal energy and entropy are given by
\begin{equation}
    E_\text{conf} = - \frac{\pi^2 \mathcal{E}}{G_N \b^2} \, , \qquad \mathcal{S}_\text{conf} = \frac{\pi \rh}{2 G_N} \, .
\end{equation}
The entropy obeys the area law of the cosmological horizon. The specific heat at fixed $K$ is given by
\begin{equation}\label{specific heat Lambda>0 non-cir}
    C_K = - \frac{2\pi^2 \mathcal{E}}{G_N \b} +\frac{4 \ell n \mathcal{E}(1-\frac{4\mathcal{E}^2}{\ell^2}+2\mathcal{E} K)\mathcal{K}(m)^2}{r_+G_N\sqrt{K^2\ell^2+4}\left(2\mathcal{E}\left(K-\frac{4\mathcal{E}}{\ell^2}\right)\mathcal{K}(m)+r_+^2\left(K^2+\frac{4}{\ell^2}\right)E(m)\right)}<0 \, ,
\end{equation}
The negative definiteness of $C_K$ implies that every non-circular cosmic patch is thermally unstable.

\subsection{Thermodynamic phase space} \label{Lambda>0 thermodynamic phase space}

Here, we combine results from the previous sections and study thermodynamic properties of the total system. The main quantity of interest is the torus partition function $\mathcal{Z}(\b,K)$ which, in the semi-classical limit, includes contributions from all permissible classical solutions. As we have seen above, there is a distinction between the cases $K\ell>0$ and $K\ell<0$.

\subsubsection{\texorpdfstring{The case with $K\ell>0$}{KL>0 thermodynamic phase space}}

The torus partition function for $K\ell>0$ is, in the saddle-point approximation, given by
\begin{equation}
    \label{partition function Lambda>0 KL>0}
    \mathcal{Z}(\tilde{\beta},K\ell>0)=e^{-I^{(\text{hom. pole})}_{\text{on-shell}}} \,+ \, e^{-I^{(\text{hom. cosmic})}_{\text{on-shell}}}+\sum_{n} e^{-I^{(\text{non-static pole})}_{n,\text{on-shell}}}+\sum_{n}e^{-I^{(\text{non-circular cosmic})}_{n,\text{on-shell}}} \, .
\end{equation}
We now provide a summary of the relevant details of each of these contributions:
\begin{itemize}
    \item The first two contributions are the pole and cosmic patches with a static and circular, or homogeneous, finite boundary. The on-shell actions, given in \eqref{eqn: Lambda>0 I pole} and \eqref{on-shell action homogeneous cosmic patch} respectively, are
    \begin{equation}
    \label{hom on-shell action summary}
        I^{(\text{hom. pole})}_\text{on-shell}=I^{(\text{hom. cosmic})}_\text{on-shell}\bigg|_{\tilde{\beta} \rightarrow \frac{4\pi^2}{\tilde{\beta}}}=  -\frac{\tilde{\beta} \ell}{16 G_N}\left(\sqrt{K^2\ell^2+4}-K\ell\right) \, .
    \end{equation}
    The solutions exist for all $\b>0$ and $K\ell>0$.
    \item The third and fourth contributions come from the non-static pole patch and non-circular cosmic patch solutions. The number of such solutions \eqref{number of non-static pole-patch sols Lambda>0 KL>0} is determined by the value of $\tilde{\beta}$,
    \begin{equation}
        \mathfrak{n}_{\text{non-static pole, }K\ell>0}=\mathfrak{n}_{\text{non-circular cosmic, }K\ell>0}\bigg|_{\b \rightarrow \frac{4\pi^2}{\b}}= \left\lceil\frac{\tilde{\beta}}{\tilde{\beta}^{(-)}_{n=1}} \right \rceil-1 \, ,
    \end{equation}
    where $\b_{n=1}^{(-)}$ is a function of $K\ell$ defined in \eqref{lower bound on beta Lambda>0}. These solutions have an on-shell action \eqref{on-shell action Lambda > 0, pole patch} given by
    \begin{equation}
    \label{non-static pole on-shell summary Lambda>0}
        I^{(\text{non-static pole})}_{\text{on-shell}}\!=\!I^{(\text{non-circular cosmic})}_{\text{on-shell}}\bigg|_{\b \rightarrow \frac{4\pi^2}{\b}}\!=\! \frac{\ell n \left(K \ell^2-2 \mathcal{E}\right)\! \left(\mathcal{K}(m) \left(r_{+}^2-\ell^2\right)+\ell^2 \Pi \left(\left.\frac{r_{+}^2-r_{-}^2}{r_{+}^2-\ell^2}\right| m \right)\right)}{2 G_N r_{+} \sqrt{K^2 \ell^2+4} \left(r_{+}^2-\ell^2\right)} \, ,
    \end{equation}
    where $(\mathcal{E},n)$ are the solutions to, in the case of the pole patch, the $\tilde{\beta}$ equation in \eqref{conformal periodicity}, and in the case of the cosmic patch, the $\b$ equation in \eqref{conformal periodicity non-cir cosmic}. The parameter $m$ is given by $m=1-\frac{r_-^2}{r_+^2}$ where $r_{\pm}$ are functions of $\mathcal{E}$ and $K\ell$ given in \eqref{R plus and minus Lambda>0}.
\end{itemize}
\begin{figure}[H]
        \centering
         
                \includegraphics[scale=0.5]{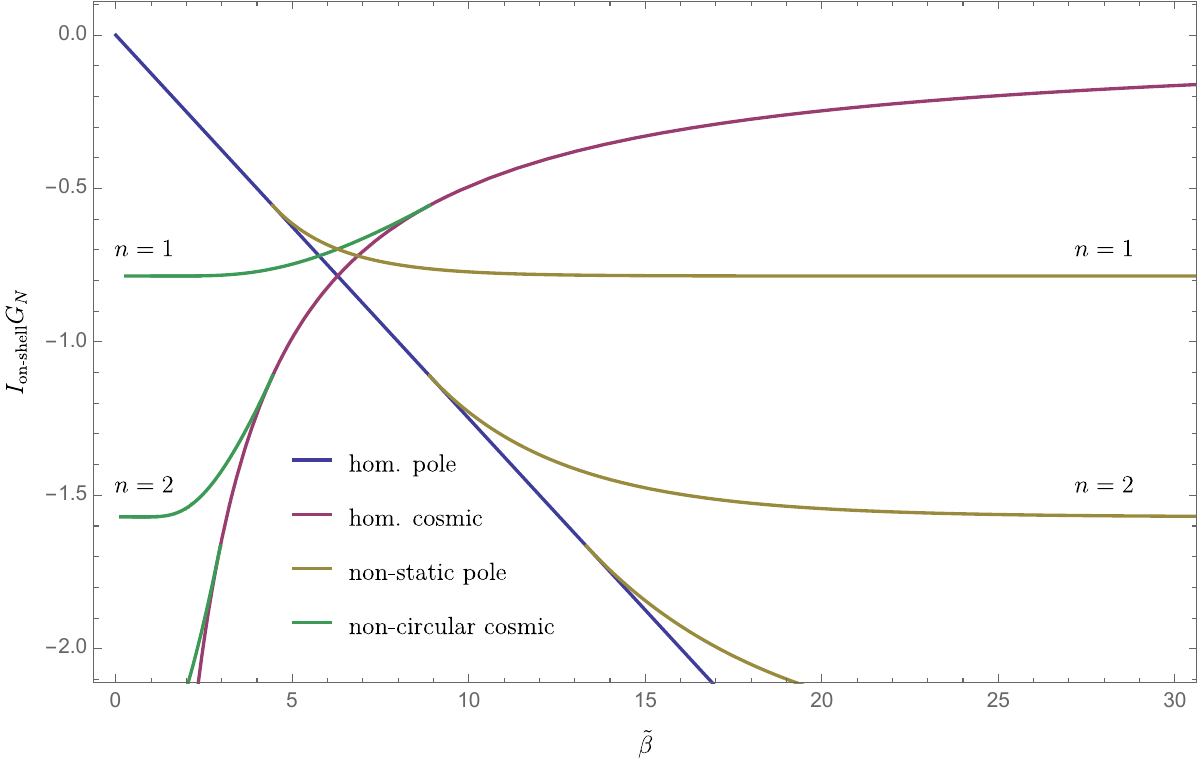}
                \caption{A plot of $I_\text{on-shell}$ of various contributions versus $\b$ for $K\ell =+10^{-3}$.} \label{fig: Ionshell dS KL>0}
\end{figure}
We now describe the different thermodynamic phases of the system.  A plot of the on-shell action of various contributions is depicted in figure \ref{fig: Ionshell dS KL>0}, which is qualitatively the same as we change the value of $K\ell>0$. Evidently, at low temperatures $\b>\b_c \equiv 2\pi$, the dominant saddle is the static and circular pole patch solution, while at high temperatures $\b < \b_c$, the dominant saddle is the static and circular cosmic patch solution. At the critical point, the system undergoes a first-order phase transition, where the energy exhibits a discontinuous jump,
\begin{equation}
    E_\text{conf}=\begin{cases}
        -\frac{\ell}{16 G_N}\left(\sqrt{K^2\ell^2+4}-K\ell\right) \, , \qquad &  \b>\b_c \, , \\
        \frac{\pi^2\ell}{4G_N \b^2}\left(\sqrt{K^2\ell^2+4}-K\ell \right) \, , \qquad &  0<\b<\b_c \, .
    \end{cases}
\end{equation}

\textbf{Zoo of unstable configurations.} It turns out that all the non-static pole patches and non-circular cosmic patches are sub-dominant with a negative specific heat, and are therefore unstable configurations. As $\tilde{\beta} \rightarrow \b_{n}^{(-)}$ (or $\tilde{\beta} \rightarrow \frac{4\pi^2}{\b_{n}^{(-)}}$), the on-shell action, energy, and entropy of these solutions match those of the homogeneous pole patch (homogeneous cosmic patch) at these temperatures. However, the specific heat does not respect this limit and is discontinuous.

\subsubsection{\texorpdfstring{The case with $K\ell<0$}{KL thermodynamic phase space}}

For negative value of the trace of the extrinsic curvature $K$ of the boundary, there are more contributions to the torus partition function in the saddle-point approximation:
\begin{equation}
    \label{partition function Lambda>0 KL<0}
    \begin{aligned}
        \mathcal{Z}(\tilde{\beta},K\ell<0)=e^{-I^{(\text{hom. pole})}_{\text{on-shell}}} \,+ \, e^{-I^{(\text{hom. cosmic})}_{\text{on-shell}}}&+\sum_{n} e^{-I^{(\text{non-static pole})}_{n,\text{on-shell}}}+\sum_{n}e^{-I^{(\text{non-circular cosmic})}_{n,\text{on-shell}}}\\
        &+\sum_{n}e^{-I^{(\text{non-circular pole})}_{n,\text{on-shell}}}+\sum_{n}e^{-I^{(\text{non-static cosmic})}_{n,\text{on-shell}}} \, .
    \end{aligned}
\end{equation}
Here is a summary of these contributions:
\begin{itemize}
    \item The first two contributions are no different from those of the $K\ell>0$ partition function, with the on-shell actions given in \eqref{hom on-shell action summary}. They exist for all $\b>0$ and $K\ell<0$.
    \item The third and fourth contributions are also the same as the ones listed in the $K\ell>0$ contributions, with on-shell actions in \eqref{non-static pole on-shell summary Lambda>0}, although the number of solutions differs in the $K\ell<0$ case:
    \begin{equation}
        \mathfrak{n}_{\text{non-static pole, }K\ell<0}= \mathfrak{n}_{\text{non-circular cosmic, }K\ell<0}\bigg|_{\b \rightarrow \frac{4\pi^2}{\b}} = \left\lceil \frac{\tilde{\beta}}{\tilde{\beta}^{(-)}_{n=1}} \right\rceil - \left \lceil \frac{\tilde{\beta}}{\tilde{\beta}_{n=1}^{(0)}} \right \rceil \, ,
    \end{equation}
    where $\b_{n=1}^{(-)}$ and $\b_{n=1}^{(0)}$ are functions of $K\ell$ given in \eqref{lower bound on beta Lambda>0} and \eqref{bounds on beta KL<0 Lambda>0}. Each solution is characterised by a pair $(\mathcal{E},n)$ that solves the $\b$ equation in \eqref{conformal periodicity}. 
    \item The fifth and sixth contributions are unique to the $K\ell<0$ case. They come from the non-circular pole patch and non-static cosmic patch solutions, the number of which depends only on the value of $K\ell$, given by
    \begin{equation}\label{eqn: number sol Lambda>0}
        \mathfrak{n}_{\text{non-circular pole}}=\mathfrak{n}_{\text{non-static cosmic}}=\left\lfloor-\frac{\pi }{\mathrm{arccot}\!\left(\frac{K \ell}{2}\right)} \right\rfloor-\left\lfloor \frac{2}{\sqrt{2+\frac{2 K \ell}{\sqrt{K^2 \ell^2+4}}}} \right\rfloor \, .
    \end{equation}
    These solutions exist for all $\b >0$ and have an on-shell action \eqref{on-shell action de Sitter non-circular pole pach} given by
    \begin{equation}
        I^{(\text{non-circular pole})}_{n,\text{on-shell}}=I_{n,\text{on-shell}}^{(\text{non-static cosmic})}\bigg|_{\b \rightarrow \frac{4\pi^2}{\b}}=\frac{2  \mathfrak{e} \ell^2 n^2 \mathcal{K}\left(\frac{2 \sqrt{-4 \mathfrak{e}^2-2 K \mathfrak{e}+\frac{1}{\ell^2}} \ell}{-\mathfrak{e} K \ell^2+\sqrt{-4 \mathfrak{e}^2-2 K \mathfrak{e}+\frac{1}{\ell^2}} \ell+1}\right)^2}{\left(1-\mathfrak{e}K\ell^2+ \sqrt{-4 \mathfrak{e}^2\ell^2-2 \mathfrak{e} K\ell^2+1}\right)G_N}\frac{\b}{4\pi^2} \, ,
    \end{equation}
    where $(\mathfrak{e},n)$ solves the second equation in \eqref{periodicities non-circular pole dS}, which depends only on $K \ell$.
\end{itemize}
\begin{figure}[H]
        \centering
         
                \includegraphics[scale=0.5]{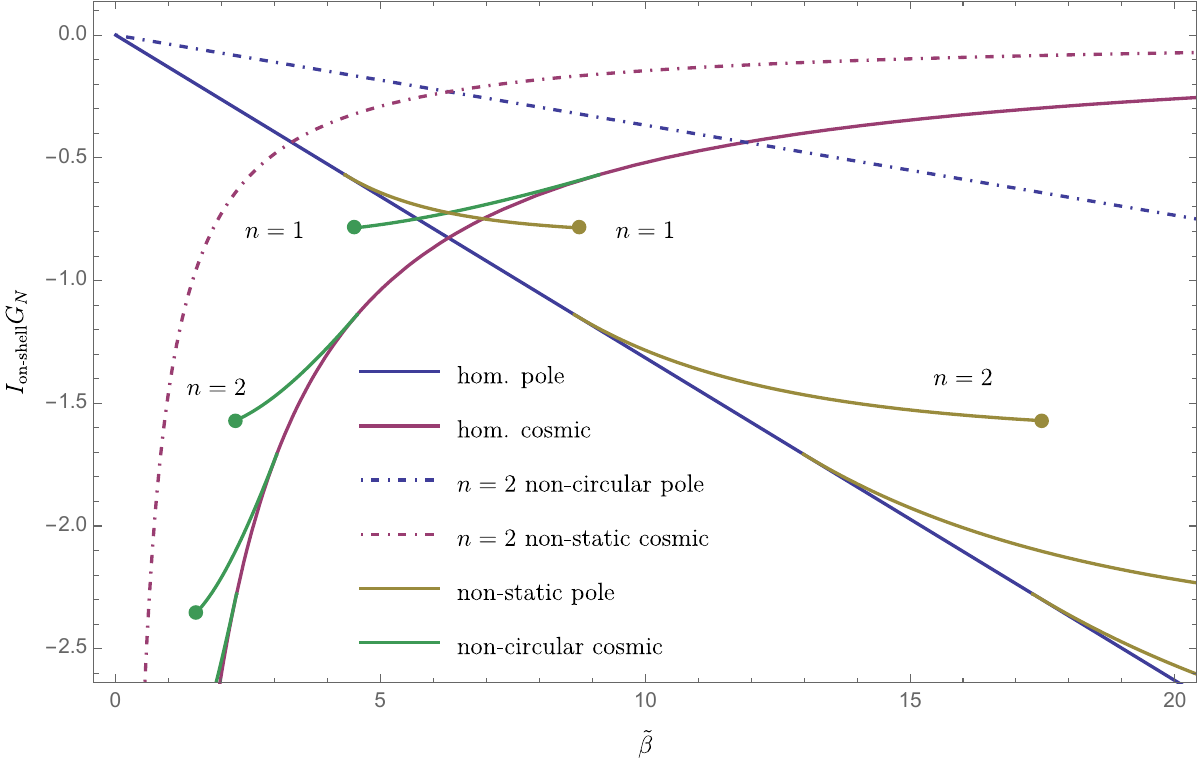}
                \caption{A plot of $I_\text{on-shell}$ of various contributions versus $\b$ for $K\ell =-0.1$.} \label{fig: Ionshell dS KL<0}
\end{figure}
We now describe the different thermodynamic phases of the system.  A plot of the on-shell action of various contributions is depicted in figure \ref{fig: Ionshell dS KL<0}.

As with the $K\ell>0$ case, there exists a first-order phase transition at $\b_c=2\pi$, where the dominant saddle transitions from the homogeneous pole patch solution at low temperatures to the homogeneous cosmic patch solution at high temperatures.

\textbf{Thermal stability.} Unlike the all the previous cases, however, there exist in the case $K\ell<0$ a set of inhomogeneous solutions with positive specific heat. These are the non-static cosmic patch solutions with $C_K$ given in \eqref{non-static cosmic patch thermo}. Nevertheless, their on-shell action is dominated by that of the homogeneous solution. All other inhomogeneous solutions with $K\ell<0$ have negative specific heat.

\subsection{The stretched horizon limit}\label{sec: stretch}

One limit of interest is to take the boundaries we found above to be very close to the cosmic horizon. In this subsection, we show that this is possible in the cases of the non-static pole patch and the non-circular pole patch.\footnote{It is, in fact, also possible for the non-circular cosmic patch solutions, but these reduce to Rindler patch solutions in the stretched horizon limit.} We briefly comment on the number of solutions and their thermodynamic properties.

\subsubsection{Non-static pole patch}
The solutions that have a stretched horizon limit in this case are those with $K\ell \rightarrow -\infty$. In particular, for values of $\mathcal{E}$ of the order
\begin{equation}
    \frac{\mathcal{E}}{\ell}=\frac{1}{2}K\ell +\frac{\delta \mathcal{E}}{K\ell} \, ,
\end{equation}
with $0<\delta \mathcal{E}<\frac{1}{2}$ fixed in this limit, the minimum and maximum boundary radii $r_{\pm}$ are given by
\begin{equation}
    \frac{r_{\pm}}{\ell}=1-\frac{1-\delta \mathcal{E}\mp\sqrt{1-2\delta \mathcal{E}}}{(K\ell)^2} \, , 
\end{equation}
and are therefore parametrically close to the horizon. 

To find the number of such solutions at some $\b$ and $|K\ell| \gg 1$, we recall that this number is determined by the number of intervals $(\b_n^{(-)},\b_{n}^{(0)})$ a given value of $\b$ belongs to. In the $K\ell \rightarrow -\infty$ limit, these bounds are given by
\begin{equation}
    \b_n^{(-)}=-\frac{2\pi n}{K\ell}+\frac{3\pi n}{(K\ell)^3}+\mathcal{O}\!\left(\! \frac{1}{(K\ell)^5}\!\right) \, , \qquad \b_{n}^{(0)}=-\frac{2\pi n}{K\ell}+\frac{2\pi n}{(K\ell)^3}+\mathcal{O}\!\left( \!\frac{1}{(K\ell)^5}\!\right) \, .
\end{equation}
We now observe that $\b_{n}^{(0)}-\b_{n+1}^{(-)}=\frac{2\pi}{K\ell}+\mathcal{O}\!\left(\!\frac{1}{(K\ell)^3}\!\right)$ is negative for any $n$ in this limit. Therefore, given large and negative $K\ell$, there exists at most one non-static pole patch solution with this boundary data. More specifically, on the $\b$ line, there exist disconnected intervals, with a width of the order $\tfrac{1}{|K\ell|^3}$, where exactly one solution exists. These intervals are located at values of $\b$ of the order $\tfrac{1}{|K\ell|}$ and are separated by a gap with a width of the order $\tfrac{1}{|K\ell|}$.

    The solutions have $\frac{\mathcal{E}}{\ell}=\frac{1}{2}K\ell+\frac{\delta \mathcal{E}}{K\ell}$, where $0<\delta \mathcal{E}<\frac{1}{2}$ and winding number $n$. Their on-shell action is given by
\begin{equation}
    I_{\text{on-shell}}=-\frac{\pi  \ell n}{4 G_N}+\frac{\pi\ell n  \delta\mathcal{E}  }{4 (K\ell)^2 G_N}+\mathcal{O}\!\left(\! \frac{1}{(K\ell)^4}\!\right) \, ,
\end{equation}
and its thermodynamic quantities are given by
\begin{equation}
    E_{\text{conf}}=\frac{\ell}{4G_N}\left(\frac{1}{2}K\ell+\frac{\delta \mathcal{E}}{K\ell}\right) \, , \qquad \mathcal{S}_{\text{conf}}=\frac{\pi  (1-2 \delta \mathcal{E} ) \ell n}{4 (K\ell)^2 G_N}+\mathcal{O}\!\left(\!\frac{1}{(K\ell)^3}\!\right) \, .
\end{equation}

\subsubsection{Non-circular pole patch}
In this case, the stretched horizon limit is also at $K\ell \rightarrow -\infty$. In particular, $r_{\pm}$ become parametrically close to $\ell$. As for the number of solutions, found in \eqref{number of non-circular pole patch solutions dS}, it diverges linearly as $K\ell \rightarrow -\infty$, independently of the value of $\b$.

These solutions are parameterised by the variable $\mathfrak{e}$, introduced in section \ref{Lambda>0 non-circular pole}, which has dimension length and is of the order
\begin{equation}
    \mathfrak{\ell e}=\frac{\delta \mathfrak{e}}{K\ell} \, 
\end{equation}
in this limit, where $0<\delta \mathfrak{e}<\frac{1}{2}$ is kept fixed. Recall that one of the global smoothness conditions, specifically the second equation in \eqref{periodicities non-circular pole dS}, determines $(\mathfrak{e},n)$ as a function of $K\ell$. While this equation remains unsolvable analytically in the stretched horizon limit $K\ell \rightarrow -\infty$, we find that $n$ must scale with $|K\ell|$.

In this limit, one may be interested in finding solutions with a bulk periodicity $\beta_\tau=2\pi \ell$ that matches the de Sitter temperature, which we have not enforced in the pole patch as the conical singularity lies outside the bounded region. However, as our boundary approaches the horizon in this limit, we look for solutions with this property.

Setting $\beta_\tau=2\pi \ell$, we use the other global smoothness condition, the first equation in \eqref{periodicities non-circular pole dS}, to find that the conformal inverse temperature $\b$ that scales with $\frac{1}{|K\ell|}$. As for the conformal energy, it scales with $|K\ell|$ just like that of the static solution, and we have verified numerically that the static solution is thermodynamically favourable, as expected.

\section{Negative cosmological constant}\label{sec: Lambda -}

In this section, we study the gravitational path integral in the case of negative cosmological constant. We first briefly review the Euclidean bulk solutions to Einstein's equations in three dimensions with a negative cosmological constant \cite{Banados:1992wn,Banados:1992gq}.

Solutions to \eqref{eqn: Einstein eqn} with $\Lambda<0$ are locally given by Euclidean AdS$_3$, or the hyperbolic three-manifold, with radius $(-\Lambda)^{-1/2}$. Working in global coordinates, we have
\begin{equation}\label{eqn: global AdS metric}
    ds^2= \frac{\ell^2+r^2}{\ell^2} \,d\tau^2 + \frac{\ell^2}{\ell^2+r^2} \,dr^2 + r^2 d\phi^2 \, , \qquad \Lambda=-\frac{1}{\ell^2}\, ,
\end{equation}
where $r\in(0,\infty)$ and $\phi\sim\phi+2\pi$. The coordinate $\tau$ is chosen to be identified under $\tau\sim\tau+\beta_\tau$, where $\beta_\tau$ is an arbitrary positive number. The conformal boundary of AdS$_3$ is situated at $\tfrac{r}{\ell}\to \infty$.

Another relevant solution is the Euclidean non-rotating BTZ solution,
\begin{equation}\label{eqn: BTZ metric}
    ds^2= \frac{\bar r^2-\rh^2}{\ell^2} \, d\bar\tau^2 + \frac{\ell^2}{\bar r^2-\rh^2} \, d\bar r^2 + \bar r^2 d\bar\phi^2 \, , \qquad \Lambda=-\frac{1}{\ell^2}\, ,
\end{equation}
where $\rh$ is an arbitrary positive constant, $\bar r\in(\rh,\infty)$, and $\bar \phi\sim \bar\phi+2\pi$. The black hole horizon is located at the radial coordinate $r=\rh$. The solution exhibits a conical defect at the horizon unless the coordinate $\bar \tau$ obeys the identification $\bar \tau\sim\bar\tau + \tfrac{2\pi \ell^2}{\rh}$.

We now consider solutions with a boundary obeying the boundary conditions \eqref{eqn: bdry metric S1xS1}, employing the embedding method. We review both pole patch and black hole patch solutions, defined as spacetimes which do and do not contain the black hole horizon, in sections \ref{Lambda<0 pole patch} and \ref{Lambda<0 BH patch} respectively. In each section, we look for non-static and non-circular boundaries. For each of these solutions, we compute their thermodynamic quantities in the conformal canonical ensemble defined in \ref{ref: thermo}. Finally, in section \ref{Lambda<0 thermodynamic phase space}, we combine these results and explore the thermodynamic phase space of the system.

\subsection{Pole patch}\label{Lambda<0 pole patch}

We first study the conformal thermodynamics of the pole patch solutions. These are solutions endowed with the bulk metric
\begin{equation}\label{eqn: global AdS metric again}
    ds^2= \frac{\ell^2+r^2}{\ell^2} \,d\tau^2 + \frac{\ell^2}{\ell^2+r^2} \,dr^2 + r^2 d\phi^2 \, , \qquad \Lambda=-\frac{1}{\ell^2}\, ,
\end{equation}
where $\tau\sim\tau+\beta_\tau$ for some arbitrary positive number $\beta_\tau$, and  $\phi \sim \phi+2\pi$. The spacetime region of interest includes the origin $r=0$ and has boundary situated at the radial coordinate $\left.r\right|_{\Gamma}$, which will be specified later. As this region does not contain the horizon, we do not fix $\beta_{\tau}$. At the boundary, we impose the conformal boundary conditions \eqref{eqn: bdry metric S1xS1}.

\subsubsection{Static and circular pole patch}\label{Lambda<0 static and circular pole}
For any constant value $K\ell>2$ for the trace of the extrinsic curvature, there exists a pole-patch solution parameterised by
\begin{equation}
    \label{Lambda<0 homogeneous boundary pole patch}
    \left.\tau\right|_{\Gamma}=\frac{\ell}{2\mathfrak{r}}\left(K\ell-\sqrt{K^2\ell^2-4}\right)u \, , \qquad \left.r\right|_{\Gamma}=\ell \sqrt{\frac{K\ell}{2\sqrt{K^2\ell^2-4}}-\frac{1}{2}} \, , \qquad \left.\phi\right|_{\Gamma} = \varphi \, ,
\end{equation}
where the bulk is defined over $r \in [\rh,\left.r\right|_{\Gamma})$. The unit normal vector $n^{\mu}$ is chosen to be pointing outward, i.e., $n^{r}>0$, using which we evaluate the induced metric on the boundary,
\begin{equation}
    \label{induced metric homogeneous pole patch Lambda<0}
    \left.ds^2\right|_{\Gamma}= \frac{\ell^2}{2\r^2}\left(\frac{K\ell}{\sqrt{K^2\ell^2-4}}-1\right)\left(du^2+\r^2d\varphi^2\right) \, ,
\end{equation}
which is intrinsically flat, and the conformal stress tensor,
\begin{equation}
    \label{conformal stress tensor homogeneous pole patch Lambda < 0}
    T_{ij}d\sigma^{i}d\sigma^{j}=-\frac{\ell\left(K\ell-\sqrt{K^2\ell^2-4}\right)}{32\pi G_N\mathfrak{r}^2} \left(-du^2+\mathfrak{r}^2d\varphi ^2\right) \, .
\end{equation}
Note that the Weyl factor in the induced metric diverges as $K\ell \rightarrow 2^{+}$ while the stress tensor remains finite. Moreover, the $K\ell \rightarrow \infty$ reproduces the flat space results in section \ref{Lambda = 0 static and circular pole patch}.

There are no homogeneous solutions with $K\ell \leq 2$.

\textbf{Conformal thermodynamics.} The periodicities of the bulk and boundary coordinates are related by
\begin{equation}
    \label{eqn: bdry data homo Lambda<0}
    \beta_{\tau}=\frac{\ell}{2}\left(K\ell-\sqrt{K^2\ell^2-4} \right)\b \, .
\end{equation}
The on-shell action of this homogeneous solution is given by
\begin{equation}
    \label{Lambda<0 homogeneous pole-patch on-shell action}
    I_\text{on-shell}= -\frac{\b\ell}{16 G_N}\left(K\ell-\sqrt{K^2\ell^2-4}\right) \, .
\end{equation}

Using the thermodynamic relations \eqref{eqn: thermo relation}, the corresponding conformal energy, entropy, and specific heat are given by
\begin{equation}
    \label{Lambda<0 homogeneous pole thermo}
    E_\text{conf}=-\frac{\ell}{16 G_N}\left(K\ell-\sqrt{K^2\ell^2-4}\right) \, , \qquad \mathcal{S}_\text{conf}=C_K= 0 \, ,
\end{equation}
reproducing the results found in \cite{Allameh:2025gsa}.

\subsubsection{Non-static pole patch}\label{Lambda<0 non-static pole}
We now consider the class of pole patch solutions with a boundary radius that varies with the thermal boundary coordinate $u$. In particular, we parameterise the boundary by
\begin{equation}
    \label{Lambda<0 bdry 1}
    \left.\tau\right|_{\Gamma}=\tau(u)\, , \qquad \left.r\right|_{\Gamma}=r(u) \, , \qquad \left.\phi\right|_{\Gamma}=\varphi\, .
\end{equation}
The bulk is defined over the coordinate range $r \in [0,\left.r\right|_{\Gamma})$. The unit normal vector is given by
\begin{equation}
    n^{\mu}=\frac{-\partial_{u}r \partial_{\tau}+\partial_{u}\tau \partial_{r}}{\sqrt{\frac{\ell^2}{\ell^2+r^2}(\partial_ur)^2+\frac{\ell^2+r^2}{\ell^2}(\partial_u \tau)^2}} \, ,
\end{equation}
and so the outward-pointing condition requires $\partial_u \tau>0$.

\textbf{Problem.} The conditions \eqref{eqn: embed conf class cond} and \eqref{eqn: embed K}, derived from the conformal boundary conditions, impose constraints on $\tau(u)$ and $r(u)$: one from the conformal structure of the induced metric,
\begin{equation}
    \label{Lambda<0 non-static pole eq1}
    \frac{1}{f}(\mathfrak{r}\partial_ur)^2+f(\mathfrak{r}\partial_u \tau)^2=r^2 \, ,
\end{equation}
and one from the constant trace of the extrinsic curvature $K$,
\begin{equation}
    \label{Lambda<0 non-static pole eq2}
    \frac{3 r (\partial_u r)^2 \partial_{u}\tau f'+r f^2 (\partial_u \tau)^3 f'+2 f \left(-r \partial_{u}^2r \partial_u \tau+r \partial_u r \partial_{u}^2\tau+(\partial_u r)^2 \partial_u \tau\right)+2 f^3 (\partial_u \tau)^3}{2 r \sqrt{\frac{(\partial_u r)^2}{f}+f (\partial_u \tau)^2} \left(f^2 (\partial_u \tau)^2+(\partial_u r)^2\right)}=K \, ,
\end{equation}
where $f(r)=\frac{\ell^2+r^2}{\ell^2}$ and its derivative are evaluated at $r(u)$.

In what follows, we first find general solutions to equations \eqref{Lambda<0 non-static pole eq1} and \eqref{Lambda<0 non-static pole eq2}, then impose the embedding condition, which restricts the space of solutions.

\textbf{Solution.} One can show, following a proof analogous to the one in appendix \ref{sec: du T flat 1}, that equations \eqref{Lambda<0 non-static pole eq1} and \eqref{Lambda<0 non-static pole eq2} can be rewritten as a first-order ODE of the form
\begin{equation}\label{first order ODE Lambda<0 pole non-static}
    \frac{d\mathcal{E}}{du} =0 \, , \qquad \mathcal{E}\equiv \frac{1}{2}K r^2 - \frac{\ell^2+r^2}{\ell^2} \, \r\p_u \tau \, ,
\end{equation}
where $\mathcal{E}$ is an integration constant with length dimension. We now use this and equation \eqref{Lambda<0 non-static pole eq1} to write an ordinary differential equation in $r(u)$, given by
\begin{equation}\label{eqn: AdS bdry eqn 2}
    (\r \p_ur)^2 + V_\text{eff}(r)=0 \, , \qquad V_\text{eff}(r)\equiv \left(\frac{K^2}{4}-\frac{1}{\ell^2}\right)r^4 - (1+\mathcal{E}K)r^2 +\mathcal{E}^2 \, .
\end{equation}
This is the equation of motion of a classical particle moving in a potential well $V_{\text{eff}}(r)$. Unlike the previous cases with zero or positive cosmological constant, there exist values of $K\ell$ for which the potential is inverted. In particular, the solution is non-oscillatory for $|K\ell| \leq 2$. As these do not contribute to the torus partition function, we do not study these cases here. In what follows, we assume $|K\ell|>2$.

The general solution $r(u)$ is given by
\begin{equation}\label{eqn: rsol AdS}
    r(u) = r_+ \text{dn}\left(r_+\sqrt{\frac{K^2}{4}-\frac{1}{\ell^2}}\frac{u-u_0}{\r} \, \bigg| \, m\right) \, ,
\end{equation}
where 
\begin{equation}\label{eqn: rpm AdS}
    r_\pm \equiv \ell \sqrt{\frac{2+2\mathcal{E}K\pm2 \sqrt{1+\frac{4\mathcal{E}^2}{\ell^2}+2\mathcal{E}K}}{K^2\ell^2-4}} \, , \qquad m\equiv 1-\frac{r_-^2}{r_+^2} \, .
\end{equation}
The introduced quantities $r_\pm$ are the maximum and minimum values of $r(u)$. Their reality condition enforces bounds on $\mathcal{E}$,
\begin{equation}\label{eqn: def epm nonsta AdS}
    \begin{cases}
        \mathcal{E}\geq \mathcal{E_-} \, &K\ell>2 \, , \\
        \mathcal{E}\leq \mathcal{E_+} \, &K\ell<-2 \, ,
    \end{cases}
    \qquad \mathcal{E}_\pm \equiv -\frac{\ell}{4}\left(K\ell\pm \sqrt{K^2\ell^2-4}\right) \, .
\end{equation}
Note that $\mathcal{E}_{-}< 0$ when $K\ell>2$, and $\mathcal{E}_{+} > 0$ when $K\ell<-2$. Moreover, the smaller root $r_{-}$ vanishes at $\mathcal{E}=0$. The roots $r_{\pm}$ coincide at $\mathcal{E}=\mathcal{E}_{\pm}$.

Plugging \eqref{eqn: rsol AdS} into \eqref{first order ODE Lambda<0 pole non-static}, we obtain
\begin{equation}\label{eqn: tsol AdS}
    \tau(u) = \tau_0 + \frac{K\ell^2(u-u_0)}{2\r} - \frac{(2\mathcal{E}+K\ell^2)\Pi\left(\frac{r_+^2-r_-^2}{r_+^2+\ell^2};\text{am}(x|m);m\right)}{\frac{r_+}{\ell}\left(\frac{r_+^2+\ell^2}{\ell^2}\right)\sqrt{K^2\ell^2-4}} \, ,
\end{equation}
where $x\equiv r_+\sqrt{\tfrac{K^2}{4}-\tfrac{1}{\ell^2}}\tfrac{u-u_0}{\r}$.

The solutions obey periodic conditions,
\begin{equation}
    \label{Lambda<0 non-static pole periodicities}
    \begin{cases}
        r\left(u+\frac{4n\ell\r \mathcal{K}(m)}{r_+\sqrt{K^2\ell^2-4}}\right)&=r(u) \, , \\
        \tau\left(u+\frac{4n\ell\r \mathcal{K}(m)}{r_+\sqrt{K^2\ell^2-4}}\right) &= \tau(u) + \frac{2n\ell^2}{r_+\sqrt{K^2\ell^2-4}}\left(K\ell \mathcal{K}(m)-\frac{\ell(K\ell^2+2\mathcal{E})}{r_+^2+\ell^2}\Pi\left(\frac{r_+^2-r_-^2}{r_+^2+\ell^2};m\right)\right)\, ,
    \end{cases}
\end{equation}
where $n$ is any positive integer, called the winding number.

The corresponding Weyl factor and conformal stress tensor are given by
\begin{equation}\label{eqn: Weyl nonsta pole AdS}
    \bomega(u) = \log \frac{r(u)}{\r} \, , \qquad T_{ij}d\sigma^i d\sigma^j = \frac{\mathcal{E}}{8\pi G_N\r^2}\left(-du^2+\r^2d\varphi^2\right) \, .
\end{equation}

\textbf{Self-intersections and regime of validity.} This solution violates the global embedding condition \eqref{eqn: global emb} for some values of $\mathcal{E}$.

For $K\ell>2$, there are no self-intersections over $\mathcal{E}_{-} \leq \mathcal{E} \leq 0$. They emerge, however, over $\mathcal{E}>0$.

For $K\ell<-2$, there are self-intersections over $\mathcal{E}<0$. They then disappear over $0 \leq \mathcal{E}< \mathcal{E}_{+}$. However, these solutions have the wrong sign of $\partial_{u}\tau$, and so we discard them.

Therefore, there are no physical solutions with $K\ell<-2$, and the physical solutions with $K\ell>2$ must obey
\begin{equation}\label{eqn: E bound non-sta AdS}
    \mathcal{E}_{-} \leq \mathcal{E} \leq 0 \, ,
\end{equation}
where $\mathcal{E}_-$ is given by \eqref{eqn: def epm nonsta AdS}.

\textbf{Simple solutions.} There are particular values of $\mathcal{E}$ where the solution simplifies.

At $\mathcal{E}=\mathcal{E}_-$, the solution is precisely the homogeneous one in \eqref{Lambda<0 homogeneous boundary pole patch}.

At $\mathcal{E}=0$, the smaller root $r_-$ vanishes, and the boundary has sphere topology. This solution therefore does not contribute to the thermal partition function we are studying.

\textbf{Conformal thermodynamics.} The global structure of the bulk must be consistent with the periodicity structure of the boundaries found above, which requires that
\begin{equation}\label{eqn: Lambda<0 periodic cond}
    \tau(u+\beta)=\tau(u)+\beta_\tau \, , \qquad r(u+\beta) = r(u) \, .
\end{equation}
This boundary data was found in \eqref{Lambda<0 non-static pole periodicities} to be
\begin{equation}
    \label{conformal periodicity Lambda<0}
    \b = \frac{4\ell n \mathcal{K}(m)}{r_{+}\sqrt{K^2\ell^2-4}} \,, \qquad \beta_\tau = \frac{2n\ell^2}{r_{+}\sqrt{K^2\ell^2-4}} \left(K\ell \mathcal{K}(m)-\frac{\ell\left(K\ell^2+2\mathcal{E}\right)}{r_+^2+\ell^2}\Pi\left(\frac{r_+^2-r_-^2}{r_+^2+\ell^2} \, \bigg| \, m\right)\right)  \, .
\end{equation}

\begin{figure}[H]
        \centering
         
                \includegraphics[scale=0.4]{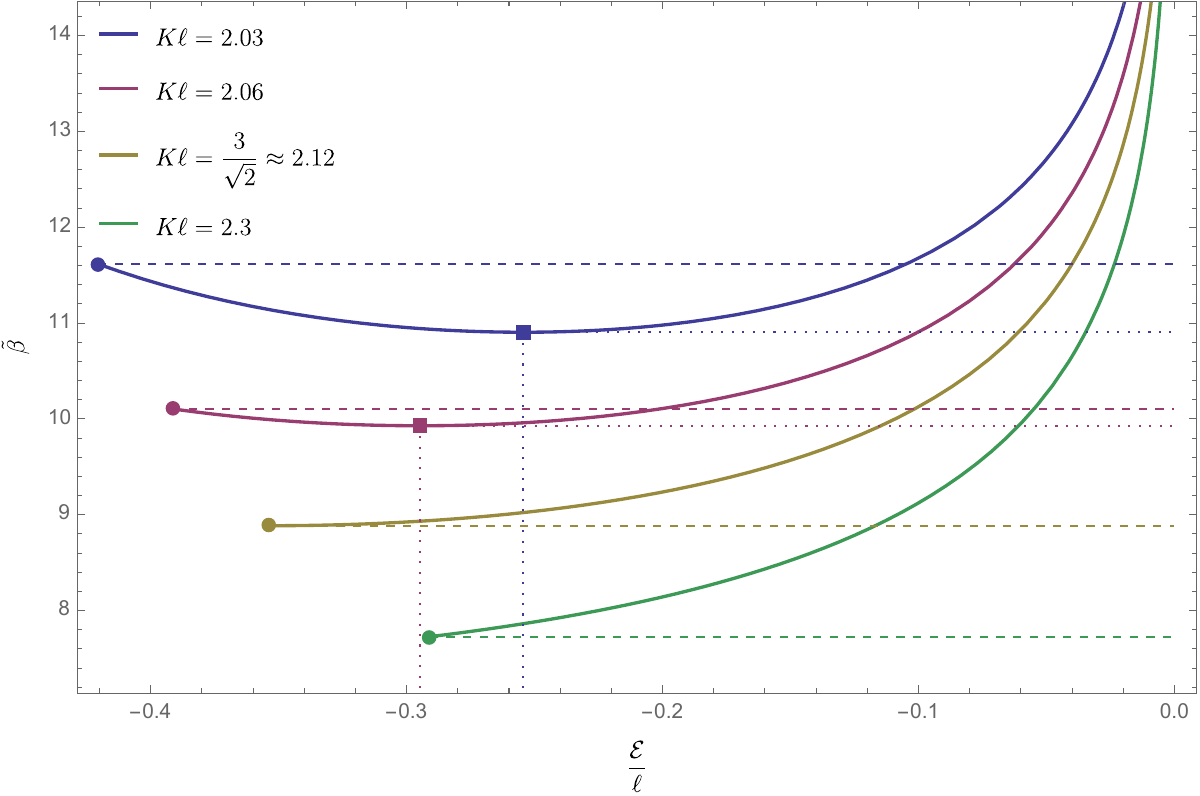}
                \caption{A plot of $\b$ versus $\tfrac{\mathcal{E}}{\ell}$ for $K\ell = 2.03$, $2.06$, $\tfrac{3}{\sqrt{2}}$, and $2.3$ and $n=1$. For the first two values of $K\ell$, $\b$ is non-monotonic with respect to $\mathcal{E}$, while it is monotonic for the latter two values. The dots denote points at which the solution approaches the static one ($\mathcal{E}=\mathcal{E}_-$). When $\b$ is non-monotonic, we depict its extremum, $(\tfrac{\mathcal{E}_*}{\ell},\b_*)$, using a square.} \label{fig: E beta AdS}
\end{figure}

Curiously, the conformal inverse temperature $\b$ behaves differently depending on whether $K\ell \geq \frac{3}{\sqrt{2}}$ or $2<K\ell<\frac{3}{\sqrt{2}}$, see figure \ref{fig: E beta AdS}.

\begin{itemize}
    \item For $2<K\ell<\frac{3}{\sqrt{2}}$, the inverse temperature $\b$ decreases over $(\mathcal{E}_-,\mathcal{E}_{*})$ then increases over $(\mathcal{E}_{*},0)$, where $\mathcal{E}_{*}>\mathcal{E}_-$ is some function of $K\ell$ that solves the equation
    \begin{equation}
        \mathcal{K}\!\left(m\right)+\ell^2\frac{\mathcal{E}K+1+\sqrt{\frac{4 \mathcal{E}^2}{\ell^2}+2 \mathcal{E}K+1}}{\mathcal{E}(4\mathcal{E}+K\ell^2)} E\!\left(m\right)=0 \, .
    \end{equation}
    Then, for values of $K\ell < \frac{3}{\sqrt{2}}$, there is a lower bound on $\b$, given by its value at $\mathcal{E}_*$,
    \begin{equation}\label{eqn: def beta star AdS}
        \b \geq \b^{*}_{n} \equiv \frac{4\ell n \mathcal{K}(m)}{r_+ \sqrt{K^2\ell^2-4}} \Big|_{\mathcal{E}=\mathcal{E}_*(K\ell)} \, .
    \end{equation}
    Accordingly, for any fixed value of $n$, there is one solution at precisely $\b^*_{n}$, characterised by $\mathcal{E}=\mathcal{E}_*$. Then, for $\b^*_{n}<\b<\b^{(-)}_n$, where
    \begin{equation}
        \b^{(-)}_{n} \equiv \frac{4\ell n \mathcal{K}(m)}{r_+ \sqrt{K^2\ell^2-4}} \Big|_{\mathcal{E}=\mathcal{E}_-}= \frac{2 \sqrt{2}n \pi }{\sqrt{-K^2\ell^2+4+K \ell \sqrt{K^2 \ell^2-4}}} \, ,
    \end{equation}
    there are two solutions, or two values of $\mathcal{E}$. Finally, for values of $\b \geq \b^{(-)}_n$, there is only one non-static solution for any fixed value of $n$. Therefore, given some values of $\b$ and $2<K\ell < \frac{3}{\sqrt{2}}$, the total number of non-static solutions is as follows:
    \begin{equation}
        \mathfrak{n}=\begin{cases}
            2\left\lfloor \frac{\b}{\b^*_{n=1}} \right\rfloor-\left\lfloor \frac{\b}{\b^{(-)}_{n=1}}\right\rfloor \, , \, \, \, \, \, \, \, \, \, \,  \qquad \frac{\b}{\b^{(-)}_{n=1}}\notin \mathbb{N}^* \, , \\
            2\left\lfloor \frac{\b}{\b^*_{n=1}} \right\rfloor-\left\lfloor \frac{\b}{\b^{(-)}_{n=1}}\right\rfloor-1 \, , \qquad \frac{\b}{\b^{(-)}_{n=1}}\in \mathbb{N}^* \, .
        \end{cases}
    \end{equation}
    \item For $K\ell \geq \frac{3}{\sqrt{2}}$, the inverse temperature $\b$ increases monotonically with respect to $\mathcal{E}$ over the interval $(\mathcal{E}_-,0)$, for any fixed value of $n$ and $K\ell$. Then, for any value of $\b > \b^{(-)}_{n}$, there is only one solution for any fixed value of $n$. Therefore, given some values of $\b$ and $K\ell \geq \frac{3}{\sqrt{2}}$, the total number of non-static solutions is
    \begin{equation}
        \mathfrak{n}=\left\lceil \frac{\b}{\b^{(-)}_{n=1}}\right\rceil-1 \, .
    \end{equation}
\end{itemize}
The on-shell action of these solutions is given by
\begin{equation}
    \label{Lambda<0 non-static pole on-shell action}
    I_{\text{on-shell}}=-\frac{\b \mathcal{E}+\beta_{\tau}}{4G_N}=-\frac{n\ell(2\mathcal{E}+K\ell^2)}{2G_Nr_+(\ell^2+r_+^2)\sqrt{K^2\ell^2-4}}\left((\ell^2+r_+^2)\mathcal{K}(m)-\ell^2\Pi\left(\frac{r_+^2-r_-^2}{\ell^2+r_+^2} \, \bigg| \, m\right)\right) \, ,
\end{equation}
from which one can derive the conformal energy and entropy,
\begin{equation}
    \label{Lambda<0 non-static pole thermo}
    E_\text{conf} = \frac{\mathcal{E}}{4G_N} \, , \qquad \mathcal{S}_\text{conf} = \frac{2\mathcal{E} \b+\beta_\tau}{4G_N}\, .
\end{equation}
The specific heat is given by
\begin{equation}
    \label{specific heat non-static pole patch Lambda<0}
    C_K=\frac{4 \ell n \mathcal{E}(1+\frac{4\mathcal{E}^2}{\ell^2}+2\mathcal{E} K)\mathcal{K}(m)^2}{r_+G_N\sqrt{K^2\ell^2-4}\left(2\mathcal{E}\left(K+\frac{4\mathcal{E}}{\ell^2}\right)\mathcal{K}(m)+r_+^2\left(K^2-\frac{4}{\ell^2}\right)E(m)\right)}  \, .
\end{equation}
For $K\ell \geq \frac{3}{\sqrt{2}}$, the specific heat is always negative. However, recall that for values $K\ell < \frac{3}{\sqrt{2}}$, there exists an interval $(\mathcal{E}_-,\mathcal{E}_*(K\ell))$ over which $\b$ decreases as a function of $\mathcal{E}$. Moreover, equation \eqref{Lambda<0 non-static pole thermo} tells us that $\mathcal{E}$ is the conformal energy in units of $4G_N$. Therefore, solutions with $K\ell<\frac{3}{\sqrt{2}}$ and $\mathcal{E}_-<\mathcal{E}<\mathcal{E}_*(K\ell)$ have positive specific heat and are therefore (meta)stable. 

\subsubsection{Non-circular pole patch}\label{Lambda<0 non-circular pole}

Now we study a class of solutions with a boundary that varies with respect to the boundary coordinate $\varphi$. We parameterise this boundary by
\begin{equation}\label{eqn: Lambda<0 bdry non-circular pole}
    \left.\tau\right|_{\Gamma}=\frac{\beta_{\tau}u}{\beta} \, , \qquad \left.r\right|_{\Gamma}=r(\varphi) \, , \qquad \left.\phi\right|_{\Gamma}=\phi(\varphi)\, .
\end{equation}
We consider the region $r \in [0,r(\varphi))$. The $n^r$ component of the unit normal vector has the same sign as $\partial_{\varphi}\phi$, and so we impose $\partial_{\varphi}\phi>0$.

\textbf{Problem.} The conformal boundary conditions impose restrictions on \eqref{eqn: Lambda<0 bdry non-circular pole} via the equations \eqref{eqn: embed conf class cond} and \eqref{eqn: embed K}. The condition on the conformal class is given by
\begin{equation}\label{eqn: Lambda<0 non-circular pole eq1}
    \frac{\beta_{\tau}^2 \mathfrak{r}^2}{\beta^2}f=\frac{ (\partial_{\varphi}r)^2}{f}+r^2(\partial_{\varphi}\phi)^2 \, ,
\end{equation}
while the condition on the trace of the extrinsic curvature gives
\begin{equation}\label{eqn: Lambda<0 non-circular pole eq2}
    \frac{2 r \partial_{\varphi}\phi (\partial_{\varphi}r)^2  f'+f \left(r^3 (\partial_{\varphi}\phi)^3 f'-2 r \partial_{\varphi}\phi \partial_{\varphi}^2r +2 r \partial_{\varphi}^2 \phi \partial_{\varphi}r+4 \partial_{\varphi}\phi (\partial_{\varphi}r)^2\right)+2 r^2 f^2 (\partial_{\varphi}\phi)^3}{2 \left(r^2 f (\partial_{\varphi}\phi)^2+(\partial_{\varphi}\phi)^2\right)^{3/2}} = K \, ,
\end{equation}
where $f(r)=1+\frac{r^2}{\ell^2}$ and its derivative are evaluated at $r(\varphi)$.

In what follows, we first find general solutions to equations \eqref{eqn: Lambda<0 non-circular pole eq1} and \eqref{eqn: Lambda<0 non-circular pole eq2}, then impose the embedding condition, which restricts the space of solutions.

\textbf{Solution.} One can show, following a proof similar to the one in appendix \ref{sec: du T flat 1} for the flat case, that equations \eqref{eqn: Lambda<0 non-circular pole eq1} and \eqref{eqn: Lambda<0 non-circular pole eq2} can be rewritten as a first-order ODE of the form
\begin{equation}\label{eqn: conserved quantity non-circular pole Lambda<0}
    \frac{d\mathcal{E}}{d\varphi}=0 \, , \qquad \mathcal{E}\equiv -\frac{K \beta_{\tau}^2\mathfrak{r}^2}{2\beta^2}\left(1+\frac{r^2}{\ell^2}\right)+\frac{\beta_{\tau}\mathfrak{r}}{\ell^2 \beta} \, r^2 \partial_{\varphi}\phi  \, ,
\end{equation}
where $\mathcal{E}$ is an integration constant of dimension length. We now use this and equation \eqref{eqn: Lambda<0 non-circular pole eq1} to write an ODE in the variable $y\equiv\tfrac{\beta_\tau}{2\pi}\sqrt{1+\tfrac{r^2}{\ell^2}}$, given by
\begin{equation}
    \frac{\beta^2}{4\pi^2\r^2}(\p_\varphi y)^2 +V_\text{eff}(y)  =0 \, , \qquad V_\text{eff}(y)\equiv\left(\frac{K^2}{4}-\frac{1}{\ell^2}\right)y^4 + \left(\frac{\beta_\tau^2}{4\pi^2\ell^2}+\frac{\mathcal{E} K \beta^2}{4\pi^2\r^2}\right)y^2 + \left(\frac{\mathcal{E}\beta^2}{4\pi^2\r^2}\right)^2 \, .
\end{equation}
This is the equation of motion of a classical particle moving in a potential well $V_{\text{eff}}(r)$. As in the non-static case, we do not consider cases with $|K\ell| \leq 2$ which have an inverted effective potential, and focus on $|K\ell|>2$. 

The solution can be written in terms of the elliptic Jacobi dn function,
\begin{equation}\label{eqn: non cir pole r sol Lambda<0}
    y(\varphi)= \frac{\beta_\tau}{2\pi}\sqrt{1+\frac{r(\varphi)^2}{\ell^2}}=y_+ \text{dn}\left(\frac{y_+\sqrt{K^2\ell^2-4}}{2\ell}\frac{2\pi \r \varphi}{\beta} \, \bigg| \, m\right) \, ,
\end{equation}
where the roots of the effective potential $y_\pm$ and the parameter $m$ are given by
\begin{equation}\label{eqn: ypm non cir pole Lambda<0}
    y_\pm \equiv \ell \sqrt{\frac{-\frac{\beta_\tau^2}{2\pi^2\ell^2}-\frac{\mathcal{E}K\beta^2}{2\pi^2\r^2}\pm\frac{\beta_\tau}{\pi\ell}\sqrt{\frac{\beta_\tau^2}{4\pi^2\ell^2}+\frac{\mathcal{E}^2\beta^4}{\pi^2\beta_\tau^2\r^4}+\frac{\mathcal{E}K\beta^2}{2\pi^2\r^2}}}{K^2\ell^2-4}} \, , \qquad m\equiv1-\frac{y_-^2}{y_+^2}\, .
\end{equation}
The introduced quantities $y_{\pm}$ are the maximum and minimum values of $y(\varphi)$. Their reality condition enforces bounds on $\mathcal{E}$,
\begin{equation}
    \begin{cases}
        \mathcal{E} \leq \mathcal{E}_{-} \equiv \frac{\beta_{\tau}^2\mathfrak{r}^2}{4\ell \beta^2}\left(-K\ell-\sqrt{K^2\ell^2-4}\right) \, , \qquad K\ell>0\\
        \mathcal{E}\geq \mathcal{E}_{+} \equiv \frac{\beta_{\tau}^2\mathfrak{r}^2}{4\ell \beta^2}\left(-K\ell+\sqrt{K^2\ell^2-4}\right) \, , \qquad K\ell<0 \, .
    \end{cases}
     .
\end{equation}
Note that $\mathcal{E}_{-}<0$ for $K\ell>0$ and $\mathcal{E}_{+}>0$ for $K\ell<0$, which means that $\mathcal{E}=0$ is never attained, unlike all the previous cases. Moreover, the two roots coincide at $\mathcal{E}=\mathcal{E}_{\pm}$. In addition, $y_{-}$ saturates its lower bound $\frac{\beta_{\tau}}{2\pi}$ at the particular value $\mathcal{E}_0 \equiv -\frac{K\beta_{\tau}^2\mathfrak{r}^2}{2\beta^2}$.

Plugging \eqref{eqn: non cir pole r sol Lambda<0} back in \eqref{eqn: conserved quantity non-circular pole Lambda<0} and integrating over $\varphi$, we obtain
\begin{equation}\label{eqn: non cir pole phi sol Lambda<0}
    \phi(\varphi)=\phi_0 +\frac{K \beta_\tau \r(\varphi-\varphi_0)}{2\beta} - \frac{\left(\frac{\mathcal{E}\beta^2}{2\pi^2\r^2}+\frac{\beta_\tau^2K}{4\pi^2}\right)\Pi\left(\frac{y_+^2-y_-^2}{y_+^2-\frac{\beta_\tau^2}{4\pi^2}};\text{am}(x|m)|m\right)}{y_+ \frac{\beta_\tau}{2\pi\ell}\left(1-\frac{4\pi^2y_+^2}{\beta_\tau^2}\right)\sqrt{K^2\ell^2-4}} \, ,
\end{equation}
where $x = \frac{\pi y_+ \mathfrak{r}  \sqrt{K^2 \ell^2-4}}{\beta \ell}\varphi$.

The solution admits a periodic structure,
\begin{equation}
    \label{Periodicity properties Lambda<0 non-circular dS pole}
    \begin{cases}
        r\!\left(\varphi+\frac{2n\beta \ell \mathcal{K}(m)}{\pi y_+ \mathfrak{r}\sqrt{K^2\ell^2-4}}\right) = r(\varphi) \, , \\
        \phi \!\left(\varphi+\frac{2n\beta \ell \mathcal{K}(m)}{\pi y_+ \mathfrak{r}\sqrt{K^2\ell^2-4}}\right) \! =\phi(\varphi)+n\left[\frac{\beta_{\tau}  K \ell \mathcal{K}\left(m\right)}{\pi  y_{+} \sqrt{K^2 \ell^2-4}}-\frac{4 \pi  \ell \Pi \left(\frac{y_{+}^2-y_{-}^2}{y_{+}^2-\frac{\beta_{\tau} ^2}{4 \pi ^2}} \, \big| \, m\right) \left(\frac{\beta ^2 \mathcal{E}}{2 \pi ^2 \mathfrak{r}^2}+\frac{\beta_{\tau} ^2 K}{4 \pi ^2}\right)}{\beta_{\tau}  y_{+} \sqrt{K^2 \ell^2-4} \left(1-\frac{4 \pi ^2 y_{+}^2}{\beta_{\tau}^2}\right)}\right] \, ,
    \end{cases}
\end{equation}
where $n$ is an arbitrary positive integer.

The Weyl factor and conformal stress tensor are given by
\begin{equation}
    \bomega(\varphi) = \log \left(\sqrt{1+\frac{r^2}{\ell^2}}\frac{\beta_\tau}{\beta}\right) \, , \qquad T_{ij}d\sigma^i d\sigma^j =\frac{\mathcal{E}}{8\pi G_N \r^2}\left(-du^2+\r^2d\varphi^2\right)\, .
\end{equation}

\textbf{Self-intersections and regime of validity.} This solution violates the global embedding condition \eqref{eqn: global emb} for some values of $\mathcal{E}$.

For $K\ell>0$, there are self-intersections over $\mathcal{E} < \mathcal{E}_{0}$. They then disappear over $\mathcal{E}_{0} < \mathcal{E} < \mathcal{E}_{-}$.

For $K\ell<0$, there are no self-intersections over $\mathcal{E}_{+}<\mathcal{E}<\mathcal{E}_0$. However, these solutions have the wrong sign of $\partial_{\varphi}\phi$, and are therefore discarded. Self-intersections then emerge over $\mathcal{E}>\mathcal{E}_0$.

Therefore, there are no physical solutions with $K\ell<0$, and the physical solutions with $K\ell>0$ have
\begin{equation}
    \mathcal{E}_{0} < \mathcal{E} < \mathcal{E}_- \, .
\end{equation}

\textbf{Global smoothness condition.} In fact, to ensure that the global structure is well-defined, there are two more conditions we need to impose: the periodicities in $\phi$ and $\varphi$, which have been evaluated in \eqref{Periodicity properties Lambda<0 non-circular dS pole} must both be equal to $2\pi$. From now on, we denote these quantities by
\begin{equation}
    \label{periodicities non-circular pole AdS}
    \beta_{\varphi} \equiv \frac{2n\beta \ell \mathcal{K}(m)}{\pi y_+ \mathfrak{r}\sqrt{K^2\ell^2-4}} \,, \qquad \beta_{\phi}=n\left[\frac{\beta_{\tau}  K \ell \mathcal{K}\!\left(m\right)}{\pi  y_{+} \sqrt{K^2 \ell^2-4}}-\frac{4 \pi  \ell \Pi \left(\frac{y_{+}^2-y_{-}^2}{y_{+}^2-\frac{\beta_\tau^2}{4 \pi ^2}} \, \big| \, m\right) \left(\frac{\beta ^2 \mathcal{E}}{2 \pi ^2 \mathfrak{r}^2}+\frac{\beta_{\tau} ^2 K}{4 \pi ^2}\right)}{\beta_{\tau}  y_{+} \sqrt{K^2 \ell^2-4} \left(1-\frac{4 \pi ^2 y_{+}^2}{\beta_{\tau}^2}\right)}\right] \, .
\end{equation}
To solve the equations $\beta_{\varphi}=2\pi$ and $\beta_{\phi}=2\pi$, we find it convenient to implement a change of variable $\mathcal{E}=\frac{\mathfrak{e}\beta_{\tau}^2 \mathfrak{r}^2}{\beta^2}$. The equation $\beta_{\varphi}=2\pi$ then determines $\beta_{\tau}$ in terms of the boundary data $\tilde{\beta}$,
\begin{equation}
    \label{beta tau in terms of beta non-circular AdS pole}
    \beta_{\tau} = \frac{\sqrt{2} \ell n \mathcal{K}\!\left(m\right)}{\pi \sqrt{-1+\left(\sqrt{1+2\ell^2\mathfrak{e} (2 \mathfrak{e}+K)}-\mathfrak{e} K \ell^2\right)}} \, \tilde{\beta} \, .
\end{equation}
Note that $m$ becomes independent of $\beta$, $\beta_{\tau}$, and $\mathfrak{r}$ when written in terms of $\mathfrak{e}$, and so $\beta_{\tau}$ is directly proportional to $\tilde{\beta}$. As for the $\beta_{\phi}=2\pi$ equation, it becomes entirely independent of $\beta_{\tau}$, $\beta$,  and $\mathfrak{r}$. We find that there are no solutions to this equation,
\begin{equation}
    \mathfrak{n}_{\text{non-circular pole}}=0 \, .
\end{equation}

\subsection{Black hole patch}\label{Lambda<0 BH patch}

We now study the conformal thermodynamics of the black hole patch solutions. These are solutions endowed with the bulk metric
\begin{equation}\label{eqn: BTZ metric again}
    ds^2= \frac{\bar r^2-\rh^2}{\ell^2} \, d\bar\tau^2 + \frac{\ell^2}{\bar r^2-\rh^2} \, d\bar r^2 + \bar r^2 d\bar\phi^2 \, , \qquad \Lambda=-\frac{1}{\ell^2}\, ,
\end{equation}
where $\bar \tau\sim\bar\tau + \tfrac{2\pi \ell^2}{\rh}$ and  $\bar\phi \sim \bar\phi+2\pi$. The spacetime region of interest includes the horizon $\bar r=r_h$ and admits a boundary situated at the radial coordinate $\left.\bar r\right|_{\Gamma}$, which will be specified later. At the boundary, we impose the conformal boundary conditions \eqref{eqn: bdry metric S1xS1}.

We recall that the metric \eqref{eqn: BTZ metric again} is related to \eqref{eqn: global AdS metric again} via the following coordinate transformation,
\begin{equation}
    \frac{\bar r^2}{\rh^2}= 1+\frac{r^2}{\ell^2} \, , \qquad \bar \tau =\frac{\ell^2}{\rh}\phi \, , \qquad \bar \phi = \frac{\tau}{\rh} \, ,
\end{equation}
with $\beta_\tau = 2 \pi \rh$. The transformation exchanges the Euclidean temporal and spatial coordinates.

In the following subsections, we implement the boundary conditions \eqref{eqn: bdry metric S1xS1} using the embedding method to obtain the boundary location profile. We study three classes of boundaries. In section \ref{Lambda<0 static and circular black hole}, we study boundaries situated at a fixed radial coordinate. Section \ref{Lambda<0 non-static black hole} involves boundaries with a radius that varies along the boundary coordinate $u$, dubbed non-static black hole patches. In section \ref{Lambda<0 non-circular black hole} we analyse boundaries with a radius that instead varies instead with respect to the boundary coordinate $\varphi$, dubbed non-circular black hole patches. Their thermodynamic properties are discussed at the end of each subsection.

\subsubsection{Static and circular black hole patch}\label{Lambda<0 static and circular black hole}

Given some constant value of $K\ell>2$, the class of static and circularly symmetric black hole patch is described by the metric \eqref{eqn: BTZ metric again} with a boundary parameterised by
\begin{equation}
    \left.\bar \tau\right|_\Gamma = \frac{\ell}{2\r}\left(K \ell + \sqrt{K^2\ell^2-4}\right)u \, , \qquad \left.\bar r\right|_{\Gamma} = \rh\sqrt{\frac{K\ell}{2\sqrt{K^2\ell^2-4}}+\frac{1}{2}}\, , \qquad \left.\bar \phi\right|_{\Gamma}= \varphi \, .
\end{equation}
From this, the induced metric can be computed straightforwardly,
\begin{equation}
    \left.ds^2\right|_{\Gamma} = 
    \frac{\rh^2}{2\r^2}\left(1+\frac{K\ell}{\sqrt{K^2\ell^2-4}}\right)\left(du^2+\r^2d\varphi^2\right) \,.
\end{equation}
The constant Weyl factor implies a flat intrinsic geometry. We note also that the conformal stress tensor of the boundary is given by
\begin{equation}
    T_{ij}d\sigma^i d\sigma^j = \frac{\rh^2\left(K\ell+\sqrt{K^2\ell^2-4}\right)}{32 \pi G_N\r^2\ell}\left(-du^2+\r^2d\varphi^2\right) \, .
\end{equation}

\textbf{Conformal thermodynamics.} The regularity at the black hole horizon fixes the periodicity of the coordinate $\bar \tau$ to be $\tfrac{2\pi \ell^2}{\rh}$. Using the boundary parameterisation, this fixes the periodicity of $u$, which in turn leads to
\begin{equation}
    \b = \frac{\pi \ell}{\rh}\left(K\ell-\sqrt{K^2\ell^2-4}\right) \, .
\end{equation}
This equation fixes $\rh$ as a function of $\b$ and $K\ell$. Evaluating the on-shell action, we obtain
\begin{equation}
    \label{BH patch homo on-shell action}
    I_\text{on-shell}=- \frac{\pi^2\ell}{4G_N\b}\left(K\ell-\sqrt{K^2\ell^2-4}\right) \, .
\end{equation}
Plugging into the thermodynamic relations, we obtain the corresponding conformal energy, entropy, and specific heat at fixed $K$,
\begin{equation}
    E_\text{conf} = \frac{\pi^2 \mathfrak{c}_\text{AdS}}{3\b^2} \, , \qquad \mathcal{S}_\text{conf} = C_K = \frac{2\pi^2 \mathfrak{c}_\text{AdS}}{3\b} \, , \qquad \mathfrak{c}_\text{AdS}\equiv \frac{3\ell}{4G_N}\left(K\ell-\sqrt{K^2\ell^2-4}\right)\,.
\end{equation}
In \cite{Allameh:2025gsa}, the dimensionless parameter $\mathfrak{c}_\text{AdS}$ is shown to capture the number of effective degrees of freedom of the dual CFT$_2$ coupled to timelike Liouville theory and deformed by a marginal deformation of the $T\bar{T}$-type.

\subsubsection{Non-static black hole patch}\label{Lambda<0 non-static black hole}

Now we consider a class of black hole patches with a non-trivial Euclidean time dependence. We describe the boundary by
\begin{equation}
    \left.\bar \tau\right|_\Gamma = \bar \tau(u) \, , \qquad \left.\bar r\right|_\Gamma=\bar r(u) \,  , \qquad \left.\bar \phi \right|_\Gamma=\varphi \, .
\end{equation}
The bulk region is defined as $\bar r \in (\rh ,\bar r(u))$. The unit normal vector has a component $n^{\bar r}$ that has the same sign as $\partial_{u}\bar \tau$, and so we require $\partial_{u}\bar \tau>0$.

\textbf{Problem.} The conformal boundary conditions impose equations \eqref{eqn: embed conf class cond} and \eqref{eqn: embed K} on $\tau(u)$ and $r(u)$. These are given by \eqref{Lambda<0 non-static pole eq1} and \eqref{Lambda<0 non-static pole eq2} with $f(r) = \tfrac{\bar r^2-\rh^2}{\ell^2}$ and replacing $(r(u),\tau(u))\to(\bar r(u), \bar \tau(u))$.

\textbf{Solutions.} The general solution is given by
\begin{equation}
    \bar r(u) = \rh \sqrt{1+\frac{r\left(\frac{u\beta}{2\pi\r^2}\right)^2}{\ell^2}}\, , \qquad \bar \tau (u) = \frac{\ell^2}{\rh}\phi\left(\frac{u\beta}{2\pi\r^2}\right),
\end{equation}
where the functions $r(\varphi)$ and $\phi(\varphi)$ are given by \eqref{eqn: non cir pole r sol Lambda<0} and \eqref{eqn: non cir pole phi sol Lambda<0} with $\beta_\tau =2 \pi \rh$. Consequently, the solutions only exist when $K\ell>2$ and are labelled by the dimensionful parameter $\mathcal{E}$. By imposing the reality and the absence of self-intersection conditions, this parameter is subject to 
\begin{equation}\label{eqn: space of E non-sta bh}
    \mathcal{E}_0 < \mathcal{E} < \mathcal{E}_- \, , \qquad 
\end{equation}
where $\mathcal{E}_0 = -\tfrac{2\pi^2K\rh^2\r^2}{\beta^2}$ and $\mathcal{E}_-=-\tfrac{\pi^2\rh^2\r^2}{\ell \beta^2}\left(K\ell+\sqrt{K^2\ell^2-4}\right)$ are negative definite functions of $K\ell$.

The Weyl factor and conformal stress tensor are given by
\begin{equation}
    \bomega(u) = \log \frac{\bar r(u)}{\r}\, , \qquad T_{ij}d\sigma^i d\sigma^j = \frac{\mathcal{E}\beta^2}{32\pi^3G_N\r^4}\left(-du^2+\r^2d\varphi^2\right) \, .
\end{equation}

Demanding that the solutions are compatible with the identifications $u\sim u+\beta$ and $\tau\sim\tau+\tfrac{2\pi\ell^2}{\rh}$ leads to
\begin{equation}
    \b = \frac{4n \ell \mathcal{K}(m)}{y_+\sqrt{K^2\ell^2-4}} \, , \qquad 2\pi = n\left[\frac{2  K\rh \ell \mathcal{K}\!\left(m\right)}{  y_{+} \sqrt{K^2 \ell^2-4}}-\frac{2  \ell \Pi \left(\frac{y_{+}^2-y_{-}^2}{y_{+}^2-\rh^2} \, \big| \, m\right) \left(\frac{\beta ^2 \mathcal{E}}{2 \pi ^2 \mathfrak{r}^2}+K\rh^2\right)}{\rh y_{+} \sqrt{K^2 \ell^2-4} \left(1-\frac{y_{+}^2}{\rh^2}\right)}\right]\,.
\end{equation}
Performing a change of variable $\mathcal{E} =\tfrac{4\pi^2\rh^2\r^2}{\beta^2}\mathfrak{e}$, the second equation becomes an equation determining $\mathfrak{e}$ for a given $K\ell>2$ and a positive integer $n$. Restricted to $\mathcal{E}$ which lies in \eqref{eqn: space of E non-sta bh}, we find that there is no $(\mathfrak{e},n)$ which satisfies this equation, and thereby no non-static black hole patch exists.

\subsubsection{Non-circular black hole patch}\label{Lambda<0 non-circular black hole}

The class of non-circular black hole patches can be described by the boundary
\begin{equation}
    \left.\bar \tau\right|_\Gamma = \frac{2\pi \ell^2 u }{\rh \beta } \, , \qquad \left.\bar r\right|_\Gamma=\bar r(\varphi) \,  , \qquad \left.\bar \phi \right|_\Gamma=\bar \phi(\varphi) \, .
\end{equation}
The outward-pointing normal vector imposes $\p_\varphi \bar \phi>0$.

\textbf{Problem.} The conformal boundary conditions lead to equations of $\bar r(\varphi)$ and $\bar \phi(\varphi)$. These are given by \eqref{eqn: Lambda<0 non-circular pole eq1} and \eqref{eqn: Lambda<0 non-circular pole eq2} with $f(r) = \tfrac{r^2-\rh^2}{\ell^2}$ and replacing $(r(\varphi),\phi(\varphi))\to (\bar r(\varphi),\bar \phi(\varphi))$. 

\textbf{Solutions.} The general solution to the above problem is then given by
\begin{equation}\label{eqn: r phi sol non-cir bh}
    \bar r (\varphi)= \rh\sqrt{1+ \frac{r\left(\frac{2\pi \r^2\varphi}{\beta}\right)^2}{\ell^2}} \, , \qquad \bar \phi(\varphi) = \frac{\tau\left(\frac{2\pi \r^2\varphi}{\beta}\right)}{\rh} \, ,
\end{equation}
where the functions $r(u)$ and $\tau(u)$ are given by \eqref{eqn: rsol AdS} and \eqref{eqn: tsol AdS}. Consequently, the solutions only exist for $K\ell>2$ and are labelled by the dimensionful parameter $\mathcal{E}$. Due to the reality and the absence of self-intersection conditions, $\mathcal{E}$ is required to obey
\begin{equation}\label{eqn: space of e non-cir bh}
    \mathcal{E}_-<\mathcal{E} <0 \, ,
\end{equation}
where $\mathcal{E}_-$ is defined in \eqref{eqn: def epm nonsta AdS}.

The Weyl factor and conformal stress tensor are given by 
\begin{equation}\label{eqn: Weyl Tij non-cir bh}
    \bomega(\varphi) = \log \frac{2\pi r\!\left(\frac{2\pi \r^2 \varphi}{\beta}\right)}{\beta} \, , \qquad T_{ij}d\sigma^i d\sigma^j = - \frac{\pi \mathcal{E}}{2\beta^2 G_N}\left(-du^2 + \r^2 d\varphi^2\right) \, .
\end{equation}

The conformal boundary data of the solution follows from the periodic structure \eqref{Lambda<0 non-static pole periodicities}. In particular, requiring that $\varphi \sim \varphi+2\pi$ and $\bar \phi\sim \bar \phi+2\pi$ leads to
\begin{equation}\label{eqn: conf data non-cir bh}
    \b = \frac{\pi^2 r_+ \sqrt{K^2\ell^2-4}}{n \ell \mathcal{K}(m)} \, , \qquad \rh =\frac{n \ell^2}{\pi r_+\sqrt{K^2\ell^2-4}}\left(K\ell \mathcal{K}(m) - \frac{\ell\left(K\ell^2+2\mathcal{E}\right)}{r_+^2+\ell^2}\Pi \left(\frac{r_+^2-r_-^2}{r_+^2+\ell^2} \, \bigg| \, m\right)\right) \, ,
\end{equation}
where $r_\pm$ and $m$ are given by \eqref{eqn: rpm AdS}. Inverting the first equation fixes $\mathcal{E}$ to be a function of $K\ell$ and $\b$. In the parameter regime $ \tfrac{3}{\sqrt{2}}<K\ell$, there exists a single branch of $\mathcal{E}$ which solves this equation, while for $2< K\ell<\tfrac{3}{\sqrt{2}}$, we obtain two different branches of $\mathcal{E}$.

\textbf{Conformal thermodynamics.} Evaluating the action on-shell, we obtain
\begin{equation}
    \label{Lambda<0 non-circular BH patch on-shell action}
    I_\text{on-shell} = - \frac{\pi^2 \mathcal{E}}{G_N \b} - \frac{\pi \rh}{2 G_N} \, .
\end{equation}
Applying the thermodynamic relations, we obtain the conformal energy and entropy,
\begin{equation}\label{eqn: thermo non-cir bh}
    E_\text{conf} = -\frac{\pi^2 \mathcal{E}}{G_N \b^2} \, , \qquad \mathcal{S}_\text{conf} = \frac{\pi \rh}{2G_N} \, .
\end{equation}
Indeed, the entropy obeys the area law of the black hole horizon. The specific heat at fixed $K$ is given by 
\begin{equation}\label{specific heat Lambda<0 non-cir}
    C_K = - \frac{2\pi^2 \mathcal{E}}{G_N \b} +\frac{4 \ell n \mathcal{E}(1+\frac{4\mathcal{E}^2}{\ell^2}+2\mathcal{E} K)\mathcal{K}(m)^2}{r_+G_N\sqrt{K^2\ell^2-4}\left(2\mathcal{E}\left(K+\frac{4\mathcal{E}}{\ell^2}\right)\mathcal{K}(m)+r_+^2\left(K^2-\frac{4}{\ell^2}\right)E(m)\right)} \, .
\end{equation}
For $\tfrac{3}{\sqrt{2}}<K\ell$, the specific heat is negative for any $\b$ implying that the solutions are thermally unstable. In contrast, for $2<K\ell<\tfrac{3}{\sqrt{2}}$ where there exists two branches of $\mathcal{E}$, we find that one branch has a positive $C_K$ implying its thermal stability, while the other one has a negative $C_K$. In particular, the thermally stable branch has a lower value of $\mathcal{E}$ than the unstable one when evaluated at the same $K\ell$ and $\b$.

\subsection{Thermodynamic phase space}\label{Lambda<0 thermodynamic phase space}

In this section, we combine results in the $\Lambda<0$ section and analyse thermodynamic phase structure of the total system. The main quantity of interest here is the torus partition function $\mathcal{Z}(\b,K)$ with $K\ell>2$ in the limit $G_N\to0$, which is given by
\begin{equation}\label{eqn: partition fn Lambda<0}
    \mathcal{Z}(\b,K) = e^{-I_\text{on-shell}^{\text{(hom. pole)}}}+e^{-I_\text{on-shell}^{\text{(hom. BH)}}}+ \sum_n e^{-I_{n,\text{on-shell}}^{\text{(non-static pole)}}}+ \sum_n e^{-I_{n,\text{on-shell}}^{\text{(non-circular BH)}}} \, .
\end{equation}
Similarly to the flat and de Sitter cases, we call the saddle which dominates the partition function and has a positive specific heat a stable configuration, while a sub-dominant saddle that has a positive specific heat is called a meta-stable configuration. A saddle which a negative specific heat is called unstable. Below, we explain each terms in \eqref{eqn: partition fn Lambda<0}.

\begin{itemize}
    \item The first and second terms come from the static and circular pole and black hole patches, which have an on-shell action, given in \eqref{Lambda<0 homogeneous pole-patch on-shell action} and \eqref{BH patch homo on-shell action},
    \begin{equation}
        I_{\text{on-shell}}^{\text{(hom. pole)}}=I_{\text{on-shell}}^{\text{(hom. BH)}}\bigg|_{\b \rightarrow \frac{4\pi^2}{\b}}=-\frac{\b\ell}{16 G_N}\left(K\ell-\sqrt{K^2\ell^2-4}\right) \, .
    \end{equation}
    These solutions exist for all $\b>0$ and $K\ell>2$.
    
    \item The third and fourth terms consist of a finite sum of non-static pole patch and non-circular black hole patch solutions, respectively. The number of solutions depends on both $\b$ and $K\ell$. Moreover, there exists a critical value $K_c\ell\equiv \tfrac{3}{\sqrt{2}}$ around which the results are qualitatively different.

    For $K_c\ell \leq K\ell$, the number of coexisting solutions is given by
    \begin{equation}
        \mathfrak{n}_\text{non-static pole}=\mathfrak{n}_\text{non-circular BH}\bigg|_{\b \rightarrow \frac{4\pi^2}{\b}}= \left\lceil\frac{\b}{\b^{(-)}_{n=1}}\right\rceil-1 \, . 
    \end{equation}

    For $2<K\ell<K_c\ell$, the number of coexisting solutions is given by
    \begin{equation}
        \mathfrak{n}_\text{non-static pole}=\mathfrak{n}_\text{non-circular BH}\bigg|_{\b \rightarrow \frac{4\pi^2}{\b}}= 2\left\lfloor\frac{\b}{\b^*_{n=1}}\right\rfloor-\left\lfloor\frac{\b}{\b^{(-)}_{n=1}}\right\rfloor \, .
    \end{equation}

    The first sum comes from non-static pole patch solutions, where these solutions exist for $\b>\b^{(-)}_n$, while the second sum comes from non-circular black hole patch solutions, where they obey $\b<\tfrac{4\pi^2}{\b^{(-)}_n}$. As such, the numbers of coexisting non-static pole patches and non-circular black hole patches are given by
    \begin{equation}
        \mathfrak{n}_\text{non-static pole}= \left\lceil\frac{\b}{\b^{(-)}_{n=1}}\right\rceil-1 \, , \qquad \mathfrak{n}_\text{non-circular BH} = \left\lceil\frac{4\pi^2}{\b^{(-)}_{n=1}\b}\right\rceil-1 \, . 
    \end{equation}
    The on-shell action of these solutions are given in \eqref{Lambda<0 non-static pole on-shell action} and \eqref{Lambda<0 non-circular BH patch on-shell action},
    \begin{equation}
    \begin{aligned}
        &I_\text{on-shell}^{(\text{non-static pole})}=\left.I_\text{on-shell}^{(\text{non-circular BH})}\right|_{\b\to\frac{4\pi^2}{\b}} = \\
        &-\frac{n\ell(2\mathcal{E}+K\ell^2)}{2G_Nr_+(\ell^2+r_+^2)\sqrt{K^2\ell^2-4}}\left((\ell^2+r_+^2)\mathcal{K}(m)-\ell^2\Pi\left(\frac{r_+^2-r_-^2}{\ell^2+r_+^2} \, \bigg| \, m\right)\right) \, ,
    \end{aligned}
    \end{equation}
    where $(\mathcal{E},n)$ solve, in the pole patch case, the $\b$ equation in \eqref{conformal periodicity Lambda<0}, and in the black hole patch case, the $\b$ equation in \eqref{eqn: conf data non-cir bh}. The parameter $m$ is defined by $m=1-\frac{r_-^2}{r_+^2}$ where $r_\pm$ are functions of $\mathcal{E}$ and $K\ell$, given in \eqref{eqn: rpm AdS}.
\end{itemize}

We now describe the different thermodynamic phases of the system. This varies depending on whether $K\ell < K_c\ell$ or $K\ell \geq K_c\ell$.

\textbf{The case of $K\ell \geq K_c\ell$.} We plot the on-shell action of the various contributions in figure \ref{fig: Ionshell AdS K>3/sqrt{2}}. At low temperatures $\b > \b_c \equiv 2\pi$, the dominant saddle is the homogeneous pole patch solution. At high temperatures $\b < \b_c$, the dominant saddle is the homogeneous black hole patch solution. At the critical point, the system undergoes a first-order phase transition where the conformal energy exhibits a discontinuous jump. Moreover, all inhomogeneous solutions have a negative specific heat and are therefore thermally unstable.

\begin{figure}[H]
        \centering
         
                \includegraphics[scale=0.5]{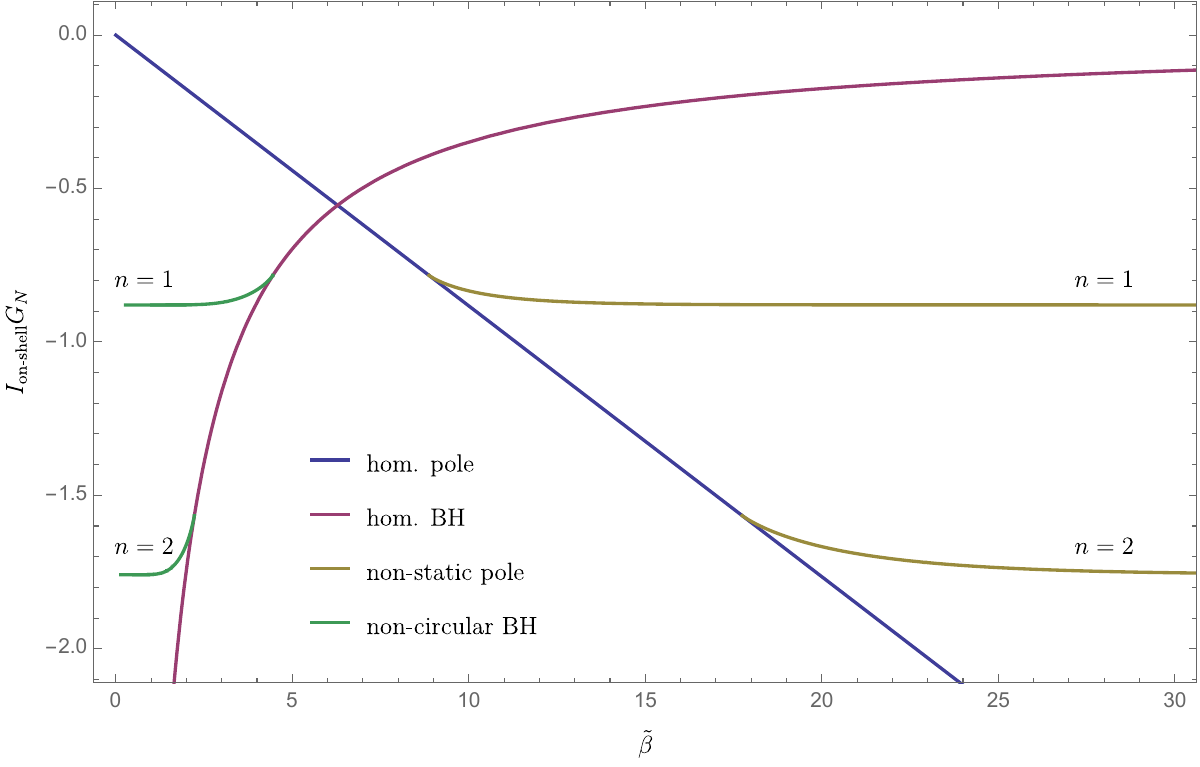}
                \caption{A plot of $I_\text{on-shell}$ of various contributions versus $\b$ for $K\ell =\tfrac{3}{\sqrt{2}}+10^{-3}$.} \label{fig: Ionshell AdS K>3/sqrt{2}}
\end{figure}

\textbf{The curious case of $2<K\ell < K_c\ell$.} We plot the on-shell action of the various contributions in figure \ref{fig: Ionshell AdS K<3/sqrt{2}}. Evidently, this is a unique case where there exist solutions that are more thermodynamically favourable than the homogeneous solutions. As one can see in figure \ref{fig: Ionshell AdS K<3/sqrt{2}}, the non-static pole patch solutions and the non-circular black hole patch solutions have an on-shell action that is lower than that of the homogeneous solutions. Moreover, the specific heat of these solutions, found in \eqref{specific heat non-static pole patch Lambda<0} and \eqref{specific heat Lambda<0 non-cir}, is positive. These solutions are therefore thermally stable as well.

It is interesting to note that the resulting semi-classical torus partition function exhibits a modular invariance while the entropy does not follow the Cardy formula at high temperature. This fact is not in tension with \cite{CARDY1986186}, since the partition function at low temperature does not take the form of $-\log \mathcal{Z} = \b E_\text{vac}$ for some $\b$-independent quantity $E_\text{vac}$.

\begin{figure}[H]
        \centering
         
                \includegraphics[scale=0.5]{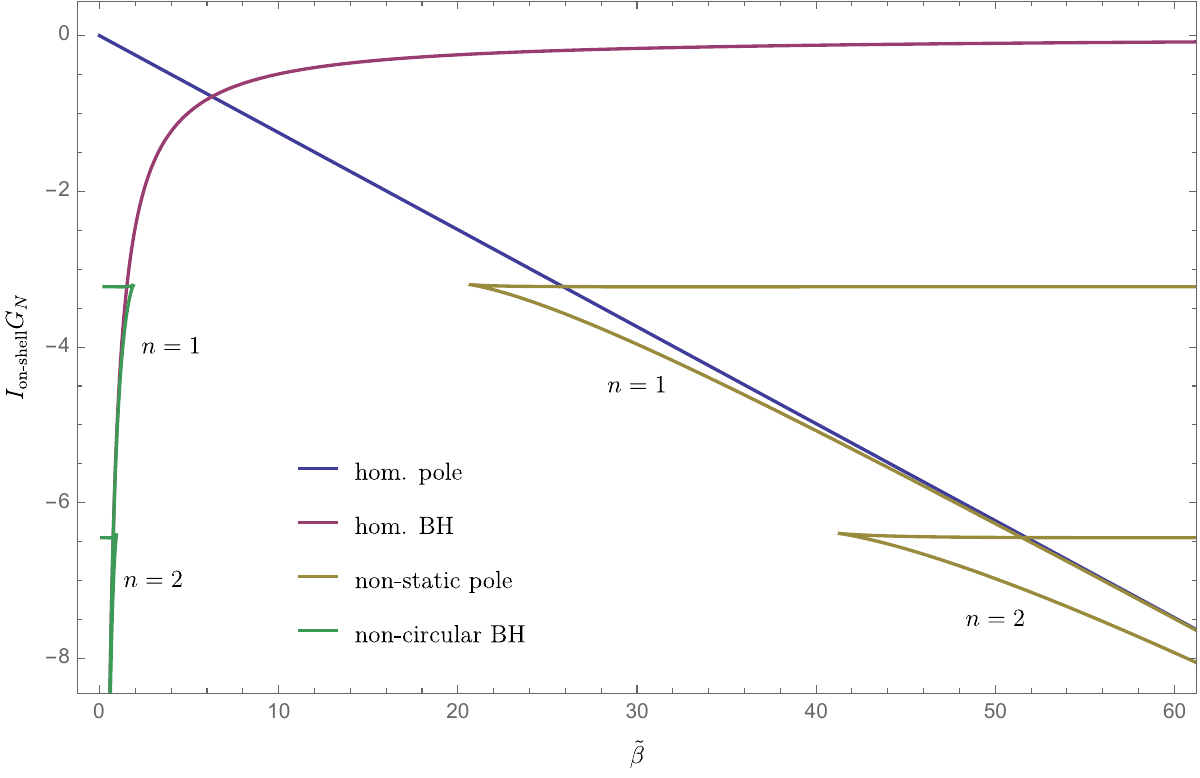}
                \caption{A plot of $I_\text{on-shell}$ of various contributions versus $\b$ for $K\ell =2+10^{-5}$.} \label{fig: Ionshell AdS K<3/sqrt{2}}
\end{figure}

\subsection{The AdS boundary limit}\label{sec: AdS limit}

A special limit unique to the case of negative cosmological constant is taking $K\ell \to 2^{+}$. For the homogeneous solutions, this limit corresponds to taking the boundary to be parametrically large and approaching the conformal boundary of AdS$_3$.

This section concerns \textit{inhomogeneous} solutions in this $K\ell \to 2^{+}$ limit. We will focus on the analysis of non-circular black hole patch solutions. In this case, the rate at which the boundary approaches conformal infinity depends on the scaling of the conformal energy. We distinguish two regimes below. 
The non-static pole patch analysis follows by performing a modular transformation, $\b\to\tfrac{4\pi^2}{\b}$.

We first recall that the non-circular black hole patch solutions are labelled by their conformal energy $E_\text{conf}$. Using \eqref{eqn: space of e non-cir bh} and \eqref{eqn: thermo non-cir bh} and taking $K\ell\to2^+$, $E_\text{conf}$ is subject to the bound
\begin{equation}
    0<E_\text{conf}<E_\text{BTZ} - E_\text{BTZ}\sqrt{K\ell-2} + \mathcal{O}(K\ell-2)\,,
\end{equation}
where $E_{\text{BTZ}}\equiv \tfrac{\ell \pi^2}{2G_N\b^2}$ is the (dimensionless) energy formula for the non-rotating BTZ black hole. The upper bound of $E_\text{conf}$ is the conformal energy of the homogeneous solution. 

There are two classes of solutions, with different scalings of the Weyl factor as $K\ell\to2^+$. The first class consists of solutions for which $E_\text{conf}$ remains finitely separated from $E_\text{BTZ}$. The second consists of solutions for which $E_\text{conf}$ lies within an order $\sqrt{K\ell-2}$ window near $E_\text{BTZ}$.  Below, we present the Weyl factor as well as the on-shell action in each case.

\textbf{Solutions with $E_\text{conf}$ finitely away from $E_\text{BTZ}$, or the single-scaling limit.} We first introduce a parameter
\begin{equation}
    \chi \equiv \frac{E_\text{conf}}{E_\text{BTZ}} \, ,
\end{equation}
which is required to obey $0<\chi<1$. Keeping $\chi$ fixed and taking $K\ell \rightarrow 2^+$, the Weyl factor \eqref{eqn: Weyl Tij non-cir bh} is approximated by the leading term
\begin{equation}\label{eqn: Weyl limit AdS sol 2}
    \bomega(\varphi)= \log \left(\frac{2\pi\ell \sqrt{1-\chi}}{\b \sqrt{K\ell-2}}\text{sech}\left(\frac{2\pi\sqrt{1-\chi}}{\b}(\varphi-\varphi_0)\right)\right)\,,
\end{equation}
where the expansion is only reliable on $\varphi\in(\varphi_0-\pi,\varphi_0+\pi)$, and the leading conformal temperature is given by
\begin{equation}
    \b^{-1}= \frac{n\log \frac{8(1-\chi)}{\chi\sqrt{K\ell-2}}}{2\pi^2\sqrt{1-\chi}} \, .
\end{equation}
Note that the Weyl factor does not diverge everywhere in this limit.

As for the on-shell action, for which the exact result is given in \eqref{Lambda<0 non-circular BH patch on-shell action}, the leading terms in the expansion are given by
\begin{equation}\label{eqn: Ionshell AdS limit 2}
    I_\text{on-shell}= -\frac{\left(2-\chi\right)\pi^2\ell}{ 2G_N \b} + \frac{n\ell}{G_N}\tanh^{-1}\sqrt{1-\chi}\,.
\end{equation}

\textbf{Solutions with $E_\text{conf}$ near $E_\text{BTZ}$, or the double-scaling limit.} We now consider solutions for which $E_\text{conf}-E_\text{BTZ}$ is of order $\sqrt{K\ell-2}$. To this end, we introduce an order one parameter
\begin{equation}
    \alpha \equiv 
    \frac{E_\text{BTZ}-E_\text{conf}}{E_\text{BTZ}\sqrt{K\ell-2}} \, .
\end{equation}
The bound on $E_\text{conf}$ imposes that $\alpha>1$, with $\alpha=1$ corresponding to the homogeneous solution. 

Taking the limit $K\ell\to2^+$ while keeping $\alpha$ fixed, the leading Weyl factor \eqref{eqn: Weyl Tij non-cir bh} is given by
\begin{equation}\label{eqn: Weyl sol AdS limit 1}
    \bomega(\varphi)= \log \frac{\sqrt{2}\pi \ell \text{dn}\left(\frac{\sqrt{2}\pi(K\ell-2)^{1/4}}{\b\zeta^{1/4}}(\varphi-\varphi_0)\Bigg|1-\zeta\right)}{\b\zeta^{1/4}(K\ell-2)^{1/4}}  \,, \qquad \zeta \equiv \frac{\alpha-\sqrt{\alpha^2-1}}{\alpha+\sqrt{\alpha^2-1}}\,.
\end{equation}
The variable $\zeta$ is introduced for convenient comparison with the boundary theory analysis below.

The expansion of the conformal temperature \eqref{eqn: conf data non-cir bh}  to the first sub-leading term is given by
\begin{equation}\label{eqn: leading beta AdS limit 1}
    \b^{-1} =\frac{n\zeta^{1/4}\mathcal{K}(1-\zeta)}{\sqrt{2}\pi^2 (K\ell-2)^{1/4}}+\frac{n\left((1+\zeta^2)E(1-\zeta)-\zeta(1+\zeta)\mathcal{K}(1-\zeta)\right)(K\ell-2)^{1/4}}{2\sqrt{2}\pi^2 (\zeta-1)^2\zeta^{1/4}} + \ldots \,.
\end{equation}
And so the Weyl factor diverges as $-\log(K\ell-2)$ everywhere on the boundary. 

The expansion of the on-shell action \eqref{Lambda<0 non-circular BH patch on-shell action} to the first sub-leading term is given by
\begin{equation}\label{eqn: Ionshell expansion ads limit1}
    I_\text{on-shell}= - \frac{\pi^2\ell}{2G_N \b} +(K\ell-2)^{1/4}\frac{n\ell}{G_N}\frac{4 E(1-\zeta)-(1+\zeta)\mathcal{K}(1-\zeta)}{4\sqrt{2} \zeta^{1/4}} + \ldots \, .
\end{equation}
In writing this expression, we have absorbed the leading term, together with part of the subleading corrections, into the linear-in-temperature term $-\tfrac{\pi^2 \ell}{2G_N \b}$. The remaining subleading terms, found in the second term of \eqref{eqn: Ionshell expansion ads limit1}, do not admit a simple expression in terms of $\b$.

\subsection{Holographic comparison}
\label{sec: holographic comparison}

In this subsection, we compare the contribution of the non-circular black hole patch solutions to the partition function with their counterparts in the boundary theory proposed in \cite{Allameh:2025gsa}, which is a holographic CFT$_2$ coupled to timelike Liouville theory and deformed by an exactly marginal $\tilde T \bar{\tilde{T}}$ operator, defined further below. In principle, the effective action of the boundary theory at finite deformation parameter is expected to be complicated due to higher derivative terms induced by the deformation. As such, we only focus on linear order in the deformation parameter. 

\subsubsection{Deformed timelike Liouville theory at linear order}

We first review the proposed holographic boundary theory, keeping only linear order in the deformation parameter, as per \cite{Allameh:2025gsa}.

Let $\tilde g_{ij}$ be a Euclidean fiducial metric, with covariant derivative $\tilde\nabla_i$ and scalar curvature $\tilde R$. The action, truncated at first order in the deformation parameter $\lambda$, is
\begin{equation}\label{eqn: total Liou action}
    S=S_m+S_{tL}+S_{def}\,.
\end{equation}
We refer to \cite{Allameh:2025gsa} for the definition of the boundary theory at finite $\lambda$. Here $S_m$ is the matter action, and the timelike Liouville action is
\begin{equation}
    S_{tL}=\frac{1}{4\pi\mathfrak b^2}\int d^2\sigma\sqrt{\tilde g}
    \left(-(\tilde\nabla\Phi)^2-\tilde R\Phi
    +4\pi\mathfrak b^2\mu e^{2\Phi}\right)\,,
\end{equation}
where $\Phi$ is the rescaled Liouville field, $\mathfrak b$ is the Liouville coupling, and $\mu$ is the Liouville cosmological constant. Moreover, the first-order deformation is
\begin{equation}\label{eqn: S def}
    S_{def}=\lambda\int d^2\sigma\sqrt{\tilde g}\,
    \tilde T\bar{\tilde T}\,e^{-2\Phi}\,,\qquad
    \tilde T\bar{\tilde T}\equiv\frac18
    \left(\tilde T^{ij}\tilde T_{ij}-(\tilde T^i{}_i)^2\right)\,,
\end{equation}
where the indices are raised and lowered using $\tilde g_{ij}$. The stress tensor entering this operator is that of the undeformed theory with the Liouville cosmological term removed, 
\begin{equation}\label{eqn: combine Tij}
    \tilde T_{ij}=\tilde T^m_{ij}+\tilde T^{WZ}_{ij}\,.
\end{equation}
The matter contribution is $\tilde T^m_{ij}$, while the Wess-Zumino contribution is
\begin{equation}\label{eqn: TWZ}
    \tilde T^{WZ}_{ij}=\frac{1}{2\pi\mathfrak b^2}
    \left(\tilde\nabla_i\Phi\tilde\nabla_j\Phi
    -\tilde\nabla_i\tilde\nabla_j\Phi
    +\tilde g_{ij}\tilde\nabla^2\Phi
    -\frac12\tilde g_{ij}(\tilde\nabla\Phi)^2\right)\,.
\end{equation}

The proposed identification between the bulk and boundary paramaters is
\begin{equation}\label{eqn: duality parameters}
    \mu=\frac{K\ell-2}{16\pi G_N\ell}\,,\qquad
    \lambda=16\pi G_N\ell\,,\qquad
    c_L=-c_m=-\frac{3\ell}{2G_N}\,,
\end{equation}
where $c_L$ is the Liouville central charge and $c_m$ is the central charge of the matter theory. We work to leading order at large positive $c_m$, for which $\mathfrak b^2\simeq6/c_m$. Importantly, the dimensionless combination controlling the deformation is $\lambda\mu=K\ell-2$.

In connection with the the bulk analysis, we choose the flat torus for the metric, 
\begin{equation}\label{eqn: fidu metric}
    d\tilde s^2=du^2+\r^2d\varphi^2\,,\qquad
    u\sim u+\r\b\,,\qquad\varphi\sim\varphi+2\pi\,.
\end{equation}
We refer to $\b$ as the inverse conformal temperature. At high temperatures, the matter theory evaluated on \eqref{eqn: fidu metric} yields
\begin{equation}\label{eqn: matter info}
    -\log\mathcal Z_m=-\frac{2\pi^2}{\mathfrak b^2\b}\,,\qquad
    \tilde T^m_{ij}\,dx^i dx^j
    =\frac{\pi}{\mathfrak b^2\r^2\b^2}(du^2-\r^2d\varphi^2)\,.
\end{equation}
The classical Liouville $\Phi^*$ is defined as a solution to
\begin{equation}\label{eqn: lin def eom}
    \tilde\nabla^2\Phi^*+4\pi\mathfrak b^2\mu e^{2\Phi^*}
    -4\pi\mathfrak b^2\lambda\,
    \tilde T\bar{\tilde T}(\Phi^*)e^{-2\Phi^*}=0\,,
\end{equation}
where $T\bar{\tilde T}(\Phi^*)$ is given by \eqref{eqn: S def}, with the matter stress tensor \eqref{eqn: matter info} and the Wess-Zumino stress tensor \eqref{eqn: TWZ} evaluated on the saddle $\Phi^*$.

Below, we solve \eqref{eqn: lin def eom} at finite $\lambda$ and take $\lambda \to 0$ only at the end. It should be noted that the following finite $\lambda$ results do not include higher-order deformation effects.

\subsubsection{Inhomogeneous solutions}

We seek $u$-independent saddles $\Phi^*=\Phi^*(\varphi)$ satisfying the periodic condition,
\begin{equation}\label{eqn: bdy perio cond}
    \Phi^*(\varphi+2\pi)=\Phi^*(\varphi)\, .
\end{equation}
Here, we report on the relevant features of the solution and refer to appendix \ref{sec: bdy sol details} for more details.

For this ansatz, \eqref{eqn: lin def eom} can be integrated to a first-order equation,
\begin{equation}\label{eqn: lin def eom 2}
    (\p_\varphi\Phi^{*})^2+4\pi\mathfrak b^2\mu \r^2e^{2\Phi^*}
    +\frac{\lambda}{16\pi\mathfrak b^2\r^2}
    \left(\frac{4\pi^2}{\b^2}-(\p_\varphi\Phi^{*})^2\right)^2e^{-2\Phi^*}
    =\frac{4\pi^2(1-\chi)}{\b^2}\,,
\end{equation}
where $\chi$ is an integration constant written this way to make connection with the bulk results. The solution oscillates between the maximum and minimum $\Phi_\pm^*$ given by
\begin{equation}\label{eqn: lin def max/min}
    e^{2\Phi_\pm^*}=\frac{1}{8\pi\mathfrak b^2\mu\r^2}
    \left(\frac{2\pi}{\b}\right)^2
    \left(1-\chi\pm\sqrt{(1-\chi)^2-\lambda\mu}\right)\,.
\end{equation}
The reality condition for $\Phi^*$ requires
\begin{equation}\label{eqn: lin def bound}
    \sqrt{\lambda\mu} \leq 1-\chi\, ,
\end{equation}
where the inequality is saturated for the homogeneous solution. The periodic condition \eqref{eqn: bdy perio cond} leads to
\begin{equation}\label{eqn: lin def beta}
    \b^{-1}=\frac{n}{2\pi^2}\sqrt{\frac{1-m'}{m'(m-m')}}
    \left(mK(m)-2(m-m')\Pi(m'|m)\right)\,,
\end{equation}
where $n \in \mathbb{Z}$ is a winding number, and the parameters $m$ and $m'$ are defined as
\begin{equation}\label{eqn: bdy m m' def}
    m'\equiv\frac{2\sqrt{(1-\chi)^2-\lambda\mu}}
    {-\chi+\sqrt{(1-\chi)^2-\lambda\mu}+\sqrt{1-\lambda\mu}}\,,\qquad m\equiv\frac{2\sqrt{(1-\lambda\mu)((1-\chi)^2-\lambda\mu)}}
    {1-\chi-\lambda\mu+\sqrt{(1-\lambda\mu)((1-\chi)^2-\lambda\mu)}}\,.
\end{equation}
As in the preceding sections, $\mathcal{K}(m)$, $E(m)$, and $\Pi(m|m')$ denote complete elliptic integrals of the first, second, and third kind respectively. Note that equation \eqref{eqn: lin def beta} gives an implicit dependence of $\chi$ on $\b$ and $\lambda\mu$. 

\subsubsection{Torus partition function}

In the saddle-point approximation, the partition function of the linear theory is given by
\begin{equation}
    -\log\mathcal Z=-\log \mathcal{Z}_m
    +S_{tL}[\Phi^*]+S_{def}[\Phi^*]\,,
\end{equation}
where the matter partition function is given in \eqref{eqn: matter info}, and $\Phi^*$ solves the equation of motion \eqref{eqn: lin def eom}. Evaluating the Liouville and deformation actions on the inhomogeneous saddle gives
\begin{equation}\label{eqn: lin def on-shell S}
    S_{tL}[\Phi^*]+S_{def}[\Phi^*]
    =\frac{n}{\mathfrak b^2}\sqrt{\frac{m-m'}{m'(1-m')}}\left(\frac{2m'}{m}\sqrt{1-\lambda\mu}\,E(m)
    +(m'-2)K(m)-2(m'-1)\Pi(m'|m)\right)\,,
\end{equation}
where $m$ and $m'$ are given by \eqref{eqn: bdy m m' def}.

It follows from \eqref{eqn: lin def beta} that the partition function is an implicit function of $\b$ and the combination $\lambda \mu$.

We observe that the partition function of the truncated theory obeys
\begin{equation}
    -\left.\frac{\partial\log\mathcal Z}{\partial\b}\right|_{\lambda\mu}
    =\frac{2\pi^2 \chi}{\mathfrak b^2\b^2}\,,\qquad
    -\left.\frac{\partial\log\mathcal Z}{\partial\lambda}\right|_{\b}
    =\frac{S_{def}[\Phi^*]}{\lambda}\,,
\end{equation}
where $S_{def}[\Phi^*]$ is given by \eqref{eqn: S def}. 
Treating $\mathcal{Z}$ as a thermal partition function, the first identity computes the thermal energy. The second identity shows that $\mathcal{Z}$ obeys a linear order flow equation; check \cite{Allameh:2025gsa} for the proper definition. We now consider the $\lambda\to0$ limit. We do this in two qualitatively distinct ways.

\textbf{Double-scaling limit.}
We first hold $\tfrac{1-\chi}{\sqrt{\lambda\mu}}>1$ fixed while taking $\lambda\to0$. Equivalently, the ratio of the squared turning-point amplitudes,
\begin{equation}
    \zeta\equiv\frac{1-\chi-\sqrt{(1-\chi)^2-\lambda\mu}}
    {1-\chi+\sqrt{(1-\chi)^2-\lambda\mu}}
    =\frac{e^{2\Phi_-^*}}{e^{2\Phi_+^*}}\,,\qquad 0<\zeta<1\,,
\end{equation}
is held fixed. In this limit, \eqref{eqn: lin def beta} becomes 
\begin{equation}
    \b^{-1}=\frac{n\zeta^{1/4}K(1-\zeta)}
    {\sqrt2\pi^2(\lambda\mu)^{1/4}}
    +\mathcal O\!\left((\lambda\mu)^{1/4}\right)\,.
\end{equation}
Thus $\b$ scales as $(\lambda\mu)^{1/4}$, while the oscillation amplitude remains finite. Taking $\zeta\to1$, or $\chi\to1-\sqrt{\lambda\mu}$, reproduces the homogeneous solution.

The partition function is given by
\begin{equation}
    -\log\mathcal Z=-\frac{2\pi^2}{\mathfrak b^2\b}+\frac{(\lambda\mu)^{1/4}}{\sqrt2\mathfrak b^2\zeta^{1/4}}
    \left(4E(1-\zeta)-(1+\zeta)K(1-\zeta)\right)
    +\mathcal O\left((\lambda\mu)^{3/4}\right)\,.
\end{equation}
Using the dictionary \eqref{eqn: duality parameters}, this agrees with the bulk on-shell action \eqref{eqn: Ionshell expansion ads limit1} to the first sub-leading order. In this limit, $\Phi^*(\varphi)$ also approaches the bulk Weyl factor uniformly.

\textbf{Single-scaling limit.}
Instead, we now hold $\chi$ fixed as $\lambda\to0$. The condition \eqref{eqn: lin def beta} now gives
\begin{equation}
    \b^{-1}=\frac{1}{2\pi^2\sqrt{1-\chi}}
    \log\frac{8(1-\chi)}{\chi\sqrt{\lambda\mu}}
    -\frac{1}{\pi^2}\operatorname{arctanh}\sqrt{1-\chi}
    +\mathcal O\left(\lambda\mu\log(\lambda\mu)\right)\,.
\end{equation}
In particular, the conformal temperature now scales logarithmically with $\lambda \mu$. The ratio of the turning-point amplitudes is
\begin{equation}
    \frac{e^{\Phi_+^*}}{e^{\Phi_-^*}}
    =\frac{1-\chi+\sqrt{(1-\chi)^2-\lambda\mu}}{\sqrt{\lambda\mu}}
    \approx\frac{2(1-\chi)}{\sqrt{\lambda\mu}}\,.
\end{equation}
The oscillation therefore becomes parametrically large.

The corresponding partition function is
\begin{equation}
    -\log\mathcal Z=-\frac{2\pi^2(2-\chi)}{\mathfrak b^2\b}
    +\frac{2\sqrt{1-\chi}+2\chi \operatorname{arctanh}\sqrt{1-\chi}}{\mathfrak b^2}
    +\mathcal O\!\left(\lambda\mu\log(\lambda\mu)\right)\,.
\end{equation}
Using the identification of parameters in \eqref{eqn: duality parameters} and comparing with \eqref{eqn: Ionshell AdS limit 2}, the bulk and boundary partition functions differ at order $(\lambda\mu)^0$ in this expansion, i.e., only the first term agrees. 

As for the saddle itself, found in \eqref{eqn: lin def sol}, the bulk and boundary profiles fail to match near the first local minimum. Evaluated at the same $\b$ and $\lambda \mu$ (or $K\ell-2$), the extrema satisfy, to leading order,
\begin{equation}
    \frac{e^{\Phi_+^{(\mathrm{bulk})}}}{e^{\Phi_+^{(\mathrm{bdry})}}}\approx 1\,,\qquad
    \frac{e^{\Phi_-^{(\mathrm{bulk})}}}{e^{\Phi_-^{(\mathrm{bdry})}}}\approx \chi\,.
\end{equation}
Here the superscripts distinguish the bulk Weyl factor from the boundary Liouville field. At fixed $\chi$, the relative discrepancy at the minimum persists as $\lambda\to0$, even though the maxima agree at leading order. Hence, $\Phi^*(\varphi)$ does not converge to the bulk Weyl factor in this limit.

Importantly, we note that the combination $\lambda \mu e^{-2\Phi^*}$ evaluated near the minimum is of order $(\lambda \mu)^0$, in contrast with the double-scaling limit, where it scales as $\sqrt{\lambda \mu}$.

\subsubsection{Using the bulk saddle}

In the previous subsections, we have shown that the inhomogeneous saddles of the proposed small-$\lambda$ effective action do not generally match those of the bulk. What follows is a mismatch of their corresponding on-shell actions. We link this to the fact that the effective coupling $\lambda e^{-2\Phi}$ is not small for the single-scaled solutions, and therefore a correct effective action must include higher-order terms in a series expansion of $\lambda e^{-2\Phi}$.

Now, we attempt to mildly remedy the situation by using the bulk saddles themselves in the boundary effective action. We will see that this does not resolve the mismatch.

First, as we have access to the finite-$\lambda$ saddles in the bulk, we check that the flow equation \cite{Allameh:2025gsa} is satisfied for all inhomogeneous solutions. Indeed, we observe that the saddles in the bulk have an on-shell action that satisfies the equation
\begin{equation}
    \label{partial K}
    \partial_KI_{\text{on-shell}}=\frac{1}{16\pi G_N} \int d^2x \sqrt{\tilde g}\, e^{2\omega} \, ,
\end{equation}
where $\omega$ is the Weyl factor and $\tilde g$ is the determinant of the flat induced metric.\footnote{In fact, equation \eqref{partial K} is true for all signs of cosmological constant $\Lambda$.} And this is precisely the boundary flow equation, as we now demonstrate.

If we focus on non-circular boundaries enclosing a BTZ background, the Weyl factor satisfies the following equation of motion, which is a rewriting of equations \eqref{Lambda<0 non-static pole eq1}, \eqref{Lambda<0 non-static pole eq2}, and \eqref{eqn: Weyl Tij non-cir bh},
\begin{equation}
    \partial_{\theta}^2\omega+\frac{K\ell-2}{\ell^2}e^{2\omega}-256\pi^2G_N^2 \underbrace{\left(T\bar T-\frac{1}{4}\left(\frac{K\ell-2}{16\pi G_N \ell} \right)^2e^{4\omega} \right)}_{\tilde T \tilde{\bar T}}e^{-2\omega}=0 \, ,
\end{equation}
where $T\bar T=\frac{\pi^2\mathcal{E}^2}{16G_N^2 \tilde \beta ^4}$ is the result from the conformal stress tensor \eqref{eqn: Weyl Tij non-cir bh}. We now use this equation of motion to rewrite \eqref{partial K} as
\begin{equation}
    \partial_KI_{\text{on-shell}}=\frac{16\pi G_N \ell^2}{K\ell-2} \int d^2\sigma \sqrt{\tilde g} \, \tilde T \tilde{\bar T}e^{-2\omega} \, .
\end{equation}
Then, using the holographic dictionary in \eqref{eqn: duality parameters}, this translates to the flow equation in \cite{Allameh:2025gsa},
\begin{equation}
    \label{boundary flow equation}
    \partial_\lambda I_{\text{on-shell}}=\int d^2\sigma \sqrt{\tilde g} \, \tilde T \tilde{\bar T}e^{-2\omega} \, .
\end{equation}
Now, we simply check whether the proposed small-$\lambda$ effective action \eqref{eqn: total Liou action}, when evaluated on the bulk saddle \eqref{eqn: Weyl Tij non-cir bh} and its different limits in section \ref{sec: AdS limit}, satisfies this flow equation.

\textbf{Homogeneous case.} In this case, the $\lambda$-derivatives of both the timelike Liouville action and the $\lambda$-linear $\tilde T \tilde{\bar T}$ action contribute at order $(\lambda \mu)^{-1/2}$. In particular, 
\begin{equation}
    \partial_{\lambda}S_{tL}\big|_{\tilde \beta \text{ fixed}}=\frac{\pi ^2 c_m \sqrt{\mu }}{12 \tilde \beta  \sqrt{\lambda }} + \ldots \, , \qquad \partial_{\lambda}S_{def}\big|_{\tilde \beta \text{ fixed}}=\frac{\pi ^2 c_m \sqrt{\mu }}{12 \tilde \beta  \sqrt{\lambda }} + \ldots \, ,
\end{equation}
and their sum precisely matches the leading-order term in the right-hand side of the flow equation \eqref{boundary flow equation}.

\textbf{Double-scaling limit.} Once again, we consider the limit where $\lambda \rightarrow 0$ while keeping $\zeta$ fixed. Here, we find that the $\lambda$-derivatives of both the timelike Liouville action and the $\lambda$-linear $\tilde T \tilde{\bar T}$ action contribute at order $(\lambda \mu)^{-3/4}$. In particular,
\begin{equation}
    \partial_{\lambda}S_{tL}\big|_{\tilde \beta \text{ fixed}}=\frac{c_m \mu^{1/4}  \left(\zeta  K(1-\zeta )^2-2 (\zeta +1) K(1-\zeta ) E(1-\zeta )+3 E(1-\zeta )^2\right)}{12 \sqrt{2} \lambda ^{3/4} \zeta^{1/4} (2 E(1-\zeta )-(\zeta +1) K(1-\zeta ))} + \ldots \, ,
\end{equation}
\begin{equation}
    \partial_{\lambda}S_{def}\big|_{\tilde \beta \text{ fixed}}=\frac{c_m \mu^{1/4} \left(E(1-\zeta )^2-\zeta  K(1-\zeta )^2\right)}{12 \sqrt{2} \lambda ^{3/4}\zeta^{1/4} (2 E(1-\zeta )-(\zeta +1) K(1-\zeta ))} + \ldots \, ,
\end{equation}
and, once again, their sum precisely matches the leading-order term on the right-hand side of the flow equation.

\textbf{Single-scaling limit.} We now consider the limit where $\lambda \rightarrow 0$ while keeping $\chi$ fixed. Here, we find that the $\lambda$-derivative of the effective action does not match the right-hand side of the flow equation. In particular,
\begin{equation}
    \label{LHS flow single-scale}
    \partial_{\lambda}S_{tL}\big|_{\tilde \beta \text{ fixed}}=\frac{c_m \sqrt{1-\chi}}{6\lambda}\left( 1+\frac{\frac{2}{\chi}-\frac{3}{2}}{\log \frac{8(1-\chi)}{\chi \sqrt{\lambda \mu}}}+\ldots \right)  \, , \qquad \partial_{\lambda}S_{def}\big|_{\tilde \beta \text{ fixed}}=\frac{c_m \sqrt{1-\chi}}{6\lambda \log (\lambda \mu)}+ \ldots \, .
\end{equation}
On the other hand, the small-$\lambda$ expansion of the right-hand side of the flow equation yields
\begin{equation}
    \label{flow single-scale}
    \int d^2 \sigma \sqrt{\gamma} \tilde T \tilde{\bar T}e^{-2\omega}=\frac{c_m \sqrt{1-\chi }}{6 \lambda }-\frac{c_m \mu  \chi ^2 \log (\lambda  \mu )}{96 (1-\chi )^{3/2}}+\ldots \, .
\end{equation}
While the leading terms match, there is a $-\frac{2c_m(1-\chi)^{3/2}}{3\lambda \chi \log(\lambda \mu)}$ term arising in the sum in \eqref{LHS flow single-scale} which does not appear in \eqref{flow single-scale}. And so the flow equation is not satisfied by the proposed effective action. It may be that an alternative effective action sourced by higher powers of $\lambda e^{-2\Phi}$ correct this mismatch; we leave this for future work.

\section{Outlook \& discussion}\label{sec: outlook}

In this paper, we consider the partition function of the theory of three-dimensional Einstein gravity with a finite boundary subject to conformal boundary conditions. We follow the Gibbons-Hawking-York prescription \cite{Gibbons:1976ue,York:1986it} of computing the gravitational path integral over smooth geometries with finite boundaries obeying the boundary conditions. Here, we choose a simple set of boundary data for this partition function, consisting of a constant extrinsic curvature $K$ and a torus-topology class of conformally flat induced metrics with a conformally invariant periodicity $\b$. For vanishing, positive, and negative cosmological constant $\Lambda$, we report novel saddle-point contributions to the partition function $\mathcal{Z}(\b,K)$ in the semi-classical limit $G_N \rightarrow 0$ and analyse their thermodynamic properties.

As there are no propagating gravitational degrees of freedom in three dimensions, the problem reduces to identifying all possible boundaries of a constant-curvature three-manifold with the appropriate boundary data. Previously studied solutions are static and circular. Here, the boundaries we choose to consider have a Weyl factor which varies non-trivially along either the spatial circle of the boundary or the thermal circle, but not both. In all these cases, the conformal stress tensor is constant. 

Interestingly, we find that the non-static solutions, including ones with a non-contractible bulk thermal circle, have a non-zero entropy $\mathcal{S}_{\text{conf}}$ that does not have an obvious geometric interpretation. Moreover, the non-trivial saddles generally have a negative specific heat $C_K$ and are therefore thermally unstable, with the exception of cases with
\begin{equation}
    \Lambda<0 \, \, \text{and} \, \, 2<K|\Lambda|^{-1/2}<\frac{3}{\sqrt{2}} \, , \qquad \text{or} \, \qquad \Lambda>0 \, \, \text{and} \, \, K|\Lambda|^{-1/2}<0 \, .
\end{equation} 
More pressingly, the inhomogeneous boundaries in the first case are more thermodynamically favourable than the homogeneous solutions with the same boundary data. 

\textbf{Holography.} We have also performed a detailed study of the $\Lambda<0$ case in the regime of small $K|\Lambda|^{-1/2}-2$, together with its proposed holographic dual \cite{Allameh:2025gsa}. On the one hand, we study the contributions of inhomogeneous saddles to the gravitational torus partition function in a small $K|\Lambda|^{-1/2}-2$ expansion. On the other, we study timelike Liouville theory coupled to a matter CFT$_2$, deformed by an exactly marginal $\tilde T \tilde{\bar T}e^{-2\Phi}$ operator, to linear order in the deformation parameter $\lambda$. We find analogous inhomogeneous Liouville saddles $\Phi$ and compute the corresponding thermal partition function. We find agreement between the gravity and boundary descriptions in the regime where $\lambda e^{-2\Phi}$ is parametrically small. In the regime where $\lambda e^{-2\Phi}$ is order one while $\lambda$ remains small, we find a disagreement. Since the analysis is only at linear order in the deformation, it is worth investigating if the discrepancy can be improved by including higher order terms.

\textbf{Self-intersecting boundaries.} Curiously, some of the boundaries we find violate the global embedding conditions and self-intersect. It requires more care to evaluate the on-shell actions of these solutions, and so we leave this for future work. In any case, it is unclear whether one should include the contributions of these exotic saddles in the partition function at all.

\textbf{Three-dimensional idiosyncrasies.} There are several features that we suspect follow from the three-dimensional nature of the spacetime. One of these is the increasing number of non-circular saddles in the high temperature regime. From the gravity analysis, it it clear that these solutions are generated from a single non-circular solution by allowing it to wind different numbers of times around the spatial circle $S^1$. In a higher-dimensional setup, where the spatial slice of the boundary manifold is instead a sphere $S^{d-1}$, such a winding mechanism is not expected to exist. It would therefore be interesting to understand whether this mechanism is replaced by another one, or whether the high temperature phase structure is simpler in higher dimensions. In the low temperature regime, we however expect the increasing number of non-static solutions to persist in higher dimensions, since these solutions can still wind around the thermal circle $S^1$.

\textbf{Missing saddles and classical strings.} Moreover, the solutions we find are not exhaustive: we do not consider boundaries with a Weyl factor that varies along both cycles of the torus. As a concrete step in this direction, we follow the embedding procedure and recast the conformal boundary conditions, \eqref{eqn: embed conf class cond} and \eqref{eqn: embed K}, in a form more accustomed to a string theory context. 

Schematically, we first take the derivative of the conformal class condition \eqref{eqn: embed conf class cond}. Combining the resulting equations with the trace $K$ condition \eqref{eqn: embed K}, we obtain equations determining the boundary location $\left.x^\mu\right|_\Gamma= X^\mu$,
\begin{equation}\label{eqn: string eom}
    \p^i \p_i X^\mu + \Gamma^\mu_{\nu\rho}\delta^{ij}\p_iX^\nu \p_j X^\rho = \frac{1}{2}H^\mu{}_{\nu\rho}\,\tilde \epsilon^{ij}\p_iX^\nu \p_j X^\rho \, , \qquad H_{\mu\nu\rho}\equiv K \,\epsilon_{\mu\nu\rho} \, .
\end{equation}
These embedding functions $X^\mu$ are subject to the conformal class conditions \eqref{eqn: embed conf class cond}, which take the form of constraint equations,
\begin{equation} \label{eqn: string const}
    g_{\mu\nu}\p_i X^\mu \p_j X^\nu - \frac{1}{2}\delta_{ij} \delta^{kl} g_{\mu\nu} \p_k X^\mu \p_l X^\nu =0\, ,
\end{equation}
Here, $\Gamma^\mu_{\nu\rho}$ is the Christoffel symbol of the bulk metric, $\epsilon_{\mu\nu\rho}$ is the bulk volume form, and $\tilde \epsilon_{ij}$ is the permutation matrix.\footnote{In these formulae, we assume that the conformal representative of the boundary metric is flat, i.e. $\tilde{g}_{ij}=\delta_{ij}$. A covariant expression for arbitrary $\tilde g_{ij}$ can be obtained straightforwardly by the standard minimal-coupling procedure.} These equations are those of a classical non-critical string in conformal gauge, coupled to a curved target-space metric $g_{\mu\nu}$ and a $B$-field with field strength $H_{\mu\nu\rho}$.\footnote{Here $g_{\mu\nu}$ and $H_{\mu\nu\rho}$ should be treated as external fields, since they are not required to obey the target-space field equations following from the vanishing of the worldsheet beta functions.}

This reformulation suggests that techniques developed in the string theory literature may be useful for finding more general solutions to the problem studied in this paper. For instance, in AdS$_3$ with $K|\Lambda|^{-1/2}=2$, equations \eqref{eqn: string eom} and \eqref{eqn: string const} in Lorentzian signature reduce to the classical equations of motion of the $SL(2,\mathbb{R})$ Wess--Zumino--Witten model, studied for example in \cite{Maldacena:2000hw}. It would be interesting to understand whether integrability methods can shed light on the problem away from $K|\Lambda|^{-1/2}=2$, perhaps along the lines of \cite{Hoare:2021dix}.

\textbf{Sum over topologies.} Despite the fact that gravity in three dimensions has no local degrees of freedom, the global structure is non-trivial. In particular, the bulk solutions with zero, positive, and negative curvature are not limited to $\mathbb{R}^3$, $S^3$, and $\mathbb{H}^3$, but also include quotients of these manifolds by discrete subgroups of their isometry group \cite{thurston1997three}. The Lorentzian interpretation of some of these manifolds is not clear, nor is it obvious that we should not include them in the path integral.

\textbf{Complex-system physics.} More speculatively, and along the lines of \cite{Anninos:2011vn,Anninos:2013mfa}, there is a regime in which gravity with finite conformal boundaries appears to resemble complex-system physics. In particular, our analysis shows that, for $2<K|\Lambda|^{-1/2}<\tfrac{3}{\sqrt{2}}$, the AdS$_3$ system at low temperature contains many thermodynamically stable configurations whose boundaries break the thermal $U(1)$ isometry. All of these saddles are thermodynamically favoured over the homogeneous saddle, suggesting a free energy landscape with many non-trivial minima. It would be interesting to understand this analogy further.

\textbf{Beyond.} It is natural to consider extensions of this problem: solving the same problem in higher dimensions, understanding the Lorentzian picture, computing one-loop corrections to the thermodynamic results, including higher-derivative terms in the action, or including matter fields, etc. Moreover, it may be interesting to understand these new solutions from the point of view of Chern-Simons formulation of three-dimensional Einstein gravity \cite{Achucarro:1986uwr}.

As an example, we present the higher-dimensional analogue of the non-static pole patch equations with $\Lambda=0$, namely \eqref{eqn: du T as e} and \eqref{eqn: flat bdry eqn 2}. We impose that the boundary metric is conformal to the product of a thermal circle and a round $S^{d-1}$ of radius $\r$, together with a constant trace of the extrinsic curvature $K$. We obtain
\begin{equation}
    (\r \p_u r)^2 + V_\text{eff}(r) =0 \, , \qquad 
    V_\text{eff}(r) 
    = 
    \frac{1}{d^2}K^2 r^4 
    - r^2 
    - \frac{\mathcal{E} K}{2}r^2
    \left(\frac{r}{\r}\right)^{2-d} 
    + \mathcal{E}^2
    \left(\frac{r}{\r}\right)^{4-2d} \, .
\end{equation}
Indeed, setting $d=2$ reproduces \eqref{eqn: du T as e} and \eqref{eqn: flat bdry eqn 2}. We leave these open questions for future work.

\section*{Acknowledgments}
We are grateful to Michael Anderson, Silvia Georgescu, Albert Law, Jake Stedman, and Andrew Svesko for useful discussions. We especially appreciate stimulating discussions with Edgar Shaghoulian and thank Dionysios Anninos and Dami\'an Galante for valuable feedback. The work of WAH is funded by the CMA CGM Excellence for Education Scholarship. CM is funded by STFC under grant number ST/X508470/1 and the Heising-Simons Foundation “Observational Signatures of Quantum
Gravity” QuRIOS collaboration grant. WAH would like to thank the organisers and participants of the \href{https://www.ggi.infn.it/showevent.pl?id=543#:~:text=Pathways%20to%20Quantum%20Black%20Holes%3A%20from%20Effective%20Theories%20to%20Exact%20Methods&text=Quantum%20Black%20Holes%20are%20characterized,Mechanics%20for%20a%20proper%20description.}{Pathways to Quantum Black Holes workshop} at the Galileo Galilei Institute for Theoretical Physics, where part of this work was completed. CM would like to thank all the participants of the workshop \href{https://scgp.stonybrook.edu/archives/45387}{Timelike Boundaries in Classical and Quantum Gravity} for stimulating discussions and the Simons Center for Geometry and Physics at Stony Brook University for its hospitality.

\appendix

\section{Elliptic integrals and functions}\label{sec: Ellip Jacobi}

In this appendix, we provide definitions and properties of some special functions used throughout the text.

The complete elliptic integral of the first kind $\mathcal{K}(m)$ is defined by
\begin{equation}
    \mathcal{K}(m)\equiv \int_0^1 \frac{dt}{\sqrt{(1-t^2)(1-mt^2)}} \, , \qquad m \in (0,1) \, ,
\end{equation}
while the complete elliptic integral of the second kind $E(m)$ is defined by
\begin{equation}
    E(m)\equiv\int_0^1 dt \frac{\sqrt{1-mt^2}}{\sqrt{1-t^2}} \, , \qquad m \in (0,1) \, .
\end{equation}
Their derivatives are given by
\begin{equation}\label{eqn: ellip int deriv}
    \p_m \mathcal{K}(m) = \frac{E(m)+(m-1)\mathcal{K}(m)}{2m(1-m)} \, , \qquad \p_m E(m) = \frac{E(m)-\mathcal{K}(m)}{2m} \, .
\end{equation}
The Jacobi elliptic function $\text{dn}(x|m)$ solves the ordinary nonlinear differential equation,
\begin{equation}
    \left(\frac{dy}{dx}\right)^2 = \left(y^2-1\right)\left(1-m-y^2\right) \, .
\end{equation}
Along the real axis, dn$(x|m)$ is periodic in $x$ with a period of $2\mathcal{K}(m)$,
\begin{equation}
    \text{dn}\left(x+2 n \mathcal{K}(m)|m\right) =\text{dn}(x|m) \, ,
\end{equation}
for any integer $n$ and $m \leq 1$. Within the interval $x\in(0,2\mathcal{K}(m))$, dn$(x|m)$ attains its maximum at $x=0$ and its minimum at the half-period $x=\mathcal{K}(m)$, at which
\begin{equation}
    \text{dn}(0|m)=1 \, , \qquad \text{dn}(\mathcal{K}(m)|m)=\sqrt{1-m} \, .
\end{equation}
The asymptotic expansion of dn$(x|m)$ near $m=0$ and $m=1$ are, respectively, given by
\begin{equation}\label{eqn: asymp dn}
\text{dn}(x|m) =
\begin{cases}
    1 - \frac{m}{2}\sin^2x+\mathcal{O}(m^2) \, , \\
    \frac{1}{\cosh{x}} +\frac{1-m}{4}\left(\frac{x}{\cosh{x}}+\sinh{x}\right)\tanh{x}+\mathcal{O}((1-m)^2)\, ,
\end{cases}
\end{equation}
The integral of the $\text{dn}$ function is called the Jacobi amplitude function
\begin{equation}
    \text{am}(x|m)=\int_{0}^{x} dt \, \text{dn}(t|m) \, \,
\end{equation}
which has a pseudo-periodic property,
\begin{equation}
    \text{am}(x+2\mathcal{K}(m)|m)=\text{am}(x|m)+\pi \, , \qquad m\in(0,1) \, .
\end{equation}
The asymptotic expansion of am$(x|m)$ near $m=0$ and $m=1$ are, respectively, given by
\begin{equation}\label{eqn: asymp am}
\text{am}(x|m) =
\begin{cases}
    x+\frac{m}{8}\left(\sin(2x)-2x \right)+\mathcal{O}(m^2) \, , \qquad  \\
    2\arctan(e^x)-\frac{\pi}{2}+\frac{1-m}{4} \left(\sinh x-\frac{x}{\cosh x} \right)+\mathcal{O}((1-m)^2) \, .
\end{cases}
\end{equation}
The incomplete elliptic integral of the second kind $E(x|m)$ is defined by
\begin{equation}
    E(x|m)=\int_{0}^{x} dt\sqrt{1-m \sin^2t} \, , \qquad x \in  \left(-\frac{\pi}{2},\frac{\pi}{2}\right) \, ,
\end{equation}
and has a quasi-periodic property,
\begin{equation}
    E(x+\pi|m)=E(x|m)+2E(m) \, .
\end{equation}
The asymptotic expansion of $E(x|m)$ near $m=0$ and $m=1$ are, respectively, given by
\begin{equation}\label{eqn: asymp ellipticE}
E(x|m) =
\begin{cases}
    x-\frac{m}{8}\left(2x-\sin(2x)\right)+\mathcal{O}(m^2) \, , \qquad  \\
    \sin x-\frac{1-m}{2}\left(\sin x-\tanh^{-1} (\sin x) \right)+\mathcal{O}((1-m)^2) \, , \qquad x\in\left(-\frac{\pi}{2},\frac{\pi}{2}\right) .
\end{cases}
\end{equation}
The incomplete elliptic integral of the third kind $\Pi(n;x|m)$ is defined by
\begin{equation}
    \Pi(n;x|m)=\int_{0}^{x}\frac{dt}{(1-n \sin^2t)\sqrt{1-m \sin^2t}} \, , \qquad x \in \left(-\frac{\pi}{2},\frac{\pi}{2} \right) \, ,
\end{equation}
and has a quasi-periodic property,
\begin{equation}
    \Pi(n;x+\pi | m)=\Pi(n;x | m)+2\Pi(n|m) \, , \qquad n \in (-1,1) \, ,
\end{equation}
where $\Pi(n|m)$ is the complete elliptic integral of the third kind, defined by $\Pi(n|m)=\Pi(n;\frac{\pi}{2}|m)$.

\section{Derivation of first-order equations} \label{sec: du T flat 1}

In this appendix, we provide a derivation of \eqref{eqn: du T as e} from \eqref{eqn: Lambda=0 eq1} and \eqref{eqn: Lambda=0 eq2}, which we repeat here for convenience,
\begin{equation}\label{eqn: Lambda=0 eq1 appen}
    (\r\p_u\tau)^2+(\r\p_ur)^2-r^2 =0\,,
\end{equation}
and
\begin{equation}\label{eqn: Lambda=0 eq2 appen}
    \frac{(\p_ur)^2 \p_u \tau + (\p_u\tau)^3+r\p_u r \p_u^2 \tau-r \p_u \tau \p_u^2 r }{r\left((\p_u\tau)^2+(\p_u r)^2\right)^{3/2}} = K \, .
\end{equation}
We start by using \eqref{eqn: Lambda=0 eq1 appen} to simplify the denominator of \eqref{eqn: Lambda=0 eq2 appen},
\begin{equation}\label{eqn: Lambda=0 eq2 appen dum1}
    \r^3\frac{(\p_ur)^2\p_u \tau + (\p_u \tau)^3+r\p_u r \p_u^2 \tau- r\p_u \tau \p_u^2 r}{r^4} = K \,.
\end{equation}
Differentiating \eqref{eqn: Lambda=0 eq1 appen} with respect to $u$, we obtain
\begin{equation}
    \r^2\p_u \tau\p_u^2 \tau + \r^2\p_u r \p_u^2 r= r \p_u r\, .
\end{equation}
Using this equation to eliminate $\p_u^2 r$ in \eqref{eqn: Lambda=0 eq2 appen dum1}, we find
\begin{equation}
    \r^3\frac{(\p_ur)^2\p_u \tau + (\p_u \tau)^3+r\p_u r \p_u^2 \tau- \frac{r^2 \p_u \tau}{\r^2} + \frac{r (\p_u \tau)^2\p_u^2 \tau}{\p_u r}}{r^4} = K \, .
\end{equation}
The first, second, and fourth terms on the left hand side cancel out by virtue of \eqref{eqn: Lambda=0 eq1 appen}. The remaining third and fifth terms can be combined to
\begin{equation}
    \r^3\frac{((\p_u r)^2+(\p_u \tau)^2)\p_u^2 \tau}{r^3 \p_u r}= K \,.
\end{equation}
Implementing \eqref{eqn: Lambda=0 eq1 appen} again, the resulting equations can be put into the following form
\begin{equation}
    \p_u \left(\frac{1}{2}Kr^2 - \r\p_u \tau\right)=0 \,,
\end{equation}
which reproduces \eqref{eqn: du T as e}.

\section{Proof of self-intersection}\label{sec: self-intersection}

In this appendix, we provide a proof of the violation of the global embedding condition \eqref{eqn: global emb} of solutions \eqref{eqn: flat R sol} and \eqref{eqn: flat T sol} for any finite $\mathcal{E}K>0$. This violation can be seen numerically from the self-intersection in the parametric plot of the boundary, see figure \ref{fig: Rsol Tsol flat}. 

Below, we will use \eqref{eqn: du T as e} together with the oscillatory behaviour of $r(u)$, specifically its maximum and minimum in \eqref{eqn: flat R_pm}, to show that the boundary always leads to self-intersections for any finite $\mathcal{E}K>0$. Hence, the solutions without self-intersections must obey
\begin{equation}\label{eqn: e true domain}
    -\frac{1}{2}<\mathcal{E}K<0 \, .
\end{equation}

The main part of the proof involves the following mathematical property. Suppose that $f(u)$ is a smooth function on a finite interval $u\in(0,\beta)$ satisfying the following three conditions. First, $f(u)$ is anti-symmetric under a mid point-reflection, i.e.,
\begin{equation}\label{eqn: flat self-int 1}
    f(u)=-f(\beta-u) \, .
\end{equation}
This implies that its derivative is symmetric under the reflection, $\p_uf(u)=\p_uf(\beta-u)$. Second, at the first endpoint, $f(u)$ satisfies
\begin{equation}\label{eqn: flat self-int 2}
    f(0)\p_uf(0)<0 \, .
\end{equation}
Combining this with the first property, it implies that $f(\beta)\p_uf(\beta)>0$. Third, $f(u)$ has only one extremum, at some value $u=u^{*}$, in the first half of the interval,
\begin{equation}\label{eqn: flat self-int 3}
    \p_uf(u^{*}) = 0 ,  \qquad \partial_{u}^2f(u^{*}) \neq 0 , \qquad u^{*}\in \left(0,\tfrac{\beta}{2}\right) \, .
\end{equation}
With these three conditions, it follows that $f(u)$ has exactly three roots: the midpoint $u=\tfrac{\beta}{2}$, somewhere in the first half-interval $(0,\frac{\beta}{2})$, and its reflection in the second half. 

Now, we use the above mathematical property to conclude that, for $\mathcal{E}K>0$, there exist two distinct points on the boundary associated with the same point in the bulk, i.e. the map $(u,\varphi) \to (\tau,r,\phi)=(\tau(u),r(u),\varphi)$ fails to be injective.

The non-injectivity of $\tau(u)$ can be seen as follows. Let $f(u)$ be the difference
\begin{equation}\label{eqn: f(u) flat}
    f(u) \equiv \tau(u+u_\text{extr}) - \tau(\beta-u+u_\text{extr}) \, , \qquad \beta\equiv \frac{4 \r}{ Kr_+}\mathcal{K}(m) \, ,
\end{equation}
where $u_\text{extr}$ is some value for which $r(u_\text{extr})$ is either maximum, $r_+$, or minimum, $r_-$. We now show that $f(u)$ defined this way obeys the three conditions specified above: \eqref{eqn: flat self-int 1}, \eqref{eqn: flat self-int 2}, and \eqref{eqn: flat self-int 3}. Accordingly, $f$ has three distinct roots. The first condition \eqref{eqn: flat self-int 1} follows straightforwardly from the definition of \eqref{eqn: f(u) flat}. As for the second condition, we use \eqref{eqn: du T as e} to find that
\begin{equation}
    f(0)\p_uf(0)= - \frac{2}{K\r} \left(\tau(\beta)-\tau(0)\right) \left(\frac{1}{2}K^2r(u_\text{extr})^2-\mathcal{E}K\right) \, .
\end{equation}
The expression in the first parentheses computes the shift of $\tau(u)$ after $u$ completes a single period and is independent of $u_\text{extr}$. Plugging $r_\pm$ from \eqref{eqn: flat R_pm} and taking $\mathcal{E}K>0$, we find that the expression in the second parentheses is positive for $r(u_\text{extr})=r_+$ and negative for $r(u_\text{extr})=r_-$. This means that the second condition \eqref{eqn: flat self-int 2} always holds for $\mathcal{E}K>0$ by choosing an appropriate $u_\text{extr}$ based on the sign of $\tau(\beta)-\tau(0)$. Finally, $r(u)$ induces an oscillatory behaviour of $\p_u \tau(u)$ via \eqref{eqn: du T as e}. In the parameter regime $\mathcal{E}K>0$, $\p_u \tau(u)$ oscillates between a positive and a negative value, implying that it vanishes at a certain value of $u$. Using \eqref{eqn: f(u) flat} together with the fact that $r(u+u_\text{extr})=r(-u+u_\text{extr})$ and hence $\p_u \tau(u+u_\text{extr})=-\p_u \tau(-u+u_\text{extr})$, we obtain
\begin{equation}
    \p_u f = 2 \p_uT(u+u_\text{extr}) \, .
\end{equation}
The existence of an extremum of $\tau(u)$ then induces that of $f(u)$, satisfying the third condition \eqref{eqn: flat self-int 3}. Hence, we conclude that \eqref{eqn: f(u) flat} has three distinct roots, two of which are at values $u \neq \beta /2$. At these roots,
\begin{equation}
    \tau(u+u_\text{extr})=\tau(\beta-u+u_\text{extr}) \, ,
\end{equation}
violating the injectivity of $\tau(u)$. As for $r(u)$, it is straightforward to see that
\begin{equation}
    r(u+u_\text{extr})=r(\beta-u+u_\text{extr}) \, ,
\end{equation}
by virtue of the periodic property. Therefore, we find self-intersection for solutions with $\mathcal{E}K>0$. 

As a final remark, we note that in the parameter regime $-\tfrac{1}{2}<\mathcal{E}K<0$, no self-intersection occurs. This follows from the fact that the right hand side of \eqref{eqn: du T as e} is, upon using \eqref{eqn: flat R_pm}, bounded by two positive numbers for any $-\tfrac{1}{2}<\mathcal{E}K<0$. Hence $\tau(u)$ is monotonic, ensuring that it is injective.

\section{Details of boundary theory computations}\label{sec: bdy sol details}

In this appendix, we derive the $u$-independent saddles $\Phi^*= \Phi^*(\varphi)$ of the first-order truncated boundary theory.

For this ansatz, the non-zero components of the combined stress tensor \eqref{eqn: combine Tij} are
\begin{equation}
    \tilde T_{uu}=\frac{1}{4\pi\mathfrak b^2\r^2}
    \left(\frac{4\pi^2}{\b^2}-(\p_\varphi\Phi^{*})^2+2\p_\varphi^2\Phi^{*}\right)\,,\qquad
    \tilde T_{\varphi\varphi}=-\frac{1}{4\pi\mathfrak b^2}
    \left(\frac{4\pi^2}{\b^2}-(\p_\varphi\Phi^{*})^2\right)\,.
\end{equation}
Since the tensor is diagonal, $\tilde T\bar{\tilde T}=-\tfrac{1}{4\r^2}\tilde T_{uu}\tilde T_{\varphi\varphi}$. The field equation therefore reduces to
\begin{equation}
\p_\varphi^2\Phi^{*}+4\pi\mathfrak b^2\mu \r^2 e^{2\Phi^*}-\frac{\lambda}{16\pi\mathfrak b^2\r^2}
    \left(\frac{4\pi^2}{\b^2}-(\p_\varphi\Phi^{*})^2\right)
    \left(\frac{4\pi^2}{\b^2}-(\p_\varphi\Phi^{*})^2+2\p_\varphi^2\Phi^{*}\right)e^{-2\Phi^*}=0\,.
\end{equation}
Multiplying by $\p_\varphi\Phi^{*}$, this equation can be integrated to a first-order equation \eqref{eqn: lin def eom 2}, repeated here for convenience,
\begin{equation}\label{eqn: lin def eom 2 appen}
    (\p_\varphi\Phi^{*})^2+4\pi\mathfrak b^2\mu \r^2 e^{2\Phi^*}
    +\frac{\lambda}{16\pi\mathfrak b^2 \r^2}
    \left(\frac{4\pi^2}{\b^2}-(\p_\varphi\Phi^{*})^2\right)^2e^{-2\Phi^*}
    =\frac{4\pi^2(1-\chi)}{\b^2}\,,
\end{equation}
where $\chi$ is an integration constant.

\textbf{Turning points and the smooth periodic branch.}
At an extremum, $\p_\varphi\Phi^{*}=0$, \eqref{eqn: lin def eom 2 appen} becomes
\begin{equation}
    4\pi\mathfrak b^2\mu\r^2\,e^{4\Phi^*}
    -\frac{4\pi^2(1-\chi)}{\b^2}e^{2\Phi^*}
    +\frac{\lambda}{16\pi\mathfrak b^2 \r^2}
    \left(\frac{4\pi^2}{\b^2}\right)^2=0\,.
\end{equation}
The roots of this equation lead to the maximum and minimum of $\Phi^*$, which are given by
\begin{equation}\label{eqn: lin def max/min again}
    e^{2\Phi_\pm^*}=\frac{1}{8\pi\mathfrak b^2\mu \r^2}
    \left(\frac{2\pi}{\b}\right)^2
    \left(1-\chi\pm\sqrt{(1-\chi)^2-\lambda\mu}\right)\,.
\end{equation}
The reality condition for $\Phi^*$ requires
\begin{equation}\label{eqn: lin def bound again}
    \sqrt{\lambda\mu} \leq1-\chi\,.
\end{equation}
When this inequality is saturated, the two roots coincide, yielding a constant solution, 
\begin{equation}
    e^{2\Phi^*}=\frac{1}{8\pi\mathfrak b^2\r^2}
    \left(\frac{2\pi}{\b}\right)^2\sqrt{\frac{\lambda}{\mu}}\, ,
\end{equation}
in agreement with \cite{Allameh:2025gsa}.

\textbf{Parametric solution and periodicity.}
For the non-constant solution, it is useful to introduce parameters
\begin{equation}\label{eqn: bdy m m' def appen}
    m'\equiv\frac{2\sqrt{(1-\chi)^2-\lambda\mu}}
    {-\chi+\sqrt{(1-\chi)^2-\lambda\mu}+\sqrt{1-\lambda\mu}}\,,\qquad m\equiv\frac{2\sqrt{(1-\lambda\mu)((1-\chi)^2-\lambda\mu)}}
    {1-\chi-\lambda\mu+\sqrt{(1-\lambda\mu)((1-\chi)^2-\lambda\mu)}} \, .
\end{equation}
The general solution admits a closed form expression by introducing an auxiliary variable $y$,
\begin{equation}\label{eqn: lin def sol}
\begin{aligned}
    e^{\Phi^*(y)}&=e^{\Phi_+^*}
    \left[\frac{m'}{m}\operatorname{dn}(y|m)
    +\frac{1-m'/m}{\operatorname{dn}(y|m)}\right]\,,\\
    \varphi(y)&=\frac{\b}{2\pi}\sqrt{\frac{1-m'}{m'(m-m')}}
    \left[my-2(m-m')\Pi(m',\operatorname{am}(y|m)|m)\right]\,,
\end{aligned}
\end{equation}
where $\Phi^*_+$ is the maximum of $\Phi^*$ given by \eqref{eqn: lin def max/min again}, $\text{dn}(y|m)$ and $\text{am}(y|m)$ are the Jacobi elliptic function and amplitude, and $\Pi(m',\alpha|m)$ is the incomplete elliptic integral of the third kind. To avoid self-intersecting and/or complex solutions, we further require
\begin{equation}
    1-\chi < \frac{1+\lambda \mu}{2} \, ,
\end{equation}
which implies that $0<m'<m<1$.

The maxima occur at $y=2nK(m)$ and the minima at $y=(2n+1)K(m)$, with $n\in\mathbb Z$. 

Imposing that $\varphi(y+2K(m))-\varphi(y)=2\pi$ leads to
\begin{equation}\label{eqn: lin def beta appen}
    \b^{-1}=\frac{n}{2\pi^2}\sqrt{\frac{1-m'}{m'(m-m')}}
    \left[mK(m)-2(m-m')\Pi(m'|m)\right]\,.
\end{equation}
This relation fixes $\chi$ implicitly in terms of $\b$ and $\lambda\mu$. 

\bibliography{bdryrefs}

\end{document}